\documentclass[aps,a4paper,onecolumn,amsmath,amsfonts,preprintnumbers,nofootinbib,superscriptaddress,eqsecnum,secnumarabic]{revtex4-2}

\usepackage[T1]{fontenc} 
\usepackage{amsmath}
\usepackage{amssymb}
\usepackage{amsfonts}
\usepackage{mathtools}
\usepackage{microtype}
\usepackage{tikz}
\usepackage{bm}
\usepackage[utf8]{inputenc}
\usepackage{blkarray}
\usepackage{bigstrut}
\usepackage{nccmath}
\usepackage{enumitem}
\usepackage{braket}
\usepackage{dsfont}
\usepackage{lipsum}
\usepackage{array,multirow}
\usepackage{slashed}
\usepackage{physics}
\usepackage{xcolor}
\usepackage{color}
\usepackage{float}
\usepackage{subfig}
\usepackage{makecell}
\usepackage[export]{adjustbox}

\usepackage{caption}
\usepackage{ragged2e}
\DeclareCaptionJustification{justified}{\justifying}
\newcommand{\Den}{\ensuremath D}
\newcommand{\nn}{\nonumber \\}

\def\eps{\epsilon}

\newcommand{\GG}{\ensuremath \text{I}}
\newcommand{\GGvec}{\ensuremath \mathbf{I}}

\newcommand{\pts}[1]{\phantom{.}\hfill(\textit{#1~point}\ifthenelse{\equal{#1}{1}}{}{\textit{s}})}

\newcommand{\Msun}{{\ifmmode{{\rm{M_{\odot}}}}\else{${\rm{M_{\odot}}}$}\fi}}

\newcommand{\beq}{\begin{equation}}
\newcommand{\eeq}{\end{equation}}
\newcommand{\bea}{\begin{eqnarray}}
\newcommand{\ena}{\end{eqnarray}}

\newcommand{\lsim}{\mathrel{\mathop{\kern 0pt \rlap
{\raise.2ex\hbox{$<$}}}
\lower.9ex\hbox{\kern-.190em $\sim$}}}
\newcommand{\gsim}{\mathrel{\mathop{\kern 0pt \rlap
{\raise.2ex\hbox{$>$}}}
\lower.9ex\hbox{\kern-.190em $\sim$}}}

\newcommand{\aem}{\alpha_{\rm em}^2}

\definecolor{hjaltegreen}{rgb}{0.0,0.5,0.0}

\def \be {\begin{equation}}
\def \ee {\end{equation}}
\newcommand{\eea}{\end{eqnarray}}
\newcommand{\bei}{\begin{itemize}}
\newcommand{\eei}{\end{itemize}}

\setlist[itemize,1]{leftmargin=1.5em}

\begin{document}

\preprint{ZU-TH 21/21}

\title{\boldmath Two-photon exchange in leptophilic dark matter scenarios}

\author{Raghuveer Garani}
\email{garani@fi.infn.it}
\affiliation{INFN Sezione di Firenze, Via G. Sansone 1, I-50019 Sesto Fiorentino, Italy}
\author{\ Federico Gasparotto}
\email{federico.gasparotto@pd.infn.it}
\affiliation{Dipartimento di Fisica e Astronomia, Universit\`a di Padova, Via Marzolo 8, 35131 Padova, Italy}
\affiliation{INFN, Sezione di Padova, Via Marzolo 8, 35131 Padova, Italy}
\author{\ Pierpaolo Mastrolia} 
\email{pierpaolo.mastrolia@pd.infn.it}
\affiliation{Dipartimento di Fisica e Astronomia, Universit\`a di Padova, Via Marzolo 8, 35131 Padova, Italy}
\affiliation{INFN, Sezione di Padova, Via Marzolo 8, 35131 Padova, Italy}
\author{\\ Henrik J. Munch}
\email{henrik.munch@pd.infn.it}
\affiliation{Dipartimento di Fisica e Astronomia, Universit\`a di Padova, Via Marzolo 8, 35131 Padova, Italy}
\affiliation{INFN, Sezione di Padova, Via Marzolo 8, 35131 Padova, Italy}
\author{\ Sergio Palomares-Ruiz}
\email{sergiopr@ific.uv.es}
\affiliation{Instituto de F{\`i}sica Corpuscular (IFIC), CSIC-Universitat de Val{\`e}ncia,
Apartado de Correos 22085, E-46071 Val{\`e}ncia, Spain}
\author{\ Amedeo Primo}
\email{aprimo@physik.uzh.ch}
\affiliation{Department of Physics, University of Z{\"u}rich, CH-8057 Z{\"u}rich, Switzerland}\textbf{}

\begin{abstract}
In leptophilic scenarios, dark matter interactions with nuclei, relevant for direct detection experiments and for the capture by celestial objects, could only occur via loop-induced processes. If the mediator is a scalar or pseudo-scalar particle, which only couples to leptons, the dominant contribution to dark matter-nucleus scattering would take place via two-photon exchange with a lepton triangle loop. The corresponding diagrams have been estimated in the literature under different approximations. Here, we present new analytical calculations for one-body two-loop and two-body one-loop interactions. The two-loop form factors are presented in closed analytical form in terms of generalized polylogarithms up to weight four. In both cases, we consider the exact dependence on all the involved scales, and study the dependence on the momentum transfer. We show that some previous approximations fail to correctly predict the scattering cross section by several orders of magnitude. Moreover, we show that form factors, in the range of momentum transfer relevant for local galactic dark matter, are smaller than their value at zero momentum transfer, which is usually considered. 
\end{abstract}

\maketitle

\section{Introduction}
\label{sec:introduction}

Precise perturbative calculations in quantum field theory, within the Standard Model (SM) and in theories beyond the SM (BSM), require the computation of loop integrals. Over the past decades, many elaborate computational tools for the calculation of multi-loop Feynman diagrams have been developed (see for reviews, e.g., Refs.~\cite{Argeri:2007up, Grozin:2011mt, Smirnov:2012gma, Henn:2014qga, Duhr:2014woa}) and have been applied for the calculation of many processes. 

\begin{figure}[t]
  \centering
  \includegraphics[width=0.43\linewidth]{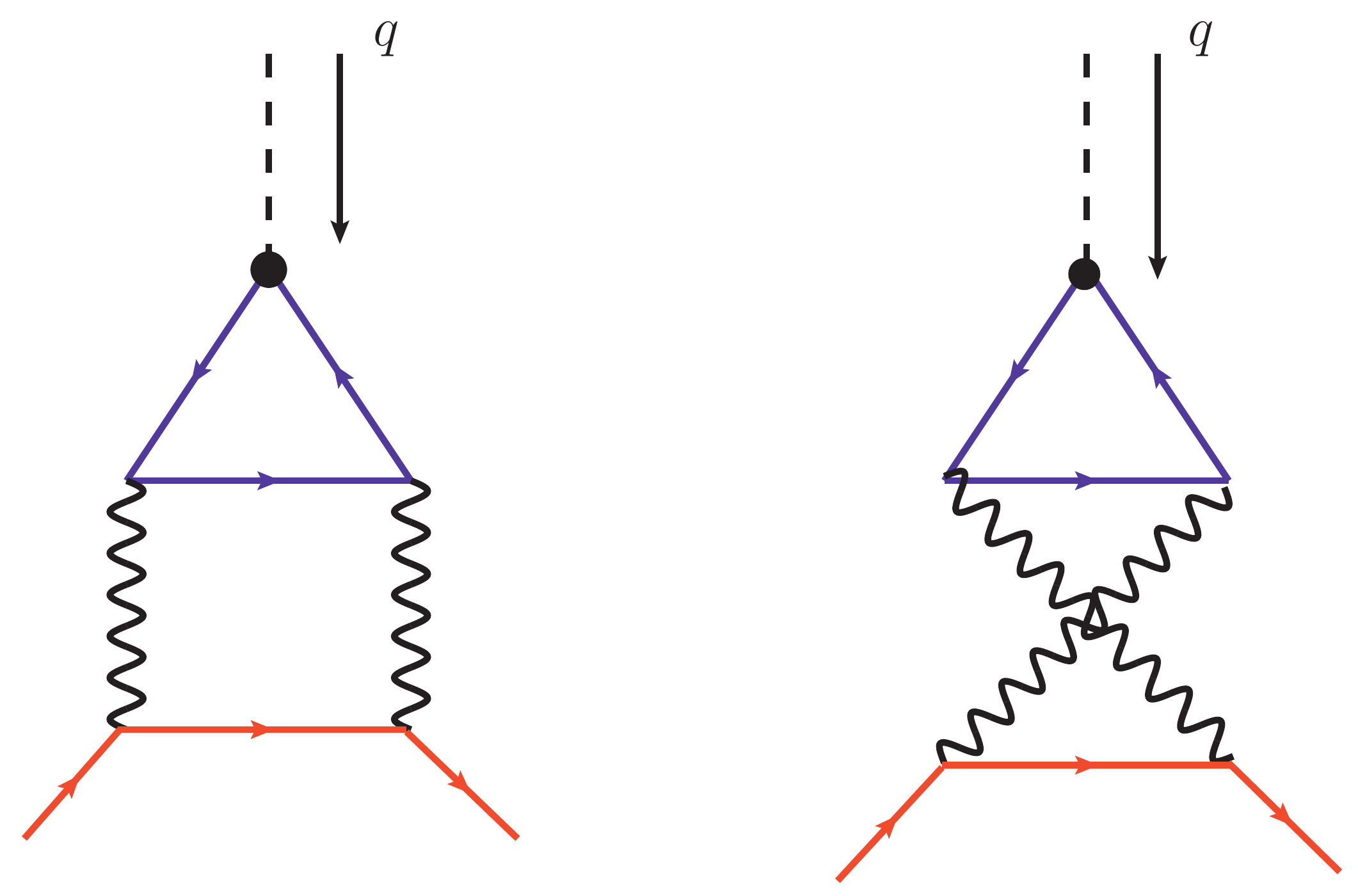} \hspace{1.5cm}
  \includegraphics[width=0.43\linewidth]{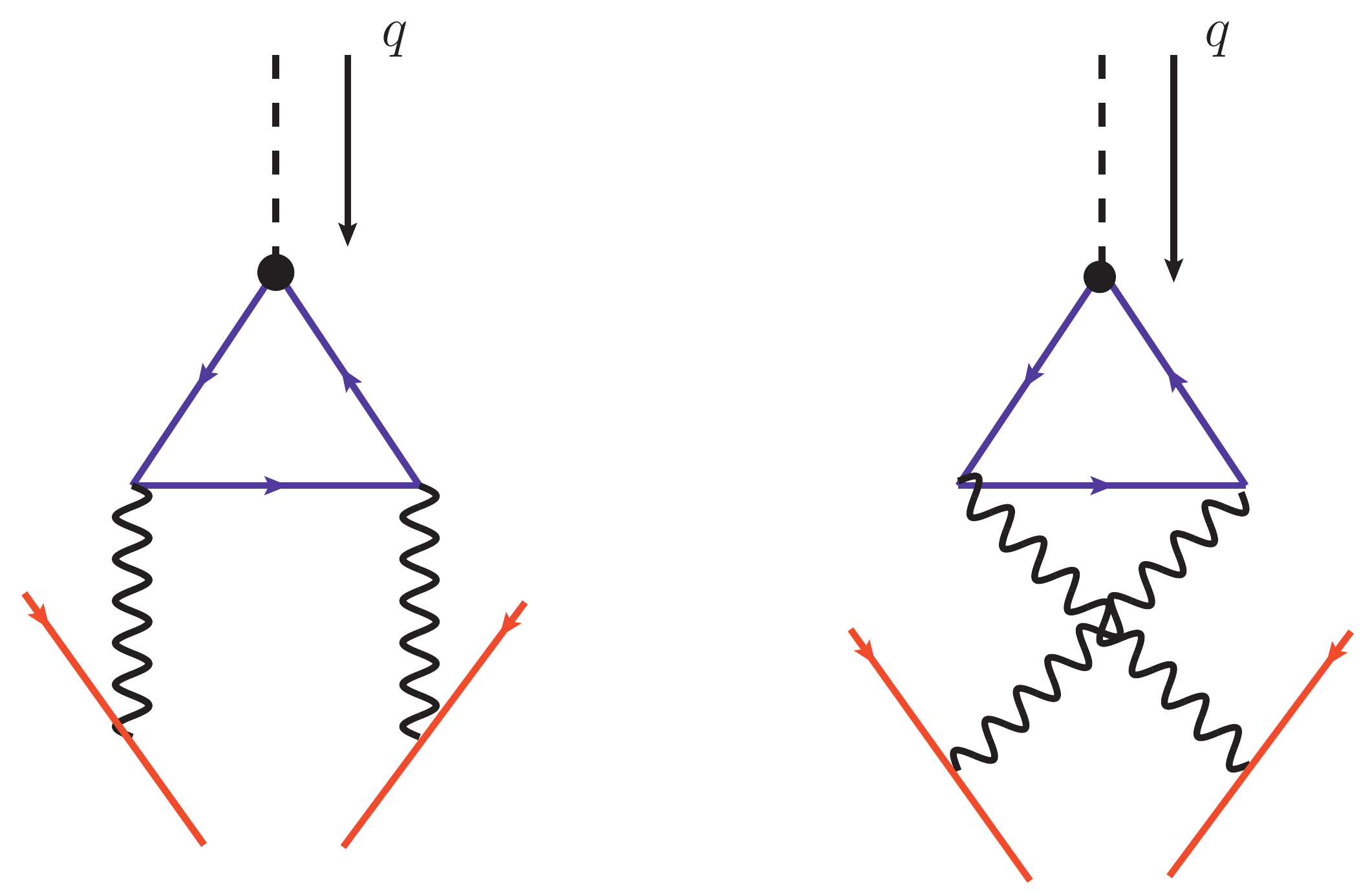}
  \caption{\textbf{\textit{Two-boson exchange diagrams for interactions between a scalar or pseudo-scalar boson and two external fermions}}, $\phi(q) \, \bar{\psi}(p') \, \Lambda_{S/P} \, \psi(p)$ ($\Lambda_S =\mathbb{I}$ and $\Lambda_P= \gamma_5$), via a lepton triangle loop which exchanges two bosons with an external fermion: one-body {\it two-loop} interactions and two-body {\it one-loop} interactions.}  	
  \label{fig:2photon}
\end{figure}

The diagrams in Fig.~\ref{fig:2photon}, with generic internal fermion and boson masses, often contribute to SM and BSM observables. Within perturbative quantum chromodynamics (QCD), the two-loop vertex graphs appear in second-order corrections to the form factors involving heavy quarks, and have been considered in the calculation of vector and axial-vector form factors in the heavy quark forward-backward asymmetry~\cite{Altarelli:1992fs, Ravindran:1998jw, Catani:1999nf,  Bernreuther:2004ih, Bernreuther:2004th, Bonciani:2008az, Ablinger:2017hst} or of the scalar and pseudo-scalar form factors in Higgs decays to heavy quarks~\cite{Bernreuther:2005rw, Bernreuther:2005gw, Ablinger:2017hst, Bernreuther:2018ynm, Primo:2018zby, Behring:2019oci, Mondini:2020uyy}. Other static quantities, such as anomalous magnetic and electric moments, also receive contributions from similar two-loop diagrams~\cite{Heinemeyer:2004yq, Stockinger:2006zn, Miller:2007kk, Davidson:2010xv, Ilisie:2015tra, Hisano:2018bpz, Dong:2019iaf} via, e.g., Barr-Zee diagrams~\cite{Barr:1990vd} (involving bosons with different masses in the loop). On the other hand, non-relativistic QCD~\cite{Caswell:1985ui}, which describes the dynamics of heavy quark-antiquark pairs at energy scales well below their masses, provides accurate predictions for, e.g., the decay rates and binding energies of heavy quarkonia and for top-pair production near threshold in electron-positron annihilation~\cite{Brambilla:2004jw, Pineda:2011dg}. Two-loop diagrams with the generic shape of those in Fig.~\ref{fig:2photon} also appear in the calculation of several matching coefficients between QCD and non-relativistic QCD at two-loop order~\cite{Gerlach:2019bso}. Within the context of the quark-loop model~\cite{Pratap:1972tb} (known to be equivalent to vector meson dominance models~\cite{Bergstrom:1983ay}), these diagrams represent the dominant contribution in the calculation of rare pseudo-scalar meson decays into leptons~\cite{Pratap:1972tb, Ametller:1983ec, Pich:1983zk}.

Moreover, the decay of a scalar or pseudo-scalar boson into quarks has been investigated in great detail, within the context of the SM Higgs boson or of BSM non-standard Higgs bosons. Early computations of second-order corrections to these decays have been obtained for the case of massless quarks~\cite{Gorishnii:1990zu} and in expansions in terms of the ratio of the external quarks and boson masses~\cite{Surguladze:1994gc, Surguladze:1994em, Chetyrkin:1995pd, Chetyrkin:1996ke, Larin:1995sq, Harlander:1997xa, Chetyrkin:1998ix}. Analytical expressions for the two-loop diagrams in Fig.~\ref{fig:2photon}, where two photons are exchanged, were earlier derived for the cases when the mass of the external fermion is either vanishing or equal to that of the particles running in the triangle loop (equal-mass case)~\cite{Kotikov:1990kg, Fleischer:1998nb, Gehrmann:1999as, Bonciani:2003hc, Davydychev:2003mv, Bernreuther:2005rw, Bernreuther:2005gw, Ablinger:2017hst}. More recently, analytical results for the different-mass case have been obtained in the context of Higgs decays into bottom~\cite{Primo:2018zby} and charm quarks~\cite{Mondini:2020uyy}.

Interestingly, the same graphs are also encountered in theoretical scenarios of particle dark matter (DM). Indeed, the diagrams shown in Fig.~\ref{fig:2photon} can represent the dominant contribution for interactions with nucleons of a particular class of DM candidates, which have no tree-level couplings to quarks, but only to leptons. These leptophilic scenarios have been proposed to reduce the tension of the DM interpretation among different direct detection experiments~\cite{Bernabei:2007gr, Dedes:2009bk, Kopp:2009et, Feldstein:2010su, Chang:2014tea, Bell:2014tta, Foot:2014xwa, Roberts:2016xfw}, to solve some cosmic-ray anomalies~\cite{Fox:2008kb, Cao:2009yy, Bi:2009uj, Ibarra:2009bm, Cohen:2009fz, Cavasonza:2016qem, Duan:2017pkq, Athron:2017drj, Duan:2017qwj, Ghorbani:2017cey, Han:2017ars, YaserAyazi:2019psw}, to explain the gamma-ray galactic center excess~\cite{Lu:2016ups}, to alleviate the tension in the measured muon magnetic moment~\cite{Agrawal:2014ufa, Bandyopadhyay:2017tlq, Chen:2018vkr}, as potential signals in collider searches~\cite{Buckley:2015cia, DEramo:2017zqw, Rawat:2017fak, Madge:2018gfl, Marsicano:2018vin, Junius:2019dci} or in gravitational wave detectors~\cite{Madge:2018gfl}, as a way to capture DM in celestial objects~\cite{Kopp:2009et, Garani:2017jcj, Liang:2018cjn, Bell:2019pyc, Garani:2019fpa, Joglekar:2019vzy, Joglekar:2020liw, Bell:2020lmm} and, within the context of sub-GeV DM, to discuss future strategies of direct detection~\cite{Essig:2011nj, Chen:2015pha, Lee:2015qva, Essig:2015cda}, as a source of distortion of the cosmic-ray electron spectrum~\cite{Cappiello:2018hsu} or as a way to produce DM up-scatterings to relativistic energies~\cite{An:2017ojc, Ema:2018bih, Cappiello:2019qsw}.

If the DM particle is leptophilic, its tree-level coupling to fermions involves just leptons, but not quarks. Therefore, the scattering cross section off nuclei (or off quarks) could only proceed via loop-induced interactions. In the case of a scalar or pseudo-scalar mediator, the one-body one-loop contribution (with a lepton loop) to the DM-quark/nucleon coupling is suppressed by the masses of electroweak gauge bosons and the leading interactions with nuclei arise from diagrams as those in Fig.~\ref{fig:2photon}, with two photons exchanged with external protons. The two left diagrams represent one-body two-loop interactions and the two right diagrams, two-body one-loop processes. In such scenarios, a mediator particle connects DM to SM particles, so DM-nucleon elastic scattering proceeds via a lepton triangle loop and two photons. Thus, in some leptophilic scenarios, those diagrams could govern the capture of DM in celestial objects and the detection rates in terrestrial experiments.

There have been attempts to compute the amplitude of the above diagrams in the context of detection of leptophilic DM. In Refs.~\cite{Kopp:2009et, Frandsen:2012db}, an effective operator product expansion approach was discussed. After integrating out the lepton loop, an effective interaction term $\bar{\chi} \chi F_{\mu \nu} F^{\mu \nu}$ was obtained (with $\chi$ being the DM field).\footnote{This coupling is analogous to that in Rayleigh DM scenarios~\cite{Weiner:2012cb}, although different UV completions can give rise to it without a fermion triangle loop~\cite{Weiner:2012gm, Kavanagh:2018xeh}.} In Ref.~\cite{Kopp:2009et}, this was then matched onto the external nucleus current to obtain an effective DM-nucleus interaction in the heavy-nucleon limit, which actually represents a two-body process~\cite{Ovanesyan:2014fha}. However, this approximation is not reliable when the momentum transfer is comparable to the fermion masses in the triangle loop~\cite{Kopp:2009et}. Furthermore, we show that this is not appropriate in general. On the other hand, in Ref.~\cite{Bell:2019pyc}, direct integration of the one-body two-loop amplitude is performed in the limit of vanishing momentum transfer and an approximate analytical expression is reported. 

In this work, we present the full analytical calculation of the one-body two-loop and two-body one loop diagrams (Fig.~\ref{fig:2photon}), keeping the complete dependence on the kinematic variables, namely the momentum transfer of the mediator and the two fermion masses. We consider Feynman diagrams in dimensional regularization, with $d = 4-2\epsilon$ dimensions (for a small, positive $\epsilon$ parameter), and exploit the integration-by-parts identities (IBPs)~\cite{Chetyrkin:1981qh}, to express the scalar and pseudo-scalar form factors in terms of a set of independent master integrals (MIs)~\cite{Laporta:2001dd, vonManteuffel:2012np}, recently computed in~\cite{Primo:2018zby, Mondini:2020uyy}. Then, we consider the four dimensional limit, by taking $\epsilon \to 0$, where the form factors are finite, and express the results of the one-body diagrams in terms of generalized polylogarithms (GPLs)~\cite{Goncharov:polylog, Remiddi:1999ew, Gehrmann:2001pz, Gehrmann:2001jv, Vollinga:2004sn,Duhr:2019tlz}. Besides the expressions corresponding to general kinematics, we provide the analytic expressions of the form factors in two limits: the {\it soft limit}, for vanishing momentum transfer of the mediator; and the {\it equal-mass} limit, for internal fermions with equal masses. We use the novel results of the two-loop one-body and of the one-loop two-body diagrams to investigate the leading contribution to the scattering of leptophilic DM off a few sample of nuclei, like helium, oxygen, iron and xenon, relevant for direct DM detection experiments and for the capture of DM by celestial objects. Finally, we critically compare our results with previous estimates, finding that the exact calculation of the form factors presented here turns out to be crucial for phenomenological analyses, and that naive approximations of the loop integrals might yield incorrect conclusions.

This article is organized as follows. We derive the relevant form factors for scalar and pseudo-scalar interactions for the one-body and the two-body processes, write down the amplitudes and obtain the analytical solution for the form factors of one-body interactions in terms of GPLs in Section~\ref{sec:1body}. The calculation of the contribution to the amplitude from two-body interactions is described in Section~\ref{sec:2body}. Thus, Section~\ref{sec:Amplitudes_and_Form_Factors} represents the core of this paper, where we describe and discuss the main results. We compare our results to the approximate results available in the literature in Section~\ref{sec:Comparison_and_discussion}. Finally, in Section~\ref{sec:conclusions} we draw our conclusions. Details about the calculations of two-loop diagrams are presented in Appendices~\ref{sec:AppendixA},~\ref{sec:FS_limit} and~\ref{sec:FP_limit}. 

\section{Amplitudes and form factors}
\label{sec:Amplitudes_and_Form_Factors}

We consider the case of a DM particle ($\chi$) that couples to a SM singlet scalar ($\phi_S$) or pseudo-scalar ($\phi_P$), which in turn, couples to SM leptons ($\ell$) at tree level, but not to quarks, through interaction terms of the form
\begin{equation}
- {\cal L}_S \supset g_S \, \phi_S \, \bar{\ell} \, \ell + g_\chi \, \phi_S \, \Gamma_\chi \hspace{1cm} \textrm{or} \hspace{1cm} - {\cal L}_P \supset i \, g_P \, \phi_P \,  \bar{\ell} \, \gamma_5 \, \ell + g_\chi \, \phi_P \, \Gamma_\chi ~.
\end{equation}
where $\Gamma_{\chi} =\{\chi^\dagger \chi\}$ (scalar DM) or $\{\bar{\chi} \, \chi, \bar{\chi} \, \gamma_5 \, \chi\}$ (fermion DM). The leading interactions with nucleons are induced by the exchange of two virtual photons. In general, the two virtual photons can scatter off a single nucleon~\cite{Kopp:2009et, Frandsen:2012db} or two nucleons~\cite{Weiner:2012cb, Ovanesyan:2014fha}, and these two types of processes govern the elastic scattering of a DM particle off a nucleus $A$, with the total amplitude given by the sum of their individual amplitudes,
\begin{equation}
{\cal M}_\beta = \langle A(p') | \, \mathcal{L} \, | A(p) \rangle = {\cal M}^{(1)}_\beta + {\cal M}^{(2)}_\beta ~,
\end{equation}
with $\beta = \{S, P\}$ for scalar or pseudo-scalar interactions. 

The computation of these contributions is the main goal of this paper and we discuss them next.

\subsection{One-body interactions} 
\label{sec:1body}

The leading one-body interactions with quarks/nucleons are induced at two-loop level through the two left diagrams shown in Fig.~\ref{fig:2photon}, where SM charged leptons that run in the triangle loop exchange two photons with an external quark/nucleon. As mentioned above, our main target is the scattering of non-relativistic DM particles off nucleons and nuclei. In principle, one could proceed in terms of interactions with quarks and then match to the effective theory that would describe the interaction of non-relativistic DM with non-relativistic nucleons. However, in order to do so, the dressed quark propagators (and vertices) within the nucleon would have to be used. Thus, for the low energies of interest, for which the internal structure of nucleons is not probed, it is more convenient to directly consider the well-known coupling of photons to nucleons. This is given by the electromagnetic current for a nucleon, which is generically written as
\begin{equation}
\langle N(p') | J^\mu (q^2) | N(p) \rangle = \bar{u}_N(p') \, \left( \gamma^\mu \, F_1(q^2) + i \, \sigma^{\mu \nu} \frac{q_\nu}{2 m_N} \, F_2(q^2) \right) \, u_N(p) ~, 
\label{eq:EMFF}
\end{equation}
where $q^2 = (p' - p)^2 \equiv t$ is the squared of the exchanged momentum, $m_N$ is the nucleon mass, and $F_1(q^2)$ and $F_2(q^2)$ are the Dirac and Pauli form factors. At $q^2 = 0$, $F_1(0)$ represents the electric charge of the nucleon and $F_2(0)$ its anomalous magnetic moment. In the non-relativistic limit (i.e., $q/m_N < 1$) relevant for the problem under discussion, $F_1(q^2) \simeq F_1(0)$ and $F_2(q^2) \simeq F_2(0)$. Moreover, the term involving $F_2(q^2)$ is proportional to $q/m_N$ and thus, can also be neglected at the lowest order. Therefore, in the non-relativistic limit, the electromagnetic coupling of photons to nucleons is analogous to that of quarks, the only difference being the charge. Obviously, at this order, only protons contribute. For the sake of comparison with other works, however, we will make this simplification for the photon-nucleon vertex, but we will keep the rest of the treatment fully relativistic.

The amplitude of the two-loop vertex diagrams in Fig.~\ref{fig:2photon} can be written, in terms of the interaction with nucleons, as
\begin{equation}
\tilde{\mathcal{M}}_{\beta}^{(1)} (t; m_N) = i \, g_\beta \, Q_N^2 \, \sum_{\ell = e, \mu, \tau} Q_\ell^2  \, \left( \bar u_N(p') \, \Gamma_\beta (t; m_N, m_\ell) \, u_N(p) \right) ~,
\label{eq:M}
\end{equation}
where $m_\ell$ ($Q_\ell = 1$) and $m_N$ ($Q_N = \{1, 0\}$, for $N = {p, n}$) are the masses (electric charges) of lepton $\ell$ and nucleon $N$, respectively, and $p^2 = p'^2 = m_N^2$. The operator $\Gamma_\beta(t, m_N, m_\ell)$ results from the sum of the two diagrams, which give identical contributions,
\begin{equation}
\Gamma_\beta (t; m_N, m_\ell) = 
- 32 \, \pi^2 \, \alpha_{\rm em}^2 \, \int \frac{{\dd^d k_1}}{(2\pi)^4} \frac{{\dd^d k_2}}{(2\pi)^4} \frac{g_{\mu \rho} \, \gamma ^{\rho} \, \left(\slashed{k_2} - m_N\right) \, g_{\nu \sigma} \, \gamma^{\sigma} \, \text{Tr}^{\mu\nu}_\beta(q, k_1,k_2; m_N, m_\ell)}{D_1 \, D_2 \, D_3 \, D_4 \, D_5 \, D_6} ~, 
\label{eq:2LAmp}
\end{equation}
where $\alpha_{\rm em} = e^2/(4 \, \pi)$ and $\text{Tr}^{\mu\nu}_\beta$ is the Dirac trace
\begin{equation}
\text{Tr}^{\mu\nu}_\beta (t, k_1, k_2; m_N, m_\ell) = 
\text{Tr}\Big\{ \left(\slashed{k_1} + \slashed{q} + m_\ell\right) 
\Lambda_\beta 
\left(\slashed{k_1} + m_\ell\right) 
\gamma ^{\mu} 
\left(\slashed{k_1} + \slashed{k_2} + \slashed{p}' + m_\ell\right)
\gamma^{\nu}\Big\} ~,
\label{eq:trace}
\end{equation}
with $\Lambda_S = \mathbb{I}$ and $\Lambda_P = \gamma_5$. The inverse propagators $D_i$ in Eq.~\eqref{eq:2LAmp} are
\begin{gather}
\Den_1 = k_1^2 - m_\ell^2 ~,\quad
\Den_2 = (k_1 + q)^2 - m_\ell^2 ~, \quad
\Den_3 = k_2^2 - m_N^2 ~,  \nn
\Den_4= (k_2 + p)^2 ~, \quad
\Den_5 = (k_2 + p')^2 ~, \quad
\Den_6 = (k_1 + k_2+p')^2 - m_\ell^2 ~.
\label{eq:Ds}
\end{gather}  

From Lorentz invariance, the operators $\Gamma_\beta$ must have the form
\begin{equation}
\Gamma_\beta (t; m_N, m_\ell) = A_\beta (t; m_N, m_\ell) \, \Lambda_\beta + B_\beta (t; m_N, m_\ell) \, \left( \slashed{p}^\prime + \slashed{p} \right) \, \Lambda_\beta + C_\beta (t; m_N, m_\ell) \, \left( \slashed{p}^\prime - \slashed{p} \right) \, \Lambda_\beta ~,
\label{eq:amp}
\end{equation}
where the form factors $A_\beta$, $B_\beta$ and $C_\beta$ are scalar functions of the square of the momentum transfer, $t$, as well as of the fermion masses $m_N$ and $m_\ell$. In order to write the effective interaction term $\phi_\beta \bar{N} \Lambda_\beta N$ in the Lagrangian, we use the Dirac equation for the external fermions. That is, we make the following identifications: 
\begin{align}
\bar{u}_N(p^\prime) \, \Lambda_\beta \, u_N(p) & \to \bar{N}(p^\prime) \, \Lambda_\beta \, N(p)  ~, \nn
\bar{u}_N(p^\prime) \, (\slashed{p}^\prime + \slashed{p}) \, \left(1 + \gamma_5\right) \, u_N(p) & \to 2 \, m_N \, \bar{N}(p')N(p) ~, \nn
\bar{u}_N(p^\prime) \, (\slashed{p}^\prime - \slashed{p}) \, \left(1 + \gamma_5\right) \, u_N(p) & \to 2 \, m_N \, \bar{N}(p') \, \gamma_5 \, N(p) ~.
\end{align}
In this way, this dark mediator-nucleon interaction can be written as 
\begin{equation}
\mathcal{L}_{\beta} \supset g_\beta \, Q_N^2 \, \left[ \sum_{\ell = e, \mu, \tau} Q_\ell^2 \, {\cal F}_\beta^{\rm 1b} (t; m_N, m_\ell) \right] \phi_\beta \, \bar{N}  \,\Lambda_\beta \, N ~,
\end{equation}
where ${\cal F}_S^{\rm 1b} = A_S + 2 \, m_N \, B_S$ and ${\cal F}_P^{\rm 1b} = A_P + 2 \, m_N \, C_P$. 

Since we are interested in the elastic scattering of DM particles off nuclei in the non-relativistic limit, we can write the one-body contribution to the DM-nucleus amplitude, for small momentum transfer, as
\begin{equation}
{\cal M}^{(1)}_\beta (\boldsymbol{q}) = \langle A(p') | \mathcal{L}_{\beta} | A(p) \rangle = G \, g_\beta \, {\cal N}  \, \left[ \sum_{\ell = e, \mu, \tau} {\cal F}_\beta^{\rm 1b} (t; m_N, m_\ell) \right] \, F^{(1)}\left(\frac{\boldsymbol{q}^2}{Q_0^2}\right) ~,
\label{eq:M1}
\end{equation}
where $G$ includes the $\phi_\beta$ propagator and the DM-$\phi_\beta$ coupling $g_\chi$, ${\cal N}$ depends on the type of interaction, and $F^{(1)}\left(-\boldsymbol{q}^2/Q_0^2\right)$ is the nuclear form factor, which takes into account the coherence of the interaction. In the non-relativistic limit, it depends on the square of the three-momentum transfer $\boldsymbol{q}^2 \simeq - t$, and throughout this work, it is approximated as
\begin{equation}
F^{(1)}(\boldsymbol{q}^2/Q_0^2) = e^{-\boldsymbol{q}^2/Q_0^2} .
\end{equation}
The latter expression of the form factor can be adapted both to the scalar (spin-independent) and to pseudo-scalar (spin-dependent) cases, by choosing a proper definition of $Q_0$ (which would be different for the two cases). For the scalar case, $1/Q_0$ represents the radius of the nucleus, $Q_0 = 0.48 \, (0.3 + 0.89 \, A^{1/3})^{-1}$~GeV. Also note that, in the non-relativistic limit, the nuclear amplitude only depends on the number of protons in the nucleus, ${\cal N} = Z$, with $Z$ the atomic number of the nucleus. 

The form factors ${\cal F}_\beta^{\rm 1b}$ can be extracted from $\Gamma_\beta$ by projection,
\begin{equation}
\label{eq:traceF}
{\cal F}_\beta^{\rm 1b} (t; m_N, m_\ell) = 
\frac{1}{2 \left(p^\prime \pm p\right)^2} \, 
\text{Tr}
\left\{
\Lambda_\beta   \left( \slashed{p} \pm m_N \right) \, 
\Gamma_\beta \, \left( \slashed{p}^\prime \pm m_N \right) 
\right\} , 
\end{equation}
where the sign $+$ corresponds to the scalar case ($\Lambda_S = \mathbb{I}$) and the sign $-$ to the pseudo-scalar case ($\Lambda_P = \gamma_5$). The two form factors are both UV and IR finite in four dimensions, so that Eq.~\eqref{eq:traceF} could be directly evaluated in $d=4$. 
However, as we describe in the following, it is technically more convenient to exploit the algebraic properties of Feynman integral in dimensional regularization, and to recover the four dimensional result at the end of the calculation.
Fermionic traces and loop integration are performed, therefore, in arbitrary $d$ dimensions, by adopting the prescription of Refs.~\cite{tHooft:1972tcz, Akyeampong:1973vj, Larin:1993tq} for the $d$-dimensional treatment of $\gamma_5$.

Upon the calculation of the fermion traces, the form factors ${\cal F}_\beta^{\rm 1b} (t; m_N, m_\ell)$ are expressed as linear combinations of two-loop Feynman integrals of the type
\begin{equation}
\label{eq:family}
{\cal F}_\beta^{\rm 1b} (t; m_N, m_\ell) = \sum_{\vec{n}} c_{\vec{n}, \beta}(m_i^2,t,\eps)\int \frac{{\dd^d k_1}}{(2\pi)^d} \frac{{\dd^d k_2}}{(2\pi)^d} \,
\frac{D_7^{n_7}}{\Den_{1}^{n_1}\Den_{2}^{n_2}\Den_{3}^{n_3}\Den_{4}^{n_4} \Den_{5}^{n_5}\Den_{6}^{n_6}}\,, \quad n_i\in \mathbb{N} ~,
\end{equation}
where $c_{\vec{n}}$ are coefficients that depend on the kinematic variables and on the dimensional regulator $\eps = (4 - d)/2$. The auxiliary inverse propagator $\Den_7 = (k_1-p)^2$ has been introduced to complement the six denominators defined in Eq.~\eqref{eq:Ds}, so that all the seven possible scalar products between the loop momenta $k_{1,2}$ and the independent external momenta $q\,,p$ are expressible as linear combinations of $D_i$.

The exact expression of ${\cal F}_S^{\rm 1b}$ contains 34 distinct Feynman integrals of the type in Eq.~\eqref{eq:family}, that of ${\cal F}_P^{\rm 1b}$ only a subset of 5 integrals. However, these 34 integrals are not all independent from each other, due to linear relations entailed by $d$-dimensional IBPs. We employ the packages {\sc Reduze2}~\cite{vonManteuffel:2012np} and {\sc LiteRed}~\cite{Lee:2012cn} for generating the IBPs needed to reduce the form factors to a combination of a minimal set of 20 independent MIs.

The set of 20 MIs have been analytically computed through the differential equation method in Ref.~\cite{Primo:2018zby} in the context of the evaluation of top Yukawa corrections to the decay of a Higgs boson into a $b\bar b$-pair, which involve two-loop triangle diagrams similar to the one considered here. Therefore, we adopt the same method of Ref.~\cite{Primo:2018zby} to evaluate the MIs, as we summarize in the following.

The integrals under consideration depend on three kinematic invariants, the square of the momentum transfer $t$, and the nucleon and lepton masses, $m_N$ and $m_\ell$. For a given $m_\ell$, the other two variables can be combined into two dimensionless parameters $x$ and $y$, which retain, up to a trivial scaling factor depending on $m_\ell$, the full dependence of the integrals on the kinematics. For the problem under consideration, it is convenient to define
\begin{equation}
\label{eq:vars}
t = - m_\ell^2 \, \frac{(1 - x^2)^2}{x^2} ~, \qquad m_N^2 = m_\ell^2 \, 
\frac{(1 - x^2)^2}{(1 - y^2)^2}\frac{y^2}{x^2} ~.
\end{equation}
Owing to IBPs, MIs fulfill systems of coupled first order differential equations in $t/m_\ell^2$ and $m_N^2/m_\ell^2$, or equivalently, in $x$ and $y$. We determine the analytic expression of the MIs by solving, as a series expansion around $\eps=0$, the corresponding system of differential equations in $x$ and $y$. The solution of such differential equations can be greatly simplified with a suitable choice of the basis of MIs. We employ the Magnus exponential algorithm~\cite{Argeri:2014qva, DiVita:2014pza} to identify a basis that obeys canonical differential equations in the sense of Ref.~\cite{Henn:2013pwa},
\begin{eqnarray}
\label{eq:sysxy}
\partial_{x}\GGvec(\eps, x, y) & = & \eps \, \mathbb{A}_{x}(x, y) \, \GGvec(\eps, x, y) ~, \nn
\partial_{y}\GGvec(\eps, x, y) & = & \eps \, \mathbb{A}_{y}(x, y) \, \GGvec(\eps, x, y) ~,
\end{eqnarray} 
where $\mathbb{A}_{x}$ and $\mathbb{A}_{y}$ are rational matrices and $\GGvec_\beta(\eps, x, y)$ is a vector that collects the basis integrals. The factorized dependence of the differential equations on the dimensional regulator $\eps$ streamlines the derivation of their general solutions in terms of well-known special functions, GPLs~\cite{Goncharov:polylog, Remiddi:1999ew, Gehrmann:2001pz, Gehrmann:2001jv, Vollinga:2004sn}. With the general solution of Eq.~\eqref{eq:sysxy} at hand, a proper choice of the boundary conditions, which is dictated by physical requirements, allows the complete determination of the required MIs and, hence, provides all the information information required to obtain the analytic expression of the form factors. For the technical details of the computation of the MIs, we refer the reader to Appendix~\ref{sec:AppendixA}. \\

\begin{figure}
	\centering
	\includegraphics[width=0.49\textwidth]{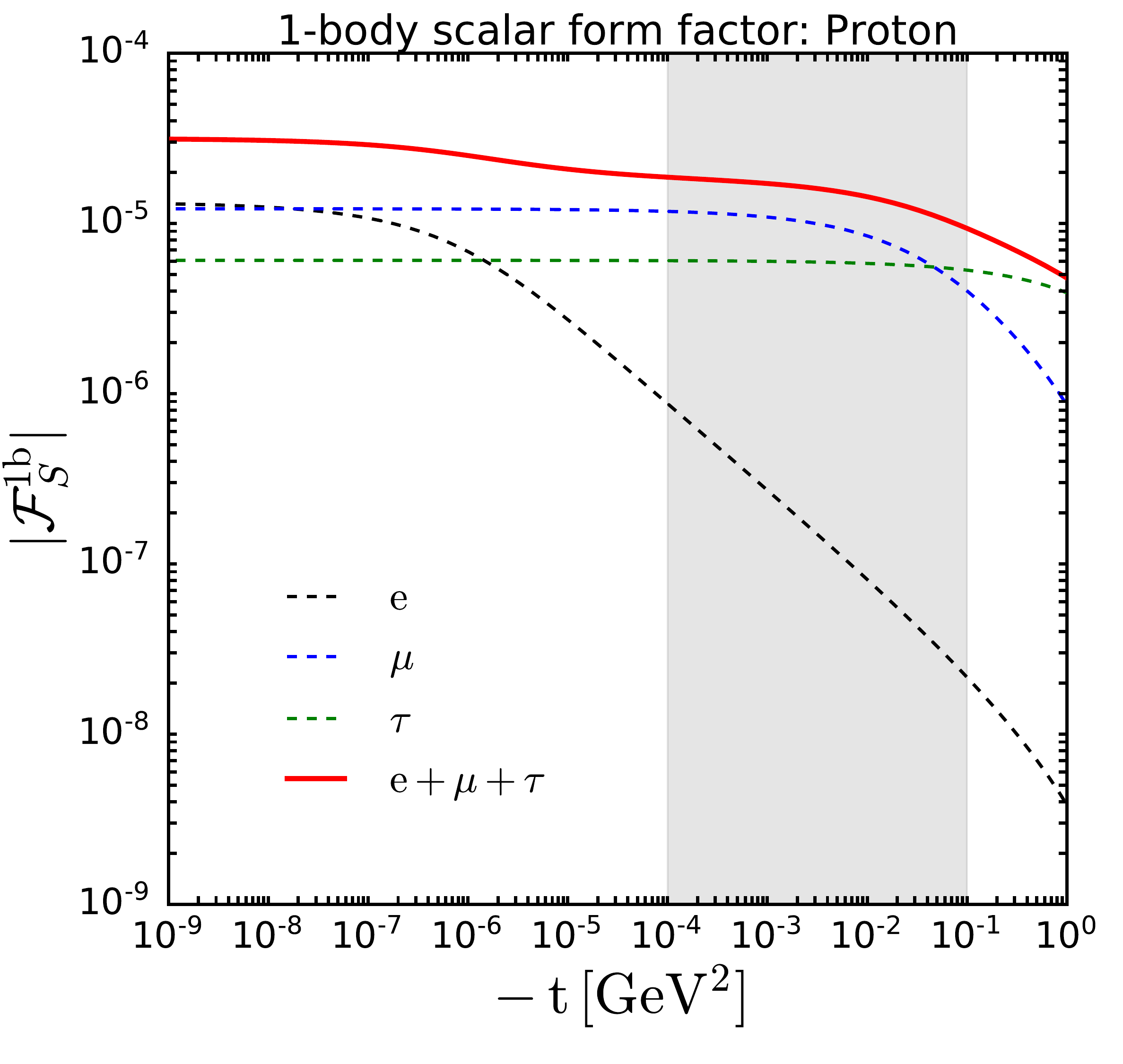}
	\includegraphics[width=0.49\textwidth]{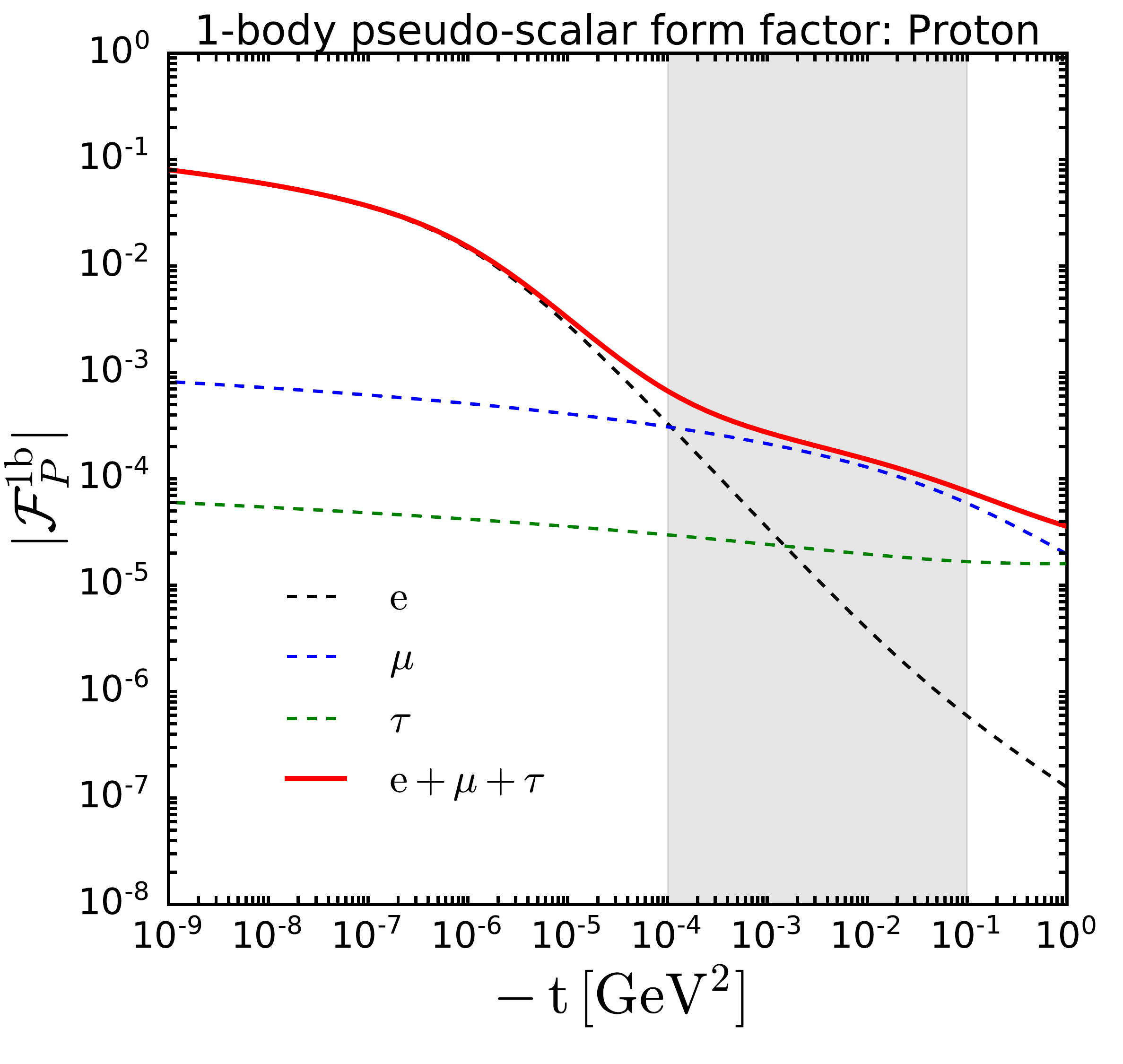} 
	\caption{\textbf{\textit{Form factors for one-body interactions of DM with protons}}, as a function of the square of the exchanged momentum, $-t$: scalar form factor, ${\cal F}_S^{\rm 1b} (t; m_p, m_\ell)$ (left panel) and  pseudo-scalar form factor, ${\cal F}_P^{\rm 1b} (t; m_p, m_\ell)$ (right panel). We show the results for interactions with a given lepton in the triangle loop: $e$ (black dashed curves), $\mu$ (green dashed curves) and $\tau$ (blue dashed curves); and for the total contribution (red solid curves). The gray shaded area indicates the range of momentum transfer relevant for scattering of local galactic DM off standard matter.}
	\label{fig:1body}
\end{figure}

After combining the analytic expression of the MIs into the form factors, we observe the expected analytic cancellation of all $\epsilon$-poles possessed by the individual integrals and obtain a finite result for both form factors in the four-dimensional limit $\eps\to 0$. The results for the scalar (${\cal F}_S^{\rm 1b}$) and pseudo-scalar (${\cal F}_P^{\rm 1b}$) form factors are presented in Fig.~\ref{fig:1body}, as a function of the squared momentum transfer $t$. We show the results for target protons, which, for the values of the considered momenta, are the relevant degrees of freedom in the scattering of local galactic DM off standard matter, and 
present the partial contributions from the diagrams with the three charged leptons in the triangle loop: $e$ (black dashed curves), $\mu$ (green dashed curves) and $\tau$ (blue dashed curves); and the sum of the three contributions (red solid curves). Note that, in the non-relativistic limit, $- t \simeq |\boldsymbol{q}|^2$, where $\boldsymbol{q}$ is the three-momentum transfer. In this limit, the maximum momentum transfer to a nucleus is $|\boldsymbol{q}|_{\rm max} = 2 \, \mu \, v_{\rm rel} < m_A \, v_0 \sim 100$~MeV, where $\mu$ is the reduced mass of the DM-nucleus system, $m_A \sim 100$~GeV is a typical target mass in direct detection experiments,\footnote{In the process of DM capture in the Sun or Earth, the target nuclei are lighter, and so is the momentum transfer. For DM capture in neutron stars, the heaviest target particles are neutrons, but the maximum momentum transfer is $\mathcal{O}(500)$~MeV.} $v_{\rm rel}$ is the relative velocity and $v_0 \sim 10^{-3}$ is the typical velocity of the DM halo in the galactic vicinity. For larger values, the velocity distribution is expected to fall off exponentially, so momentum transfers larger than $|\boldsymbol{q}|_{\rm max}$ are unlikely.

Let us first discuss the behavior of the form factors at vanishing square momentum transfer ($t = 0$), which is the usual limit adopted when studying local DM scattering, in the calculation of nuclear recoils within direct detection experiments, and in the calculation of capture of DM particles by celestial objects. At zeroth order in the momentum transfer, the one-body interaction only takes place with protons, so we will use $m_N = m_p$. For the scalar form factor, we have obtained a closed expression, which reads
\begin{eqnarray}
\label{eq:formSt0}
{\cal F}_S^{\rm 1b} \left(t = 0; z_{p\ell} \equiv \frac{m_p}{ m_\ell} \right) & = & -\frac{2 \, \alpha_{\rm em}^2}{\pi^2\, z_{p\ell}} \Big\{ 1  - \frac{\ln (z_{p\ell})}{2} + f_S(z_{p\ell}) + f_S(-z_{p\ell}) \Big\} ~, \\[2ex]
\textrm{with} \hspace{7mm} 
f(\tau) & \equiv & \frac{1}{4 \,\tau^2} \left(4 + 3 \, \tau + \tau^3\right) \Big[ \ln|\tau| \ln(1 + \tau) + {\rm Li}_2(-\tau) \Big]  ~.  \nonumber
\end{eqnarray}
Details of this calculation are provided in Appendix~\ref{sec:FS_limit}.

Two interesting limits can be studied in terms of the ratio of the proton mass and the mass of the leptons in the triangle loop, $z_{p\ell} \equiv m_p/m_\ell$. For lepton masses larger than the proton mass, $z_{p\ell} \ll 1$, 
\begin{equation}
\label{eq:formSt0series}
{\cal F}_S^{\rm 1b}(t = 0; z_{p\ell} \ll 1) \simeq \frac{\alpha_{\rm em}^2}{\pi^2} \, z_{p\ell} \, \left[ \, \ln \left(z_{p\ell}^{-1}\right) + \frac{13}{12}\right] ~.
\end{equation}	
This limit only (approximately) applies to the tau lepton loop. For the electron and muon loops, one should consider the opposite limit, $z_{p\ell} \gg 1$, for which the scalar form factor approximates as
\begin{equation}
\label{eq:formStinfseries}
{\cal F}_S^{\rm 1b}(t = 0; z_{p\ell} \gg 1) \simeq \frac{\alpha_{\rm em}^2}{4} \, \left( 1 + \frac{3}{z_{p\ell}^2} \right)~.
\end{equation}
At first order, this limit is independent of the proton-lepton mass ratio and provides an accurate estimate for the electron- and muon-loop contributions at $t = 0$. It also helps to understand the behavior of the contribution in the opposite limit, Eq.~(\ref{eq:formSt0series}), which is larger the larger the ratio $z_{p\ell}$,  but it has a limiting value given by Eq.~(\ref{eq:formStinfseries}). In short, it turns out that for small $|t|$, the electron and muon loops contribute approximately the same to the scalar form factor, whereas the tau-loop contribution is smaller, as can be seen from Fig.~\ref{fig:1body}.

For the pseudo-scalar form factor, an analytical expression for the soft asymptotic limit is presented in Appendix~\ref{sec:FP_limit}. The result varies little at small $|t|$ in the region depicted in Fig.~\ref{fig:1body}, and the contributions from the three lepton loops are larger than in the scalar case. Nevertheless, unlike the scalar case, there is no analogous saturation behavior in the soft limit, but the pseudo-scalar form factor slowly diverges as $t \to 0$. In this limit, $\cal{F}_P^{\rm 1b}$ scales as $z_{p\ell} \simeq m_p/m_\ell$ (see Appendix~\ref{sec:FP_limit}), so the muon- and tau-loop contributions are always smaller than the electron one. This is in contrast to the results at larger $|t|$ for any of the two form factors, as we shall see next.

Let us now consider the behavior of the form factors as a function of the square of the exchanged momentum. In both, the scalar and pseudo-scalar cases, the form factor starts decreasing above $|t| \gtrsim m_\ell^2$. As a result, around $|t| \sim (10^{-4} - 10^{-1})~\textrm{GeV}^2$, the relevant range of exchanged momentum for scatterings of local galactic DM off nucleons/nuclei (gray area), the contribution from the electron lepton loop is suppressed with respect to the other two. In that range, the scalar form factor is approximately constant and equal to the sum of the contributions from the diagrams with the muon and tau loop at $t = 0$, being the former the larger one.

The behavior of the pseudo-scalar form factor is qualitatively similar to the scalar case, although the former is one-two orders of magnitude larger than the latter in the relevant range of $t$ for local galactic DM-nucleon scattering. The contribution of each lepton loop decreases from a similar value of $t$ as in the scalar case, although the dependence on the lepton mass in the loop at small $|t|$ is different, and for the pseudo-scalar form factor, the diagram with the electron loop is the largest one at $|t| \lesssim 10^{-4}~\textrm{GeV}^2$. As the muon-loop contribution is larger than the tau-loop one up to $|t| \sim (1 - 10)~\textrm{GeV}^2$, the former is the most important one in the relevant range of $t$ (gray area in Fig.~\ref{fig:1body}). Moreover, the muon-loop contribution does not change much up to $|t| \gtrsim 10^{-1}~\textrm{GeV}^2$, so its value is similar to that at small $|t|$. Note, however, that at small $|t|$ the dominant contribution is always that with the electron loop. Thus, similarly to the scalar case, using the value of the form factor at small $|t|$ is not a good approximation for DM scattering off nucleons/nuclei. Finally, note that the pseudo-scalar coupling would induce spin-dependent interactions which, in the non-relativistic limit, reduce to a one-nucleon potential $\propto \boldsymbol{q} \cdot \boldsymbol{s}$, and thus, depend on the spin content of the nucleus.

Finally, let us note that we have also performed the loop integrals, Eq.~(\ref{eq:2LAmp}), using the {\sc SecDec} code~\cite{Carter:2010hi, Borowka:2012yc, Borowka:2015mxa, Borowka:2017idc} and we have checked that the numerical results so obtained match very well with the analytical results presented above, in regions where we observe numerical cancellation of all $\epsilon$-poles. This occurs for relatively large momentum transfers, larger than the the lepton mass (and for larger momentum transfers for the scalar case). For relatively small momentum transfers, however, numerical results from {\sc SecDec} do not converge at the required precision, so we cannot reliably compare them to the analytical results we obtain in those regions.

\subsection{Two-body interactions}
\label{sec:2body}

Let us now consider two-body DM interactions within a nucleus for the scalar case. Within leptophilic scenarios, for local DM-nucleon scatterings, lepton masses could be comparable to the typical momentum transfer. Thus, we need to retain the full form of the loop function without integrating out heavy leptons and matching onto the Rayleigh operator, $\bar \chi \chi F_{\mu \nu} F^{\mu \nu}$. The amplitude of the two-body two-boson exchange diagrams in Fig.~\ref{fig:2photon} (including the external DM particles) takes the form
\begin{equation}
{\cal M}_{S, ij} = \sum_{\ell=e,\mu,\tau}\frac{\Delta_S^{\mu\nu}(q_i, q_j, m_\ell)}{q^2_i \, q^2_j} \, \left( \bar{u}_i \gamma_\mu u_i  \, \bar{u}_j \gamma_\nu u_j \right) ~,
\label{eq:amp_scalar}
\end{equation}
where $u_i$ and $u_j$ are the two proton spinors, and $\Delta_S^{\mu\nu}(q_i, q_j, m_\ell)$ is the result of the loop integrals, which are summed up. Note that $q_i$ above are 4-momenta of the photons that connect the vertices of the loop to the external protons, so that $q = q_i + q_j$. The function $\Delta_S^{\mu\nu}(q_i, q_j, m_\ell)$ has the generic form
\begin{equation}
\Delta_S^{\mu\nu} (q_i, q_j, m_\ell) = \alpha \,g^{\mu \nu}  + \beta \, q^\mu_i \, q^\nu_j + \gamma \, q^\mu_j \,  q^\nu_i +  \delta \, q^\mu_j \, q^\nu_j + \kappa \, q^\mu_i \, q^\nu_i ~,
\end{equation}
where the coefficients $\alpha,\beta,\gamma,\delta,\kappa$ are functions of the photon momenta $q_i$, the external momentum transfer $q$ and the lepton masses $m_\ell$. In the non-relativistic limit, up to $\mathcal{O}(\boldsymbol{q}^2/m_p^2)$, the term proportional to $g^{00}$ is the most important one,\footnote{We have checked numerically that the contributions from the terms proportional to $\beta,\gamma,\delta, \kappa$ are, indeed, smaller.} and the amplitude then reduces to
\begin{equation}
{\cal M}_{S, ij} = G \, g_S \, \sum_{\ell=e,\mu,\tau} \frac{\boldsymbol{q}_i \cdot \boldsymbol{q}_j}{\boldsymbol{q}_i^2 \, \boldsymbol{q}_j^2} \,V_\alpha(\boldsymbol{q}_i, \boldsymbol{q}_j, m_\ell) \, \left( 1 + {\cal O} (\boldsymbol{q}^2/m_p^2) \right) ~,
\label{eq:amp_full}
\end{equation}
where,
\begin{eqnarray}
V_\alpha(\boldsymbol{q}_i, \boldsymbol{q}_j, m_\ell) & = & - V_\ell  \, \frac{3 \, m_\ell^2}{\boldsymbol{q}_i \cdot \boldsymbol{q}_j} \left\{ 1 +  \frac{\left(\boldsymbol{q}^2_j - \boldsymbol{q}^2_i + \boldsymbol{q}^2\right)}{\lambda(\boldsymbol{q}^2_i, \boldsymbol{q}^2_j, \boldsymbol{q}^2)} \, \Omega(\boldsymbol{q}_i^2, m_\ell) + \frac{\left(\boldsymbol{q}^2_i - \boldsymbol{q}^2_j + \boldsymbol{q}^2\right)}{\lambda(\boldsymbol{q}^2_i, \boldsymbol{q}^2_j, \boldsymbol{q}^2)} \, \Omega(\boldsymbol{q}_j^2, m_\ell) \right. \nonumber \\
& & 
\hspace{2.2cm} + \left(\frac{1}{\boldsymbol{q}^2} + \frac{\boldsymbol{q}^2_i + \boldsymbol{q}^2_j -\boldsymbol{q}^2}{\lambda(\boldsymbol{q}^2_i, \boldsymbol{q}^2_j, \boldsymbol{q}^2)}\right) \Omega(\boldsymbol{q}^2, m_\ell) \nonumber \\ 
& & 
\hspace{2.2cm} \left. + \, 2 \, \left( m_\ell^2 - \frac{\boldsymbol{q}_i^2 + \boldsymbol{q}_j^2 - \boldsymbol{q}^2}{4} - \frac{\boldsymbol{q}_i^2 \, \boldsymbol{q}_j^2 \, \boldsymbol{q}^2}{\lambda(\boldsymbol{q}^2_i, \boldsymbol{q}^2_j, \boldsymbol{q}^2)} \right) \text{C}_0\left(q^2_i, q^2_j, q^2, m_\ell, m_\ell, m_\ell\right) \right\} ~, \nonumber \\
\Omega(x, m_\ell) & \equiv  &  \sqrt{x \, (x + 4 \, m^2_\ell)} \, \ln\left[\frac{2 \, m^2_\ell + x + \sqrt{x \, (x + 4 \, m^2_\ell) }}{2 \, m^2_\ell}\right] ~. 
\end{eqnarray}

The Kallen function is denoted by $\lambda(\boldsymbol{q}^2_i, \boldsymbol{q}^2_j, \boldsymbol{q}^2) = \boldsymbol{q}^4_i + \boldsymbol{q}^4_j  + \boldsymbol{q}^4 - 2 \, \boldsymbol{q}^2_i \, \boldsymbol{q}^2_j - 2 \, \boldsymbol{q}^2_i \, \boldsymbol{q}^2 - 2 \, \boldsymbol{q}^2_j \, \boldsymbol{q}^2$,  $\text{C}_0(q^2_i, q^2_j, q^2, m_\ell, m_\ell, m_\ell)$ is the 3-point scalar Passarino-Veltman loop function, and we define $V_\ell$ as a dimensionful constant,
\begin{equation}
V_\ell \equiv - \frac{8 \, \alpha_{\rm em}^2}{3 \, m_\ell} ~.
\end{equation}
This is the value of $V_\alpha$ in the heavy-lepton limit~\cite{Kopp:2009et}. The effective ``potential'' induced by these two-nucleon interactions is given by the Fourier transform of the amplitude,
\begin{equation}
\tilde{V}_{ij} = - \int \frac{d\boldsymbol{q}_i}{(2 \pi)^3} \int \frac{d\boldsymbol{q}_j}{(2 \pi)^3} \, e^{-i \boldsymbol{q}_i \cdot \boldsymbol{x}_i - i \boldsymbol{q}_j \cdot \boldsymbol{x}_j} (2 \pi)^3 \, \delta(\boldsymbol{q} - \boldsymbol{q}_i - \boldsymbol{q}_j) \, {\cal M}_{ij}(\boldsymbol{q}_i, \boldsymbol{q}_j)  ~,
\label{eq:Vij}
\end{equation}

The total contribution to the two-body amplitude is given by the sum over all (identical, i.e., $V_{ij} \equiv V$ for all pairs) proton pairs in the nucleus, accounting for the nuclear density by means of the nuclear form factor~\cite{Cirigliano:2012pq, Ovanesyan:2014fha},
\begin{eqnarray}
{\cal M}_S^{(2)} & = & - \frac{Z (Z - 1)}{2} \int d\boldsymbol{x}_i \int d\boldsymbol{x}_j \, \tilde{V}_{ij} \, \rho(\boldsymbol{x}_i, \boldsymbol{x}_j) \nonumber\\
 & = & - \frac{Z (Z - 1)}{2} \int d\boldsymbol{x}_i \int d\boldsymbol{x}_j \, \tilde{V}_{ij} \, \int \frac{d\tilde{\boldsymbol{q}}_i}{(2 \pi)^3} \int \frac{d\tilde{\boldsymbol{q}}_j}{(2 \pi)^3} e^{-i \tilde{\boldsymbol{q}}_i \cdot \boldsymbol{x}_i - i \tilde{\boldsymbol{q}}_j \cdot \boldsymbol{x}_j}  \, F^{(2)}(\tilde{\boldsymbol{q}}_i, \tilde{\boldsymbol{q}}_j) \nonumber\\
  & = & \frac{Z (Z - 1)}{2} \, \int \frac{d\boldsymbol{k}}{(2 \pi)^3} \, {\cal M}_{S, ij}(\boldsymbol{q}_i, \boldsymbol{q}_j) \, F^{(2)}(\boldsymbol{q}_i, \boldsymbol{q}_j) ~, 
\label{eq:M2def} 
\end{eqnarray}
where $Z (Z-1)/2$ is the number of proton pairs and $F^{(2)}(\boldsymbol{q}_i, \boldsymbol{q}_j)$ is the Fourier transform of the two-proton nuclear density matrix. In what follows we neglect possible (but unknown) correlations within proton pairs and express $F^{(2)}$ in terms of the one-proton charge form factor,
\begin{equation}
F^{(2)}(\boldsymbol{q}_i, \boldsymbol{q}_j) \simeq F^{(1)}(\bar{q}_i) \, F^{(1)}(\bar{q}_j) = e^{- \bar{q}_i^2} \, e^{- \bar{q}_j^2} \hspace{5mm} , \hspace{5mm} \bar{q}_i \equiv |\boldsymbol{q}_i|/Q_0 ~.
\end{equation}

\begin{figure}
	\centering
	\includegraphics[width=0.49\textwidth]{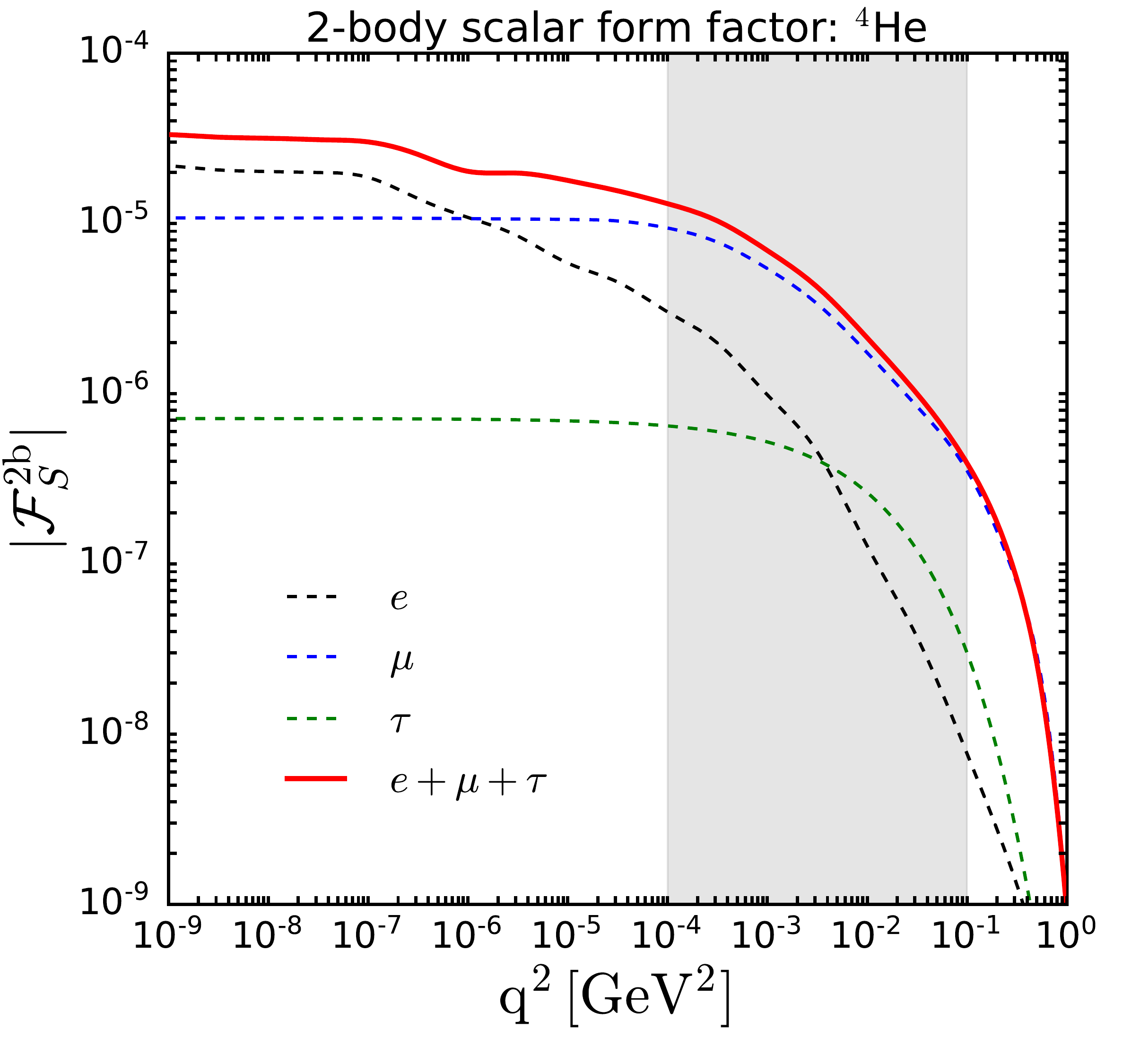} 
	\includegraphics[width=0.49\textwidth]{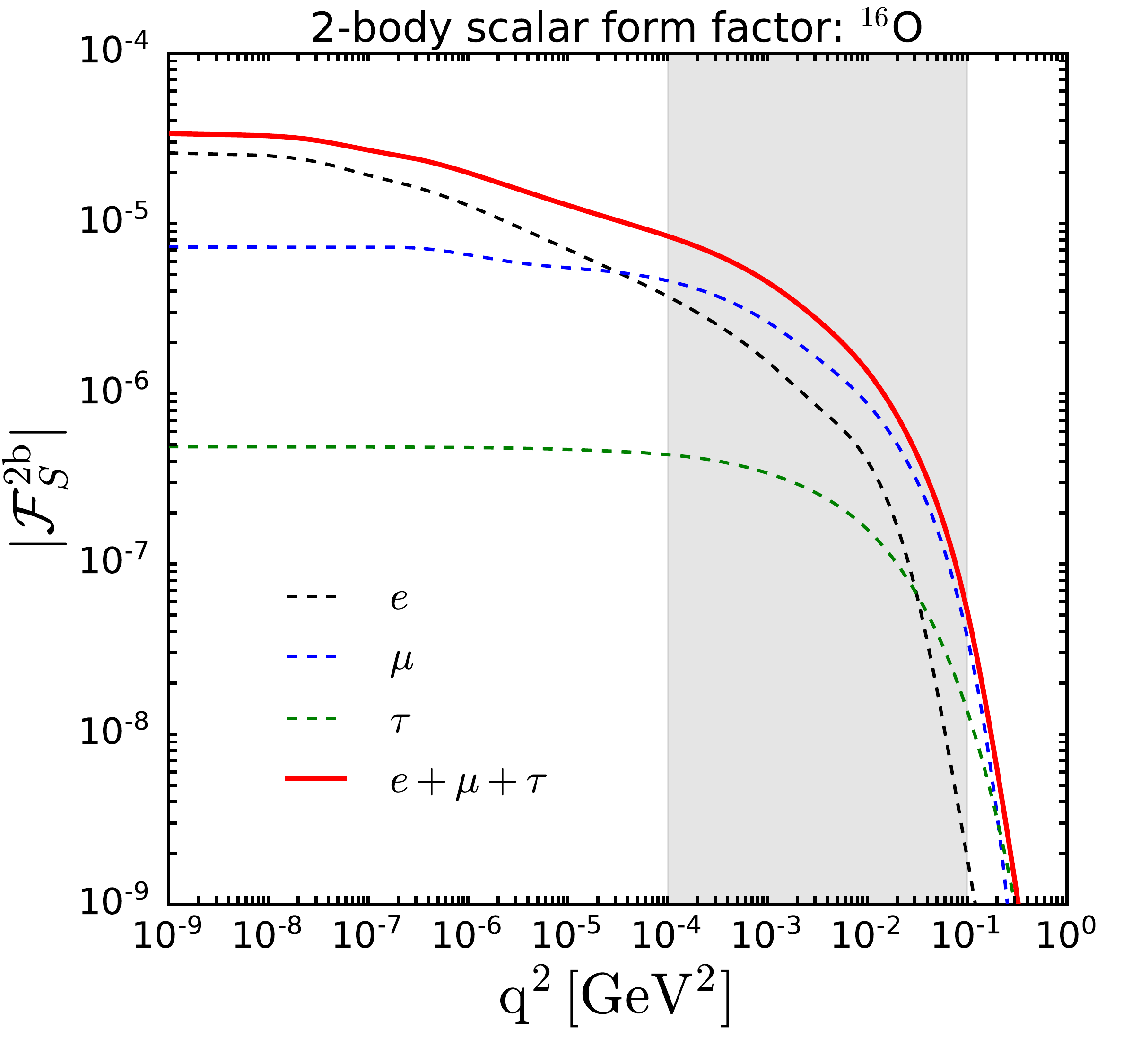} \\
	\includegraphics[width=0.49\textwidth]{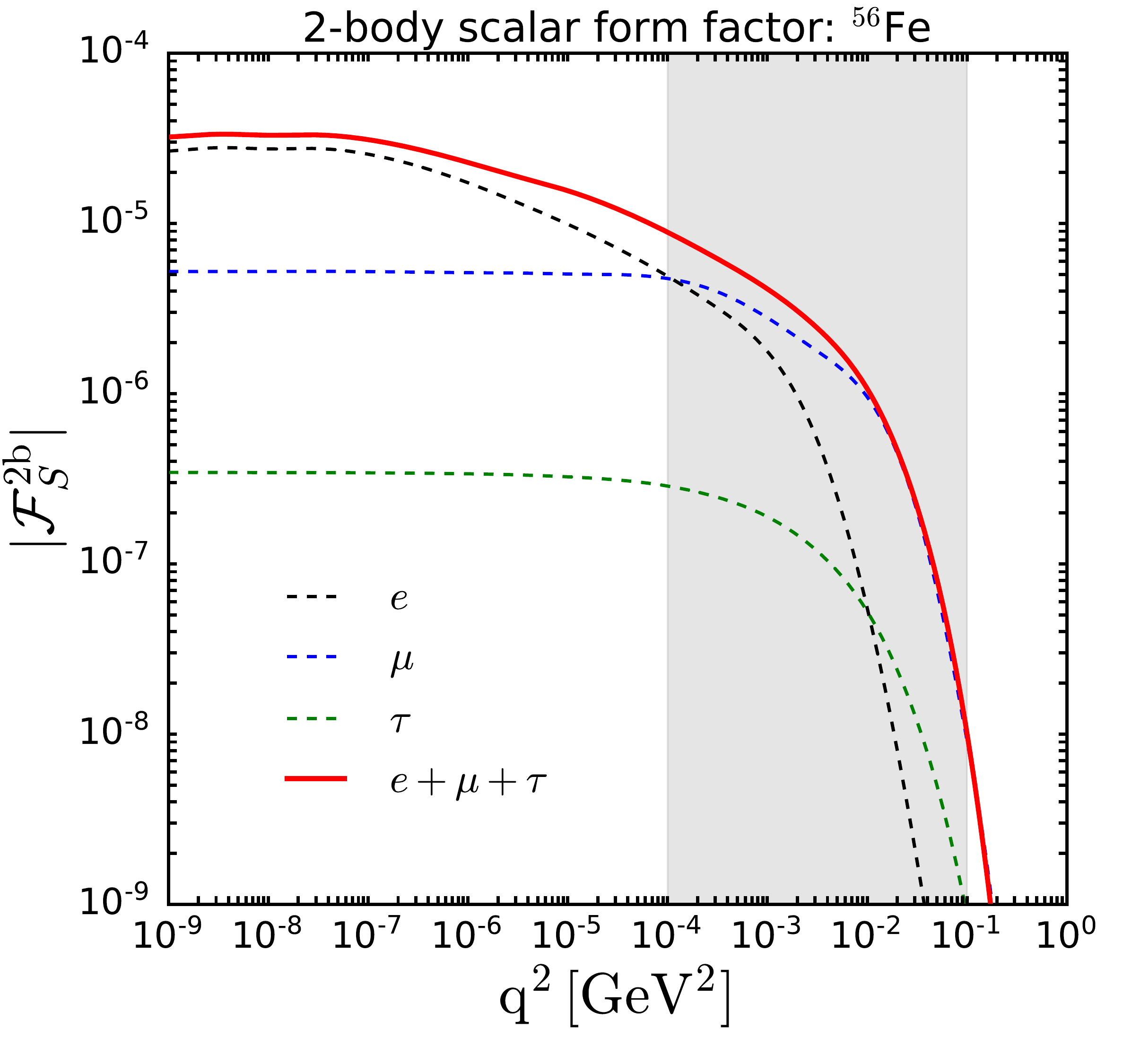} 
	\includegraphics[width=0.49\textwidth]{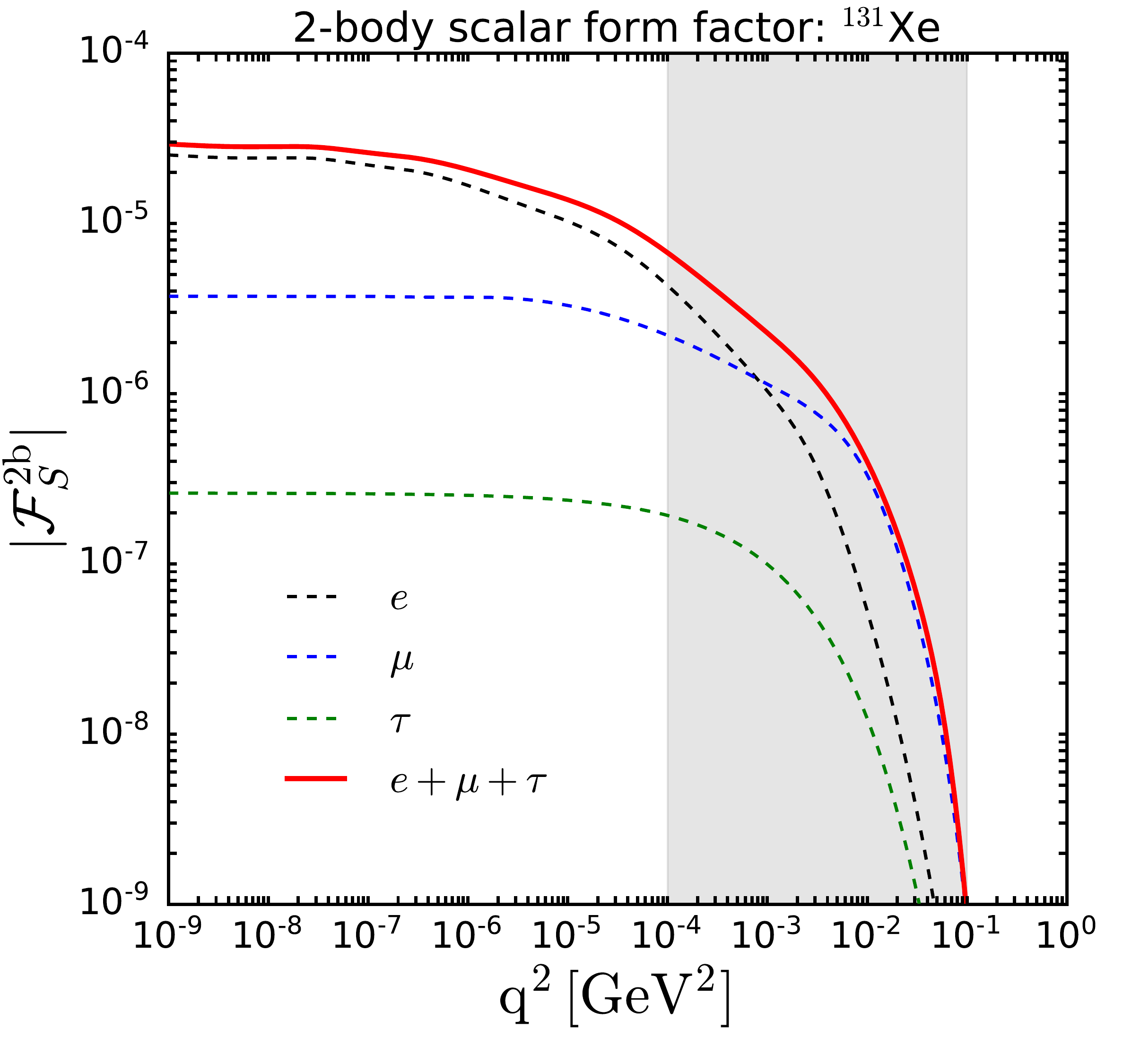} 
	\caption{\textbf{\textit{The scalar form factor, $\boldsymbol{{\cal F}_S^{\rm 2b} (t; m_p, m_\ell)}$, for two-body interactions}}, as a function of the square of the exchanged momentum, $\boldsymbol{q}^2 \simeq - t$. We show results for four different nuclei: helium (left upper panel), oxygen (right upper panel), iron (left lower panel) and xenon (right lower panel). Note that, although the factor $Z \, (Z-1)/2$ is not included, the result depends on the target. The gray shaded area indicates the range of momentum transfer relevant for scattering processes of local galactic DM off standard matter.}
	\label{fig:2b_formS}
\end{figure}

\begin{figure}
	\centering
	\includegraphics[width=0.49\textwidth]{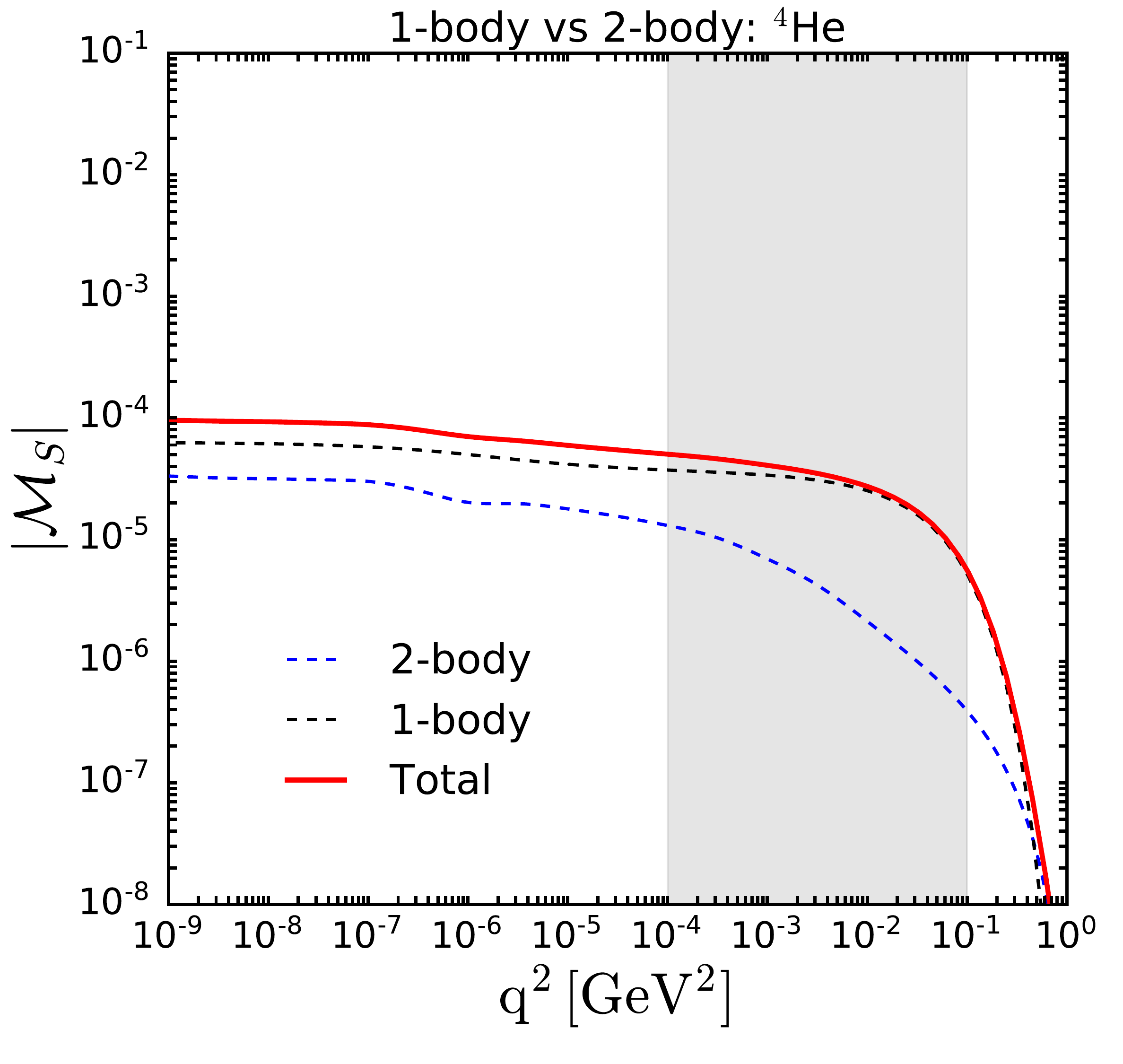} 
	\includegraphics[width=0.49\textwidth]{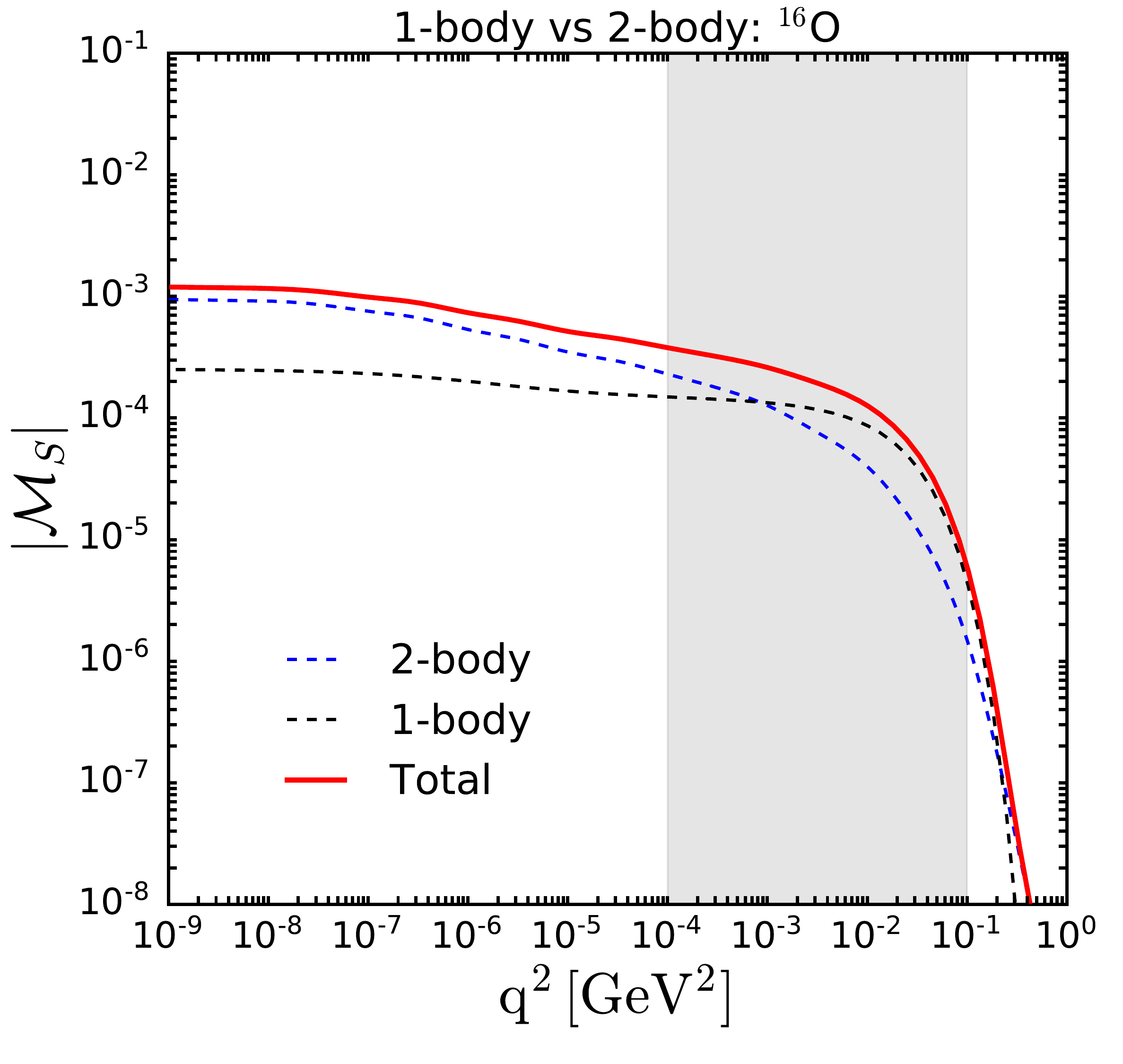} \\
	\includegraphics[width=0.49\textwidth]{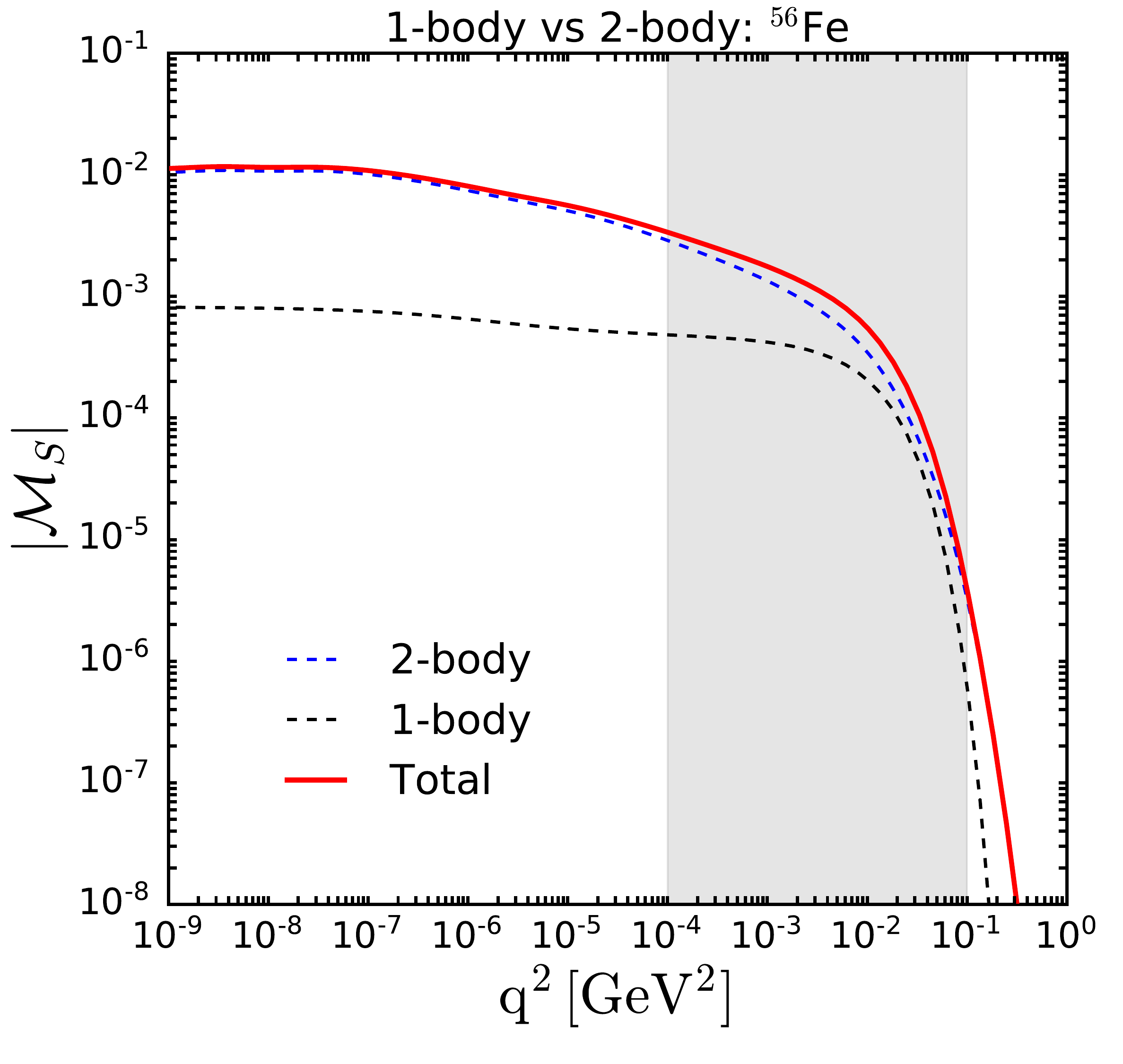} 
	\includegraphics[width=0.49\textwidth]{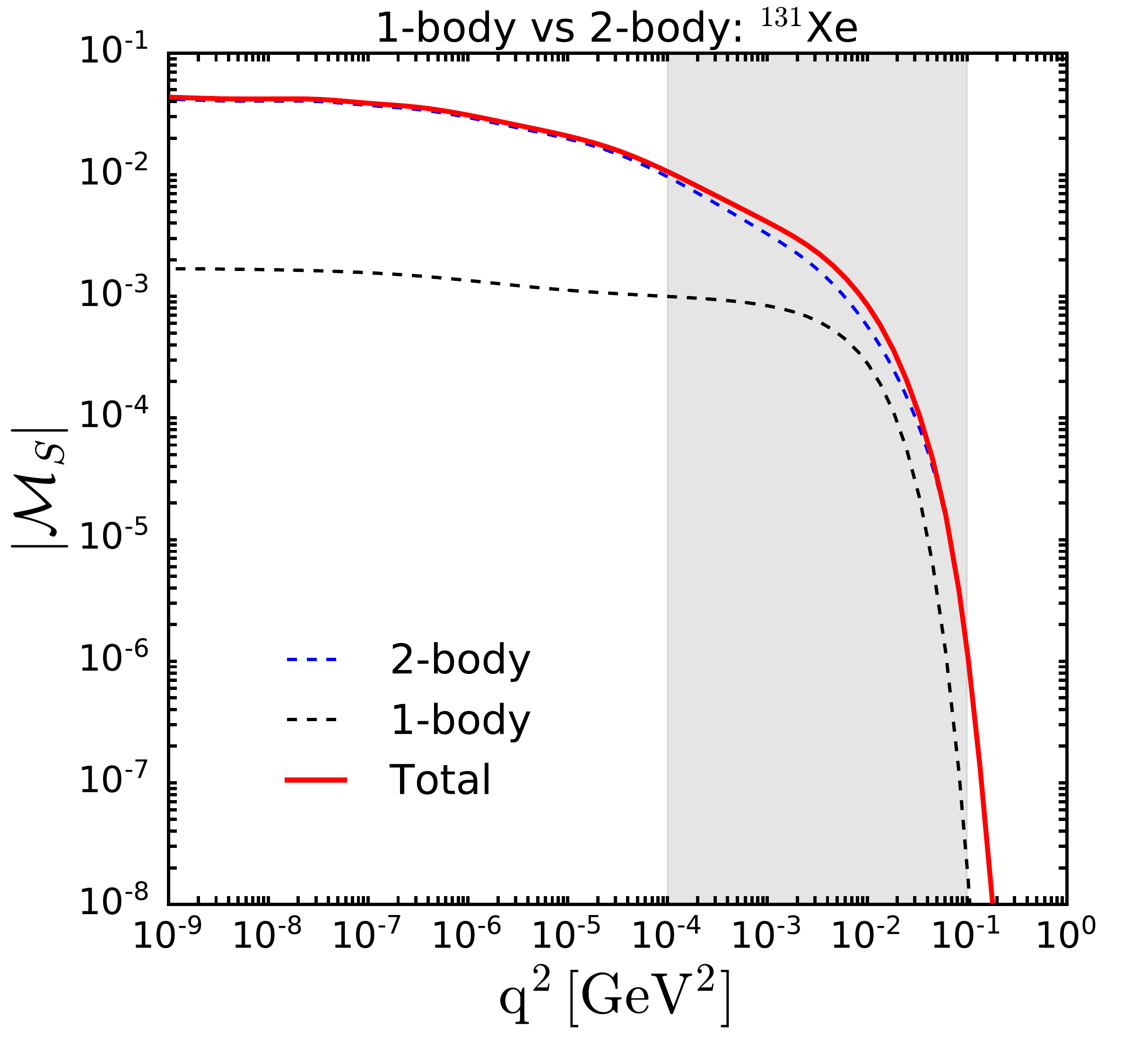} 
	\caption{\textbf{\textit{Contributions to the DM-nucleus scalar amplitude}}, as a function of the squared of the exchanged momentum $\boldsymbol{q}^2 \simeq - t$: one-body (black curves), ${\cal M}_S^{\rm 1b}$ and two-body (blue curves), ${\cal M}_S^{\rm 2b}$.  The total amplitude is also depicted (red curves). We show results for four different nuclei: helium (left upper panel), oxygen (right upper panel), iron (left lower panel) and xenon (right lower panel). The gray shaded area indicates the range of momentum transfer relevant for scattering processes of local galactic DM off standard matter.}
	\label{fig:2bvs1b_formS}
\end{figure}

Using Eq.~(\ref{eq:amp_full}), the two-body amplitude can be written as
\begin{eqnarray}
{\cal M}_S^{(2)}(\boldsymbol{q}) & = & - \frac{Z (Z - 1)}{2} \, G \, g_S \sum_{\ell=e,\mu,\tau} \int \frac{d\boldsymbol{k}}{(2 \pi)^3} \, \frac{\left(\boldsymbol{k} - \frac{\boldsymbol{q}}{2}\right) \cdot \left(\boldsymbol{k} + \frac{\boldsymbol{q}}{2}\right)}{\left(\boldsymbol{k} - \frac{\boldsymbol{q}}{2}\right)^2 \, \left(\boldsymbol{k} + \frac{\boldsymbol{q}}{2}\right)^2} \, V_\alpha(-\boldsymbol{k} + \boldsymbol{q}/2, \boldsymbol{k} + \boldsymbol{q}/2, m_\ell) \, F^{(2)}(-\boldsymbol{k} + \boldsymbol{q}/2, \boldsymbol{k} + \boldsymbol{q}/2) \nonumber \\
 & \equiv & - \frac{Z (Z - 1)}{2} \, G \, g_S \, \sum_{\ell = e, \mu, \tau} V_\ell \  F_{pp}(\bar{q}, m_\ell) \equiv \frac{Z (Z - 1)}{2} \, G \, g_S \, \sum_{\ell = e, \mu, \tau} {\cal F}_S^{\rm 2b} (\bar q, m_\ell) ~,
\label{eq:M2}
\end{eqnarray}
where we have made the substitution $\boldsymbol{q}_i = - \boldsymbol{k} + \boldsymbol{q}/2$ and $\boldsymbol{q}_j = \boldsymbol{k} + \boldsymbol{q}/2$, $F_{pp}(\bar{q})$ is defined as the two-proton form factor in the heavy-lepton limit, and ${\cal F}_S^{\rm 2b}$ is the two-body scalar form factor, which is shown in Fig.~\ref{fig:2b_formS} for different nuclei. 

For ease of understanding the $m_\ell$ dependence and for comparison with previous works, we also compute explicitly the $\boldsymbol{q}^2 \to 0$ limit, 
\begin{equation}
V_\alpha (\boldsymbol{q}_i, \boldsymbol{q}_j, m_\ell)|_{\boldsymbol{q}^2 \to 0} = V_\ell \, \frac{6 \, m_\ell^2}{\boldsymbol{k}^2} \, \left(1 - \frac{2 \, m_\ell^2 \, \Omega(\boldsymbol{k}^2, m_\ell)}{\boldsymbol{k}^2 \, (\boldsymbol{k}^2 + 4 \, m_\ell^2)} \right) ~.
\label{eq:Vq0}
\end{equation}
Note that the heavy-lepton limit, $V_\alpha \simeq V_\ell$, is easily recovered. Using Eq.~(\ref{eq:Vq0}), one can see that ${\cal F}_S^{\rm 2b} (0, m_\ell)$ is a decreasing function with $m_\ell$, with well-defined limits. In the heavy-lepton limit, ${\cal F}_S^{\rm 2b} (0, m_\ell \gg Q_0) \simeq - \frac{Q_0}{4 \, \pi \sqrt{2 \, \pi}} \, V_\ell$, whereas in the opposite limit, ${\cal F}_S^{\rm 2b} (0, m_\ell \ll Q_0) \simeq - \frac{3 \, m_\ell}{16} \, V_\ell = \alpha_{\rm em}^2/2$ tends to a constant.  Although we do not have a close form for the amplitude in the $\boldsymbol{q}^2 \to 0$, using these limiting values, it is very well approximated (using two parameters and with an accuracy better than 0.5\%) by
\begin{eqnarray}
{\cal M}_S^{(2)} \left({\boldsymbol{q}^2 \to 0}\right) & = & \frac{Z (Z - 1)}{2} \, G \, g_S \, \sum_{\ell = e,\mu,\tau} {\cal F}_{S}^{\rm 2b} (0,m_\ell) \nonumber \\
& \simeq & \frac{Z (Z - 1)}{2} \, G \, g_S \, \sum_{\ell = e,\mu,\tau} \frac{3 \, m_\ell}{16} \, V_\ell \, \frac{- 1 - 1.672 \, \left(\frac{m_\ell}{Q_0}\right)}{1 + 5.589 \, \left(\frac{m_\ell}{Q_0}\right) + 1.672 \, \frac{3 \, \pi^{3/2}}{2 \, \sqrt{2}} \, \left(\frac{m_\ell}{Q_0}\right)^2} ~.
\label{eq:M2q0}
\end{eqnarray}

Thus, the largest contribution at $\boldsymbol{q}^2 = 0$ to two-body interactions comes from the diagram with the electron loop. This is clearly seen in all panels of Fig.~\ref{fig:2b_formS}, where we depict the two-body form factor for four different nuclei as a function of $\boldsymbol{q}^2$. We have numerically evaluated these form factors with the use of the publicly available packages {\sc Package-X}~\cite{Patel:2015tea} and {\sc Collier}~\cite{Denner:2002ii, Denner:2005nn, Denner:2010tr, Denner:2016kdg}. As a consequence of the $m_\ell/Q_0$ dependence, for light nuclei (large $Q_0$), the contribution from the diagram with the muon loop is more important in relative terms than for heavier nuclei. Similarly to the one-body contribution, the form factor decreases with $\boldsymbol{q}^2$ and in the relevant range of momentum transfer, the diagram with the electron loop gets very suppressed. Indeed, except for very heavy targets, the diagram with the muon loop is the most important one in the range $\boldsymbol{q}^2 \sim \left( 10^{-4} - 10^{-1}\right)~{\rm GeV}^2$. Nevertheless, unlike what happens for one-body interactions (c.f., Figs.~\ref{fig:1body} and~\ref{fig:2b_formS}), the two-body contribution in that range of momentum transfer is significantly smaller than at $\boldsymbol{q}^2 = 0$. This can be understood from two features. On one hand, at $\boldsymbol{q}^2 = 0$, the main contribution to the two-body amplitude comes from a single diagram, which is subdominant at larger $\boldsymbol{q}^2$. On another hand, the drop in the two-body form factor occurs at smaller values of the exchanged momentum than in the one-body case due to the term $F^{(2)}(\boldsymbol{q}_i,\boldsymbol{q}_j)$.

In Fig.~\ref{fig:2bvs1b_formS}, we compare these two contributions to the total amplitude of the DM-nuclei elastic scattering cross section. From the results above, it is easy to see that, at $\boldsymbol{q}^2= 0$, the ratio of the two-body to one-body amplitudes scales as $(Z - 1)/2$.\footnote{Note that, in the limit $m_\ell \gg m_N$, we recover the scaling as $(Z - 1) \, Q_0/m_N$, discussed in Ref.~\cite{Ovanesyan:2014fha}.} At $\boldsymbol{q}^2 \sim (10^{-4} - 10^{-2})~{\rm GeV}^2$ both contributions are dominated by the diagrams with the muon loop, and the ratio approximately scales as $(Z-1) \, Q_0/m_\mu$. Indeed, for light nuclei like helium, the dominant contribution comes from one-body interactions at all relevant $\boldsymbol{q}^2$. Obviously, for scattering off hydrogen only one-body interactions take place. For heavier nuclei, however, the two-body contribution becomes relatively larger and for nuclei heavier than oxygen, it is the dominant one. Thus, within certain leptophilic scenarios considered here, one-body interactions are particularly important when computing the DM capture rate by solar nuclei, but two-body interactions become comparable or even the largest contribution to event rates in terrestrial direct detection experiments. \\

Let us finally comment on the two-body amplitude for the pseudo-scalar case. This has been briefly discussed in the context of DM scattering induced by $F_{\mu\nu} \tilde{F}_{\mu\nu}$~\cite{Ovanesyan:2014fha}. Note that this effective operator is obtained in the heavy-lepton limit within our scenario. In that work, the dominant contribution was considered to be the one from the coherent scattering of one of the two photons on the proton charge and the incoherent spin-dependent scattering of the second photon on the nucleon magnetic moment. Nevertheless, the contribution from the scattering of the two photons on the protons charge is of similar order and should be included too. These contributions are suppressed with respect to the scalar case by factors of order $\boldsymbol{q}/m_N$. Here, we just write down the form of the amplitude, including these two contributions. In analogy with the scalar case, for pseudo-scalar couplings the two-body amplitude takes the form
\begin{equation}
{\cal M}_{P, ij} = \frac{\Delta_P^{\mu\nu}(q_i, q_j, m_\ell)}{q^2_i \, q^2_j} \, \left( \bar{u}_i \gamma_\mu u_i  \, \bar{u}_j \Gamma_\nu u_j \right) ~,
\label{eq:amp_pseudoscalar}
\end{equation}
where $\Gamma_\nu = \{Q_j \, \gamma_\nu,  i \, \kappa_j \, \sigma_{\nu\mu} \, \frac{q^{\mu}}{2 m_N} \}$, with $Q_j$ the electric charge (in units of $e$) and $\kappa_i$ the anomalous magnetic moment (in units of Bohr magneton) of nucleon $j$, and
\begin{equation}
\Delta_P^{\mu\nu} (q_i, q_j, m_\ell) = G \, g_P \, 128 \, \pi^2 \, \alpha_{\rm em}^2 \, \left( i \, m_\ell \, \epsilon^{\mu\nu\rho\sigma} \, q_{i,\rho} \, q_{j,\sigma}\right) \, \text{C}_0\left(q^2_i, q^2_j, q^2, m_\ell, m_\ell, m_\ell\right) ~.
\end{equation}
For $\Gamma_{\nu} = i \, \kappa_j \, \sigma_{\nu\mu} \, \frac{q^\mu}{2 m_N}$ we get an identical result to that in Ref.~\cite{Ovanesyan:2014fha}.\footnote{Note, however, that $\kappa$ is the nucleon anomalous magnetic moment, not the nucleon magnetic moment as quoted in Ref.~\cite{Ovanesyan:2014fha}.} In the non-relativistic limit, the two-body amplitude can be written as
\begin{equation}
{\cal M}_P^{(2 {\rm em})}(\bar q) = G \, g_P \, \sum_{i < j} \, \sum_{\ell=e,\mu,\tau} \int \frac{d\boldsymbol{k}}{(2 \pi)^3} \, \frac{1}{\left(\boldsymbol{k} - \frac{\boldsymbol{q}}{2}\right)^2 \, \left(\boldsymbol{k} + \frac{\boldsymbol{q}}{2}\right)^2} \, V_P^{\rm em} (\boldsymbol{k}, \boldsymbol{q}) \, S^{(2)}(-\boldsymbol{k} + \boldsymbol{q}/2, \boldsymbol{k} + \boldsymbol{q}/2) ~,
\label{eq:M2em}
\end{equation}
with
\begin{equation}
V_P^{\rm em} (\boldsymbol{k}, \boldsymbol{q}) = 128 \, \pi^2 \, \alpha_{\rm em}^2 \, m_\ell\, \left( \left[ \boldsymbol{k}^2 \, \boldsymbol{q} - (\boldsymbol{k} \cdot \boldsymbol{q}) \, \boldsymbol{k}\right] \cdot \vec{\mu}_{ij}^{\, +} + \left[ \boldsymbol{q}^2 \, \boldsymbol{k} - (\boldsymbol{k} \cdot \boldsymbol{q}) \, \boldsymbol{q} \right] \cdot \frac{\vec{\mu}_{ij}^{\, -}}{2} \right) \, \text{C}_0\left(q^2_i, q^2_j, q^2, m_\ell, m_\ell, m_\ell\right) ~, 
\end{equation}
where $\vec{\mu}_{ij}^{\, \pm} = \boldsymbol{s}_i \, \kappa_i\, Q_j/m_i \pm \boldsymbol{s}_j \,  \kappa_j \, Q_i/m_j$, with $\boldsymbol{s}_i = \boldsymbol{\sigma}/2$ the spin of nucleon $i$. Analogously to the scalar case, $S^{(2)}$ accounts for the two-nucleon spin distribution. On another hand, for $\Gamma_\nu = Q_j \, \gamma_\nu$, we obtain a very similar expression,
\begin{equation}
{\cal M}_P^{(2 {\rm ee})}(\bar q) = G \, g_P \, \sum_{i < j} \, \sum_{\ell=e,\mu,\tau} \int \frac{d\boldsymbol{k}}{(2 \pi)^3} \, \frac{1}{\left(\boldsymbol{k} - \frac{\boldsymbol{q}}{2}\right)^2 \, \left(\boldsymbol{k} + \frac{\boldsymbol{q}}{2}\right)^2} \, V_P^{\rm ee} (\boldsymbol{k}, \boldsymbol{q}) \, S^{(2)}(-\boldsymbol{k} + \boldsymbol{q}/2, \boldsymbol{k} + \boldsymbol{q}/2) ~,
\end{equation}
with
\begin{equation}
V_P^{\rm ee} (\boldsymbol{k}, \boldsymbol{q}) = 128 \, \pi^2 \, \alpha_{\rm em}^2 \, m_\ell \, \left( \left[ \boldsymbol{k}^2 \, \boldsymbol{q} - (\boldsymbol{k} \cdot \boldsymbol{q}) \, \boldsymbol{k}\right] \cdot \vec{\varepsilon}_{ij}^{\, +} + \left[ \boldsymbol{q}^2 \, \boldsymbol{k} - (\boldsymbol{k} \cdot \boldsymbol{q}) \, \boldsymbol{q} \right] \cdot \frac{\vec{\varepsilon}_{ij}^{\, -}}{2} \right) \, \text{C}_0\left(q^2_i, q^2_j, q^2, m_\ell, m_\ell, m_\ell\right) ~, 
\end{equation}
where $\vec{\varepsilon}_{ij}^{\, \pm} = \left( \boldsymbol{s}_i/m_i \pm \boldsymbol{s}_j/m_j \right) \, Q_i \, Q_j$. Note that, as for scalar interactions, only protons contribute to ${\cal M}_P^{(2 {\rm ee})}$.

The leading contribution to the two-body scattering amplitude in the pseudo-scalar case is the sum of the two terms above and can be written as
\begin{equation}
{\cal M}_P^{(2)}(\bar q) = {\cal M}_P^{(2 {\rm em})}(\bar q) + {\cal M}_P^{(2 {\rm ee})}(\bar q) = G \, g_P \, \sum_{\ell = e, \mu, \tau} {\cal F}_P^{\rm 2b} (\bar q, m_\ell) ~. 
\end{equation}
Notice that the total amplitude has the same form of Eq.~(\ref{eq:M2em}), but substituting the anomalous magnetic moment by the magnetic moment. The ratio of the one-body to two-body contributions is expected to go as in the scalar case~\cite{Ovanesyan:2014fha}, so the amplitude is expected to be dominated by the former for interactions off light nuclei and by the latter for interactions off heavy nuclei. We do not explicitly compute numerically this contribution.

\section{Comparison with previous results and discussion}
\label{sec:Comparison_and_discussion}

In the literature, there exist various estimates of the two-photon exchange diagrams in Fig.~\ref{fig:2photon} (with a lepton triangle loop and two exchanged photons), in particular, within the context of leptophilic DM scenarios~\cite{Kopp:2009et, Bell:2019pyc}. In those scenarios, the DM particle would only couple to leptons at tree level, but not to quarks. Therefore, the scattering off nuclei could only take place via loop-suppressed processes. If the mediator particle couples to charged leptons via scalar or pseudo-scalar couplings, the loop diagrams here discussed would be the dominant ones in DM-nuclei scattering, and could determine the capture rate in astrophysical objects (like the Sun or the Earth) and the event rates at direct detection experiments. In this work, we have computed the elastic scattering amplitudes from one- and two-body processes with two-photon exchange. We have kept all mass dependencies within the loops and, in the one-body case, we have obtained analytical results in terms of GPLs. Now, we compare our results to estimates in the literature and to different approximations.

A first attempt to evaluate the DM-nucleus scattering cross section in the case of scalar DM-lepton interactions only considered the two-loop diagrams shown in Fig.~\ref{fig:2photon}~\cite{Kopp:2009et}, but with the entire nucleus as the target. An effective operator product expansion was used to compute the two-loop diagram, which was approximated by first integrating out the lepton triangle-loop in the heavy-lepton limit, then matching it onto the local Rayleigh operator and finally, attaching the nucleus electromagnetic current. Thus, integrating out the lepton loop results in the effective (Rayleigh-like) interaction with photons
\begin{equation}
\label{eq:LRay}
\mathcal{L}_{\rm Ray} = \frac{8 \, \alpha^2_{\rm em}}{3 \, e^2} \, \sum_{\ell=e,\mu,\tau} \frac{G \, g_S}{4\,m_\ell} \, \Gamma_\chi F_{\mu\nu} F^{\mu\nu} = - G \, g_S \, \sum_{\ell=e,\mu,\tau} V_\ell \, \left(\Gamma_\chi \, \frac{1}{4} F_{\mu\nu} F^{\mu\nu} \right)~,
\end{equation}
where $G \, g_S$ is the dimensionful effective coupling of DM particles to leptons, and $\Gamma_\chi$ is the DM bilinear, with $\Gamma_{\chi} =\{\chi^\dagger \chi\}$ (scalar) or $\{\bar{\chi} \chi, \bar{\chi} \gamma_5 \chi\}$ (fermion).  We have factored out the coupling $g_\beta$ of the mediator to leptons, so that we can reabsorb it into $V_\ell$.

In the long wavelength limit (small momentum transfer), leptophilic DM would couple to nuclei through two photons via coherent one- and two-body\footnote{Multi-body interactions with more than two exchanged photons would also contribute, although at a suppressed level.} interactions~\cite{Ovanesyan:2014fha}. Whereas the amplitude of one-particle interactions is proportional to the number of nucleons, the one for two-particle interactions is proportional to the number of nucleon pairs, and hence to $Z (Z - 1)/2$. In practice, the latter acts as a coupling to the total electromagnetic current\footnote{At low momentum transfer and at order $\mathcal{O}(q^2/2m_N)$, this interaction is only sensitive to the total electric charge of the nucleus, $Z e$, whereas corrections from the nuclear magnetic moment are suppressed by the nucleus mass, $q/m_A$.} of the nucleus~\cite{Weiner:2012cb}. Therefore, attaching the two-photon interaction to the nucleus electromagnetic current effectively accounts for two-body interactions~\cite{Ovanesyan:2014fha}. In the heavy nucleus limit, Ref.~\cite{Kopp:2009et} obtained the amplitude
\begin{equation}
\mathcal{M}^{(2)}_{\rm Ray} = - G \, g_S\, Z^2 \, \sum_{\ell=e,\mu,\tau}  \frac{V_\ell}{4} \, \frac{2 \, \pi^2 |\boldsymbol{q}|}{16 \,\pi^2} \, \tilde F(\bar{q}) \, \left(\tilde \Gamma_{\chi} \bar{u}_A u_A \right) ~, 
\end{equation}
where $\tilde \Gamma_{\chi} = \{1, \bar{u}_\chi u_\chi\}$ for \{scalar, fermion\} DM, and $A$ and $Z$ are the mass and atomic numbers, respectively. The form factor $\tilde F(\bar{q})$ was not computed and its $\boldsymbol{q}^2$ dependence was neglected by setting $\tilde F(\bar{q}) = 1$. The expression above obviously vanishes at $\boldsymbol{q}^2 = 0$. Nevertheless, it should not be so. Starting from Eq.~(\ref{eq:LRay}), and using an effective theory approach for the nucleus, the amplitude of the one-loop contribution with the two photons attached to the entire nucleus reads~\cite{Weiner:2012cb, Frandsen:2012db, Ovanesyan:2014fha}
\begin{eqnarray}
\label{eq:MRay}
\mathcal{M}^{(2)}_{\rm Ray} (\bar q) & = & - G \, g_S \, Z^2 \, \sum_{\ell=e,\mu,\tau}  V_\ell \,  \frac{Q_0}{4 \, \pi \, \sqrt{2 \, \pi}} \, F_{\rm Ray}(\bar q) \, \left( \tilde \Gamma_{\chi} \bar{u}_A u_A \right) \equiv - G \, g_S \, Z^2 \, \sum_{\ell=e,\mu,\tau} V_\ell \, F^{\rm eff}_{pp}(\bar q) \, \left( \tilde \Gamma_{\chi} \bar{u}_A u_A \right) ~, \\
F_{\rm Ray}(\bar{q}) & = & - \sqrt{\frac{2}{\pi}} \, \int^1_0 dx \int^\infty_0 dl \, \frac{l^2}{\left(l^2 + \bar{q}^2 \, (1 - x) \, x \right)^2} \, \exp\left[-2 \, \left(l^2 - \bar{q}^2 \, \left(x \, (1 - x) -\tfrac{1}{2}\right)\right) \right] \nonumber \\
& &\times \Big\{\cosh\left[2 \, l \, \bar{q} \, (1 - 2 \, x) \right] - \frac{l^2 - \bar{q}^2 \, (1 - x) \, x + \tfrac{1}{2}}{l \, \bar{q} \, (1 - 2 \, x)} \sinh \left[2 \, l \, \bar{q} \, \left(1 - 2 \, x\right) \right] \Big\} ~,
\label{eq:FR}
\end{eqnarray}
where $F_{\rm Ray}$ is defined so that $F_{\rm Ray}(0) = 1$. Up to order ${\cal O}(\bar{q}^3)$, the two-proton form factor in this limit is given by~\cite{Ovanesyan:2014fha}
\begin{equation}
F_{pp}^{\rm eff}(\bar{q}) \simeq \frac{Q_0}{4 \, \pi \sqrt{2 \, \pi}} \, e^{-\bar{q}^2/2} \, \left[ 1 - \frac{\pi^{3/2}}{2 \, \sqrt{2}} \, \bar{q} + \frac{5}{3} \, \bar{q}^2\right] ~.
\end{equation}
This is the result for $F_{pp}$ we also obtain in the limit when the Wilson coefficient corresponding to the triangle loop is taken to be momentum-independent~\cite{Weiner:2012cb, Ovanesyan:2014fha}, as is the case of the heavy-lepton limit, $m_\ell \gtrsim Q_0$, for which $V_\alpha = V_\ell$~\cite{Kopp:2009et}. The ratio of our result to the above estimate is given by
\begin{equation}
R^{(2)} (\bar q) = \frac{{\cal M}_S^{(2)}(\bar q)}{{\cal M}^{(2)}_{\rm Ray}(\bar q)} = - \frac{Z- 1}{2 \, Z} \, \frac{\sum_{\ell = e, \mu, \tau} {\cal F}_{S}^{\rm 2b}(\bar{q}, m_\ell)}{\sum_{\ell = e, \mu, \tau} V_\ell \, F^{\rm eff}_{pp}(\bar q)} ~.
\end{equation}	
Let us first focus on the $\boldsymbol{q}^2 = 0$ case. For $m_\ell \gtrsim Q_0$, the heavy-lepton limit is a very good approximation for tau and even for muon leptons in the loop. However, it significantly overestimates the (absolute) value of ${\cal M}^{(2)}$ with electrons running in the loop. Note that this is particularly relevant, as the contribution from the loop with the lightest lepton is the dominant one. For $m_\ell \gtrsim Q_0$, our two-body amplitude, Eq.~(\ref{eq:M2q0}), is smaller than Eq.~(\ref{eq:MRay}) by a factor $(Z - 1)/(2 \, Z)$. Nevertheless, for $m_\ell \ll Q_0$, 
\begin{equation}
R^{(2)}_\ell (0) = \frac{{\cal M}_{S, \ell}^{(2)}(0)}{{\cal M}^{(2)}_{{\rm Ray}, \ell}(0)} \, \stackrel{{m_\ell \ll Q_0}}{\simeq} \, \frac{Z- 1}{2 \, Z} \, \frac{3 \, \pi \sqrt{2 \, \pi}}{4} \, \frac{m_\ell}{Q_0} \, \stackrel{m_\ell = m_e}{\sim} \, (0.003 - 0.015) ~.
\end{equation}	

Therefore, given the typical size of nuclei, $Q_0 \simeq (0.1 - 0.3)$~GeV, the heavy-lepton limit overestimates the amplitude by two-three orders of magnitude (the difference being larger for lighter nuclei) and thus, the cross section by four-five orders of magnitude. As discussed above, at higher momenta, the contribution from the diagram with electrons in the loop gets suppressed and the diagram with muons in the loop becomes the most important one (see Fig.~\ref{fig:1body}). However, with the approximation in Eq.~(\ref{eq:MRay}), the relative importance of the different lepton loops does not change with the value of the momentum transfer and the diagram with the electron loop is always the dominant one. As a result, the ratio $R_\ell^{(2)}$ decreases with $\boldsymbol{q}^2$, and at $\boldsymbol{q}^2 \sim (10^{-3}-10^{-2})~{\rm GeV}^2$ it is up to an order of magnitude smaller than at $\boldsymbol{q}^2 = 0$. In summary, the heavy-lepton approximation overestimates the two-body contribution to the scattering cross section by more than five orders of magnitude in the relevant range of momentum transfer.

The one-body interaction was not considered for leptophilic DM models in Ref.~\cite{Kopp:2009et}, but it was the only contribution considered in Ref.~\cite{Bell:2019pyc}. Moreover, it was accounted for within more general approaches~\cite{Frandsen:2012db, Ovanesyan:2014fha}, which start from an effective description of DM interactions with two photons via a Rayleigh-like operator, $\Gamma_{\chi} F_{\mu\nu} F^{\mu \nu}$. The one-body amplitude was taken to be $(Z \, f_ p + (A - Z) \, f_n) \, F^{(1)} (t/Q_0^2)$, with $f_p$ and $f_n$ the proton and neutron couplings. Whereas Ref.~\cite{Ovanesyan:2014fha} followed a purely phenomenological approach without specifying $f_p$ or $f_n$, Ref.~\cite{Frandsen:2012db} considered the contribution to the quark $\Gamma_{\chi} \bar q q$ and gluon $\Gamma_{\chi} G_{\mu\nu} G^{\mu\nu}$ operators, via mixing with the Rayleigh operator. Then, the matrix element with nucleons was connected to the amplitudes at the quark and gluon level with the usual procedure at vanishing momentum transfer. In this paper, we have followed a different approach. Given that in the scenario under consideration, interactions with nucleons occur via photon exchange, there is no need to go through couplings to quarks. We have computed one-body interactions directly using the electromagnetic coupling to nucleons. In this way, we save one step and avoid having to deal with dressed quark propagators in the two-loop diagrams in Fig.~\ref{fig:2photon}. Our result is given in Eq.~(\ref{eq:M1}), and inserting the external DM particles implies multiplying by $G$. We compare our result for the scalar mediator case to~\cite{Frandsen:2012db, Bell:2019pyc}
\begin{equation}
{\cal M}^{(1)}_{qq} = (G \, g_S) \, \left[ \sum_{N = p, n} \, {\cal N}_N \, \left(\sum_{q = u, d, s} \frac{m_N}{m_q} \, f_{q}^N \, {\cal C}_q + \frac{2}{27} \, f_{G}^N \sum_{Q = c, b, t}  \frac{m_N}{m_Q} \, \Delta_Q \, {\cal C}_Q \right) \right] \, F^{(1)}(\bar q)
\end{equation}
where ${\cal N}_p = Z$ and ${\cal N}_n = A - Z$, $\Delta_Q = 1 + 11 \,  \alpha_s(m_Q)/(4 \pi)$ with $\alpha_s(\mu)$ the strong coupling at scale $\mu$, and the assumed values of the hadronic matrix elements at $t = 0$ are $\{f_{u}^p, \, f_{d}^p, \, f_{s}^p, \, f_{G}^p \}  = \{0.016, \, 0.029, \, 0.043, \, 0.912\}$ and $\{f_{u}^n, \, f_{d}^n, \, f_{s}^n, \, f_{G}^n\} = \{0.014, \, 0.034, \, 0.043, \, 0.909 \}$. For the quark masses we use the values quoted at the Particle Data Group~\cite{Zyla:2020zbs}. The terms $(G \, g_S) \, {\cal C}_q$ are the Wilson coefficients of the operator $\Gamma_{\chi} \bar q q $.

In Ref.~\cite{Bell:2019pyc}, the integration of the two-loop diagram seems to be performed in the limit of vanishing square momentum transfer ($t = 0$). Implicitly, it is claimed that the scalar and pseudo-scalar form factors are equal in that limit.\footnote{In Ref.~\cite{Bell:2019pyc} only fermionic leptophilic DM is considered. In the case of scalar or pseudo-scalar mediators, DM bilinears including $\gamma_5$ are claimed not to induce couplings to quarks at one or two loops. We do not agree with this statement. Our results also apply to those cases.} Although details about the followed steps are not provided, we can compare our results with the approximate analytical expression that is obtained (Eq.~(4.14) in Ref.~\cite{Bell:2019pyc}), which we reproduce here in slightly different notation,
\begin{equation}
\label{eq:CBBR}
{\cal C}_{q}^{\rm BBR}  \left(z_{q\ell} \equiv \frac{m_q}{ m_\ell}\right) = Q_q^2 \,  \left(\frac{\alpha_{\rm em}^2}{2 \, \pi^2}\right) \, \sum_{\ell = e, \mu, \tau} \, \frac{z_{q\ell}}{1 - \left(\frac{z_{q\ell}}{2}\right)^2} \, \ln\left(\frac{2}{z_{q\ell}}\right) \hspace{1cm} ~, 
\end{equation}
where $Q_q$ is the electric charge of quark $q$. On another hand, the Wilson coefficients obtained from Ref.~\cite{Frandsen:2012db}, after computing the effective coupling of the Rayleigh operator in the heavy-lepton limit ($V_\alpha = V_\ell$) within our leptophilic scenario, can be written as
\begin{equation}
\label{eq:CEFT}
{\cal C}^{\rm EFT}_q (z_{q\ell}) = - \left(\frac{1}{4 \, \pi \, \alpha_{\rm em}} \, \sum_{\ell = e, \mu, \tau} V_\ell \, m_q \right) \, 3 \, Q_q^2 \, \left(\frac{\alpha_{\rm em}}{\pi}\right) \, \ln\left( \frac{\Lambda^2}{m_y^2}\right) = Q_q^2 \, \left(\frac{\alpha_{\rm em}^2}{2 \, \pi^2}\right) \, \sum_{\ell = e, \mu, \tau} 4 \, z_{q \ell} \, \ln\left( \frac{\Lambda^2}{m_y^2}\right) ~,
\end{equation}
where $m_y = m_N$ for $q = \{u, d, s\}$ and $m_y = m_Q$ for $Q = \{c, b, t\}$, and we will take $\Lambda = 1$~TeV, although the value of the UV scale does not affect our conclusions. Note that, in addition to neglecting the momentum dependence of the hadronic matrix elements, these effective couplings are not valid for $\boldsymbol{q}^2 \gtrsim m_\ell^2$, so we can only strictly compare this estimate to our result at very low values of the momentum transfer. Therefore, we will only compare the above two estimates to our results at $t = 0$. For the scalar case, the ratio of our calculation to the ones above is
\begin{equation}
R^{(1)} (0) \equiv \frac{{\cal M}^{(1)} (0)}{{\cal M}^{(1)}_{qq} (0)} = \frac{Z \,  \sum_{\ell = e, \mu, \tau} {\cal F}_S^{\rm 1b} (t = 0; z_{p\ell})}{\sum_{N= p, n} {\cal N}_N \, \left(\sum_{q = u, d, s} \frac{m_N}{m_q} \, f_{q}^N \, {\cal C}_q + \frac{2}{27} \, f_{G}^N \sum_{Q = c, b, t}  \frac{m_N}{m_Q} \, \Delta_Q \, {\cal C}_Q\right)} \simeq 
\begin{cases} 
1.5 & \text{with ${\cal C}_{q}^{\rm BBR}$} ~, \\[2ex]
6 \times 10^{-4} & \text{with ${\cal C}^{\rm EFT}_q$} ~. \\
\end{cases}
\label{eq:R1}
\end{equation}
Note that if we compute the one-body two-loop interaction at the quark level and then we connect it with the scattering off nuclei as done above (i.e., we set ${\cal C}_q = \sum_{\ell = e, \mu, \tau} \, {\cal F}_S^{\rm 1b}(t = 0; m_q, m_\ell)$), the ratio we obtain is $R^{(1)} (0) \simeq 0.35$. This implies that the treatment at the quark level for these type of interactions overestimates the value of the amplitude (at $t = 0$) by a factor of a few, unless the up and down quark masses were actually larger by a similar factor. 

Clearly, the approach using ${\cal C}^{\rm EFT}_q$ overestimates the one-body contribution to the amplitude by about three orders of magnitude. Like in the two-body processes, the discrepancy has to do with the heavy-lepton limit being not appropriate for the dominant diagram (the one with the electron loop) and largely overestimating the correct result. 

Let us now briefly discuss the similar results obtained with our approach at the nucleon level and the estimate using ${\cal C}_{q}^{\rm BBR}$ at the quark level. First, note that when comparing the results using ${\cal C}^{\rm BBR}$ at the nucleon and quark levels, a similar ratio ($\sim 0.2$) as the one within our approach is obtained. Therefore, comparing the result in Ref.~\cite{Bell:2019pyc} to ours, both either at quark or nucleon level, the ratio is $\sim 4.4$ and $\sim 7.6$, respectively. Nevertheless, when comparing our result at nucleon level to that of Ref.~\cite{Bell:2019pyc} at quark level, numerically the sum of all one-body diagrams is similar to our result, Eq.~(\ref{eq:R1}). Their relative contributions, however, are different. Whereas in our case the contributions with the electron and muon loop are similar and that with the tau loop is about a factor of three smaller, using Eq.~(\ref{eq:CBBR}) the electron loop is much larger than the other two contributions. The relative importance of diagrams with a different lepton in the loop can be understood from the calculation at the quark or nucleon level (i.e., depending on the internal propagator). Note that when comparing the calculations at nucleon and at quark level, one should compare ${\cal F}_S^{\rm 1b} (z_{p\ell})$ to $(z_{p\ell}/z_{q\ell}) \, {\cal C}_q (z_{q\ell})$ (or to ${\cal F}_S^{\rm 1b}(z_{q\ell}) \, (z_{q\ell})$). The former is a growing function with $z_{p\ell}$ which flattens at $z_{p\ell} \gtrsim 1$ and the latter is a decreasing function with $z_{q\ell}$. This explains the relative importance of the three diagrams in each case.

Nevertheless, the major difference between our one-body form factor and that in Ref.~\cite{Bell:2019pyc} is the result of the two-loop integral itself, Eqs.~(\ref{eq:formSt0}) and~(\ref{eq:CBBR}). For lepton masses in the loop larger than the mass of the external fermion, $z \ll 1$, Eq.~(\ref{eq:CBBR}) is smaller than Eq.~(\ref{eq:formSt0}) by a factor of $\sim 2$,
\begin{equation}
\frac{{\cal F}_S^{\rm 1b}(t = 0; z \ll 1)}{{\cal C}^{\rm BBR}(z \ll 1)} \simeq 2 \, \left(\frac{\frac{13}{12} - \ln(z)}{\ln 2 - \ln(z)} \right) ~.
\end{equation}
This limit applies to the diagrams with the muon and tau loop, for light external quarks, which are the ones that contribute the most. In the opposite limit, $z \gg 1$,  the disagreement is larger and grows with $z$, 
\begin{equation}
\frac{{\cal F}_S^{\rm 1b}(t = 0; z \gg 1)}{{\cal C}^{\rm BBR}(z \gg 1)} \simeq \frac{\pi^2}{8} \, \frac{z}{\ln\left(z/2\right)} ~.
\end{equation}
This limit always applies to the diagram with the electron loop, but depending on the mass of the external fermion, it might or not apply to the other two diagrams. In particular, for light external quarks, it does not apply to the diagrams with the muon or tau loop. On the contrary, it is a good approximation for all diagrams when the external fermion is considered to be a nucleon. In summary, it is evident that these results are quite different, mainly for $z \gtrsim 1$ and it is a bit surprising the contributions to the amplitude in our case and in Ref.~\cite{Bell:2019pyc}, when summed over all diagrams, come up being similar. 

Finally, let us mention that in Ref.~\cite{Kopp:2009et} the pseudo-scalar form factor is claimed to be zero at all orders. This is also a consequence of integrating out first the lepton loop. By doing so, other unitarity cuts contributing to the amplitude are neglected~\cite{Pratap:1972tb, Ametller:1983ec, Hoferichter:2020lap}. As we have shown by performing the one-body two-loop (and have outlined for the two-body one-loop) calculation, the pseudo-scalar form factor is not zero, and can actually be larger than the scalar one. We do not further comment on the contribution from two-body interactions in this case, which is proportional to $Z$ and to an spin-dependent part (see also Ref.~\cite{Ovanesyan:2014fha} for a discussion).

\section{Conclusions}
\label{sec:conclusions}

In this work, we have considered the scattering of dark matter (DM) off standard matter nuclei, within leptophilic scenarios, by considering loop-induced interactions via a scalar and a pseudo-scalar mediator. For these cases, the leading coupling to nucleons would arise from Feynman diagrams that contain a lepton triangle loop and two photons exchanged with the external nucleons, either via one-body two-loop or via two-body one-loop interactions. For some kinematic regimes, the considered loop-induced processes could become the dominant contributions (with respect to tree-level DM-electron interactions) to the rates in direct detection experiments and to the capture of DM particles in celestial objects.

We evaluated analytically the form factors coming from one- and two-loop diagrams containing an internal triangle loop of a (massive) lepton and external (massive) fermion currents, with different masses, and for arbitrary value of the momentum transfer. Our analytical calculations show that results obtained by not including the proper dependence on the kinematic variables may lead to incorrect conclusions.

In particular, the two-loop form factors have been evaluated analytically by making use of advanced multi-loop calculus techniques, based on IBPs, and Magnus exponential method for differential equations. The calculations have been carried out within the dimensional regularization scheme, and the final expressions of the form factors have been obtained by taking the smooth four dimensional limit, verifying the finiteness of the results. We have expressed the form factors in terms of GPLs and transcendental constants (up to weight four), by keeping the complete dependence on the lepton mass, on the nucleon mass, and on the momentum transfer. The two-loop scalar form factor had already been considered in recent studies involving the Higgs boson decay into massive quark pairs, with a quark triangle loop and two-gluon exchange~\cite{Primo:2018zby, Mondini:2020uyy}. We recomputed it here for leptophilic DM scenarios. The results of the pseudo-scalar form factors are presented here for the first time. Within leptophilic DM scenarios, previous estimates of the amplitude of the same two-loop diagrams had been obtained under different approximations. 
Two-body one-loop processes had also been considered, under different approximations, in effective DM scenarios, where DM couples to photons through a Rayleigh-like operator~\cite{Weiner:2012cb, Frandsen:2012db, Ovanesyan:2014fha}. We have evaluated the two-body contributions in the non-relativistic limit, but keeping the dependence on all the relevant scales and computing the lepton triangle loop without any further approximation. 

Notably, we show that loop-induced one-body and two-body interactions could both significantly contribute to the scattering of DM particles off nuclei in certain leptophilic scenarios. For one-body interactions, the largest contributions at small momentum transfer come from the diagrams with the electron and muon loops. For two-body processes, the dominant diagram at small momenta is that with the electron loop. At larger momenta, in particular at momenta relevant for local galactic DM interactions, $\boldsymbol{q}^2 \sim (10^{-4} - 10^{-1})~{\rm GeV}^2$, the diagram with the electron loop is exponentially suppressed and the dominant contribution is from the diagram with muon in the loop. The relative importance of the three different diagrams (with different lepton loops) varies as a function of the momentum transfer, but also depends on several other factors. Whereas in the case of one-body interactions the three diagrams are comparable at $t = 0$, for two-body interactions the diagram with the electron loop is significantly larger than the other two at $t = 0$, and is larger for heavier targets. At higher momentum transfer, the scalar one-body form factor is just slightly smaller than at $t = 0$, so the $t = 0$ approximation provides a reasonable result. For two-body interactions, the diagram with the tau loop is always subdominant at all relevant momentum transfer. The differences in relative importance between the other two diagrams is larger than for one-body interactions, so using the $t = 0$ form factor at other $t$ could result in overestimating this contribution to the amplitude by about an order of magnitude. 

In previous studies of these type of leptophilic DM scenarios~\cite{Kopp:2009et, Bell:2019pyc}, only one of these contributions was considered and several approximations were included. We have compared our results to those estimates and conclude that the predicted estimate for the amplitude of one-body interactions at $t = 0$ in Ref.~\cite{Bell:2019pyc} (performed at the quark level) has approximately the same numerical value as that of our full two-loop calculation (performed at nucleon level), although the analytical result is quite different. Thus the previously estimated scattering cross section for light nuclei, which is dominated by one-body interactions, is numerically off just by a factor of a few units. Note, however, that by following the effective approach of Ref.~\cite{Frandsen:2012db}, the one-body amplitude is overestimated by about three orders of magnitude. On another hand, for nuclei heavier than oxygen, for which the amplitude is dominated by the two-body contribution, the heavy-lepton approximation~\cite{Kopp:2009et, Weiner:2012cb, Frandsen:2012db, Ovanesyan:2014fha} overestimates the scattering cross section by no less than four orders of magnitude, in the relevant range of momentum transfer. Moreover, we have showed that the two-loop pseudo-scalar form factor has a non-vanishing expression, although it had been claimed to be vanishing at all orders~\cite{Kopp:2009et}.

In summary, we have computed and critically examined two-photon exchanged diagrams, required for the calculation of the DM-nucleus scattering cross section in some leptophilic DM scenarios. We have presented new analytical calculations for one-body two-loop and two-body one-loop DM-nucleus interactions, which are needed when evaluating event rates in direct detection experiments and in the computation of the DM capture rate by celestial objects. The calculations in this work can be further extended to consider the form factors for axial-vector mediators, where anomalous terms might also play a role. We leave this application to future studies.

\section*{Acknowledgments}

We thank Seva Chestnov and Paride Paradisi for clarifying discussions.
RG is supported by MIUR grant PRIN 2017FMJFMW. 
The work of PM is partially supported by the Supporting TAlent in ReSearch at Padova University (UniPD STARS Grant 2017 ``Diagrammalgebra").
SPR is supported by the Spanish FEDER/MCIU-AEI grant FPA2017-84543-P, and partially, by the Portuguese FCT (UID/FIS/00777/2019 and CERN/FIS-PAR/0004/2019). SPR also acknowledges support from the European ITN project HIDDeN (H2020-MSCA-ITN-2019//860881-HIDDeN).
AP is supported by the Swiss National Science Foundation under grant number 200020-175595. FG is supported by Fondazione Cassa di Risparmio di Padova e Rovigo (CARIPARO).
SPR and RG also thank the Galileo Galilei Institute for hospitality.
Finally, PM and SPR acknowledge Silvia Pascoli's encouragement and stimulating discussions over the years.

\newpage
\appendix
\section{Differential equations for one-body interactions}
\label{sec:AppendixA}

In this appendix, we provide the details about the calculation of the Master Integrals (MIs) required for the determination of the one-body form factors $\mathcal{F}_{S}^{1b}(t; m_N, m_\ell)$ and $\mathcal{F}_{P}^{1b}(t; m_N, m_\ell)$. We consider the generic case, with complete dependence on the variables $\{t, m_N, m_\ell\}$, and discuss, as well, two limiting behaviors: the {\it soft limit} (i.e., $t \to 0$), and the {\it equal-mass limit} (i.e., $m_N \to m_\ell$). In particular, in the soft limit, $\mathcal{F}_{S}^{1b}(t \to 0; m_N, m_\ell)$ is evaluated by setting $t = 0$ in the integrand, which greatly simplifies the integral evaluation; whereas $\mathcal{F}_{P}^{1b}(t \to 0, m_N, m_\ell)$ admits an asymptotic expansion, performed after integration. The equal-mass limit is studied by verifying that the expressions derived by expanding $\mathcal{F}_{S}^{1b}(t; m_N, m_\ell)$ and $\mathcal{F}_{P}^{1b}(t; m_N, m_\ell)$ agree with the ones obtained from a calculation where the choice $m_N = m_\ell$ is set ${\it ab \ initio}$ in the integrands. The evaluation of the MIs for the equal-mass case is considered separately and the agreement between the two ways of computing the form factors in this limit provides a check on the expressions obtained for arbitrary values of the kinematic variables.

\subsection{Master integrals for the general kinematic case}
\label{sec:AppendixA:different mass}

\begin{figure}
	\centering
	\captionsetup[subfigure]{labelformat=empty}
	\subfloat[\hspace{1.1cm}$\mathcal{T}_1$]{%
		\includegraphics[width=0.14\textwidth]{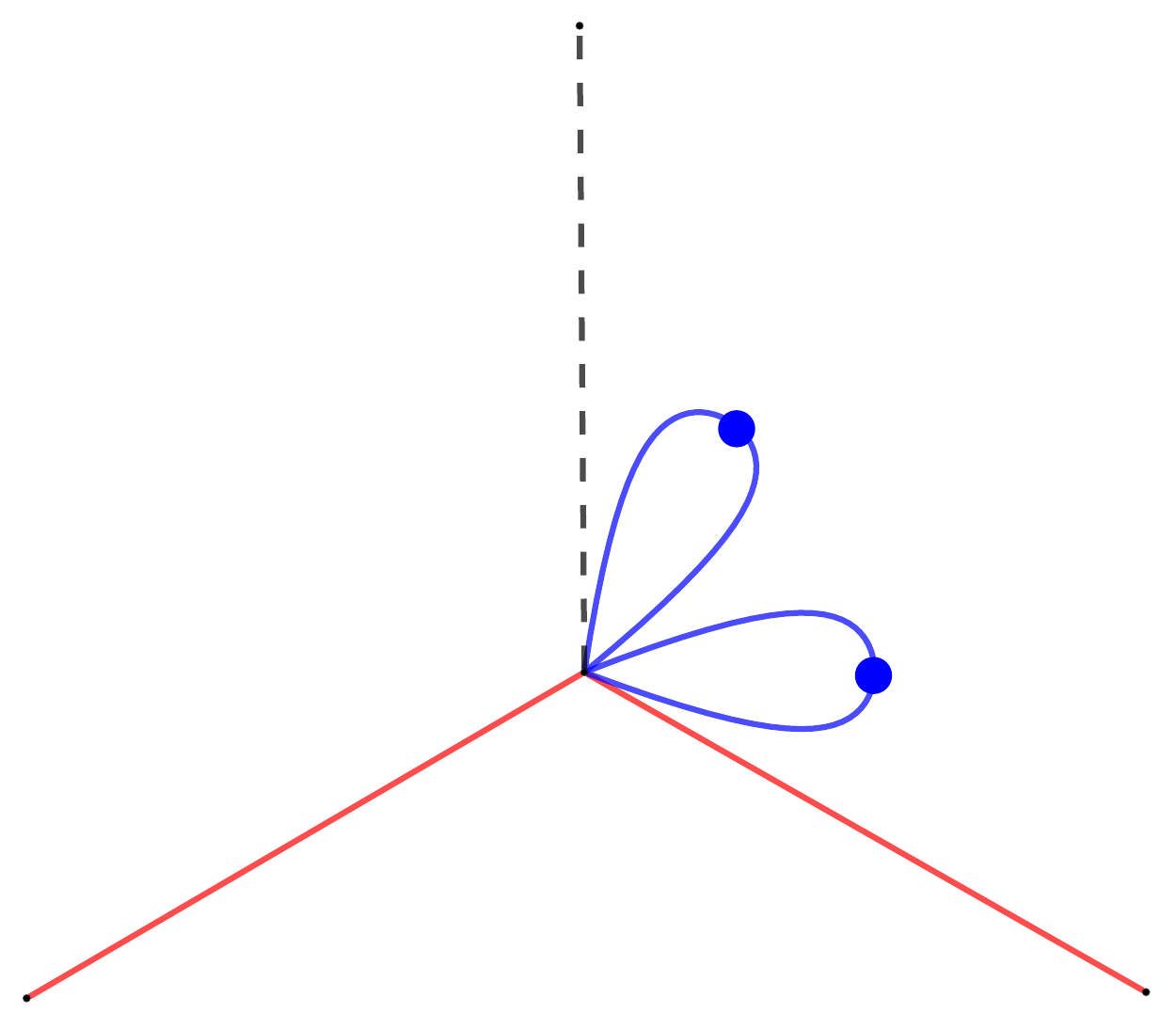}
	}
	\subfloat[\hspace{1.1cm}$\mathcal{T}_2$]{%
		\includegraphics[width=0.14\textwidth]{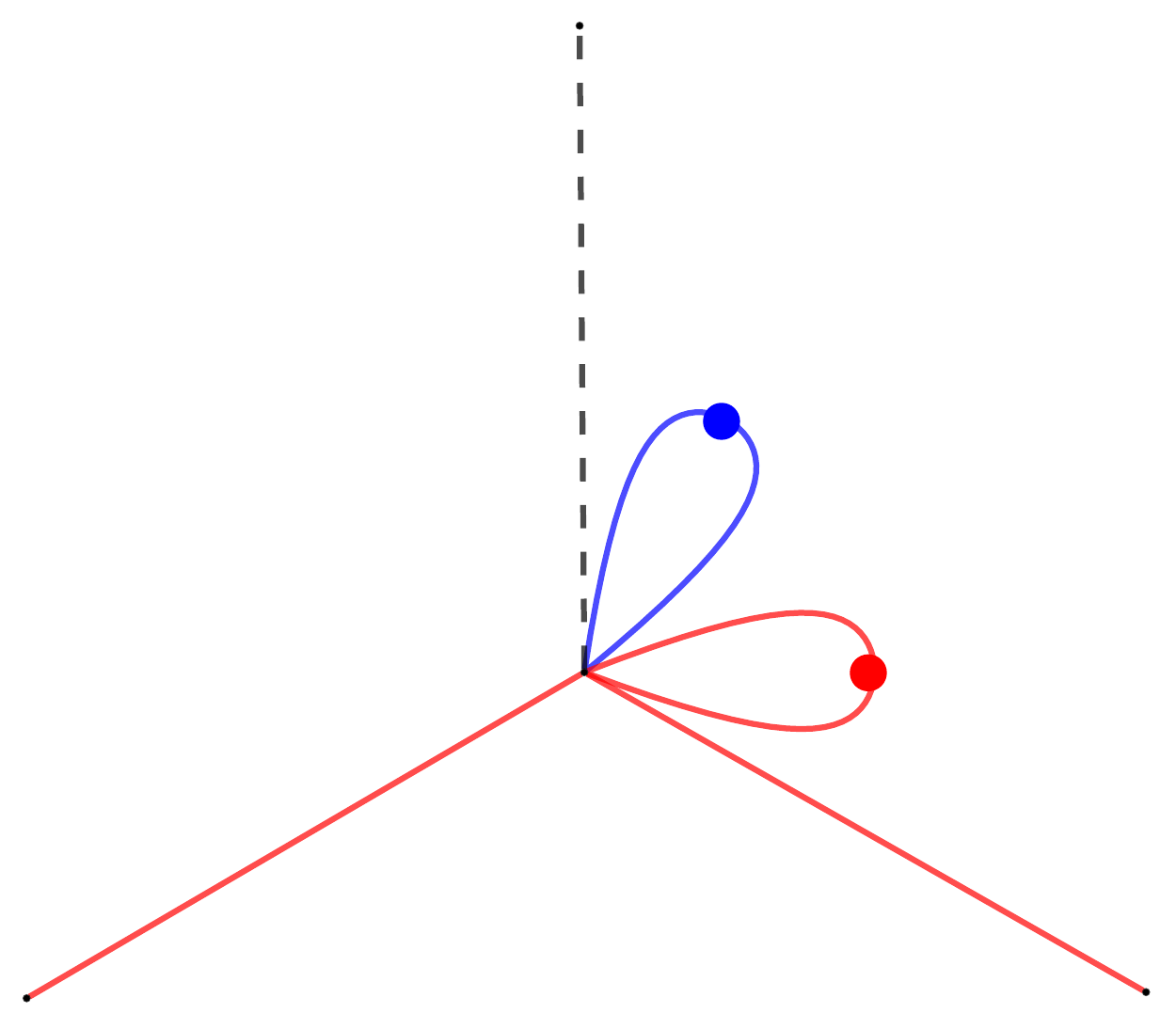}
	}
	\subfloat[\hspace{1.1cm}$\mathcal{T}_3$]{%
		\includegraphics[width=0.14\textwidth]{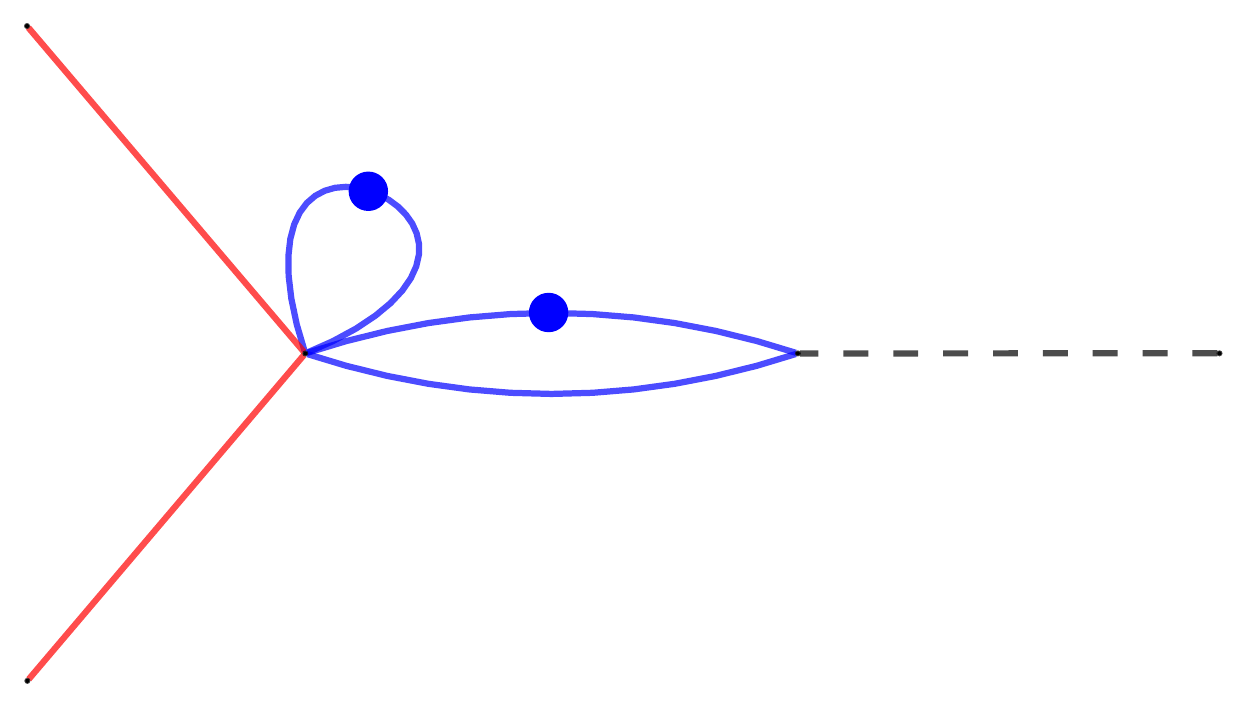}
	}
	\subfloat[\hspace{1.1cm}$\mathcal{T}_4$]{%
		\includegraphics[width=0.14\textwidth]{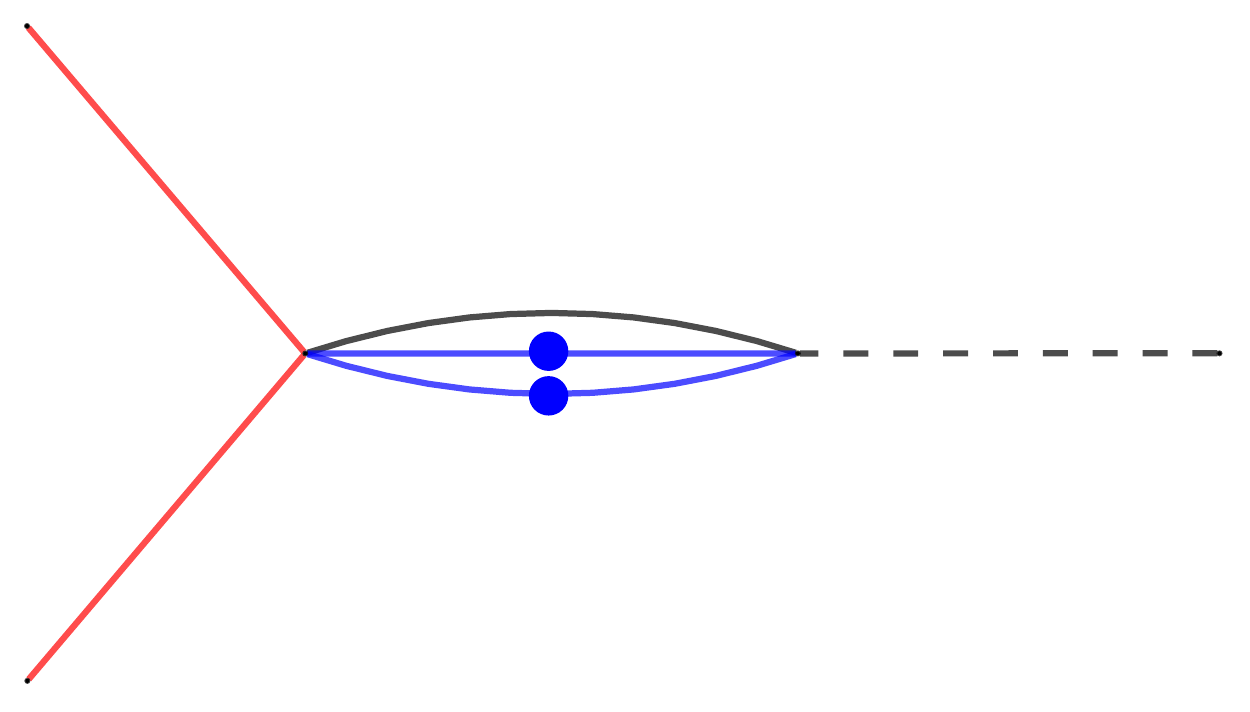}
	}
	\subfloat[\hspace{1.1cm}$\mathcal{T}_5$]{%
		\includegraphics[width=0.14\textwidth]{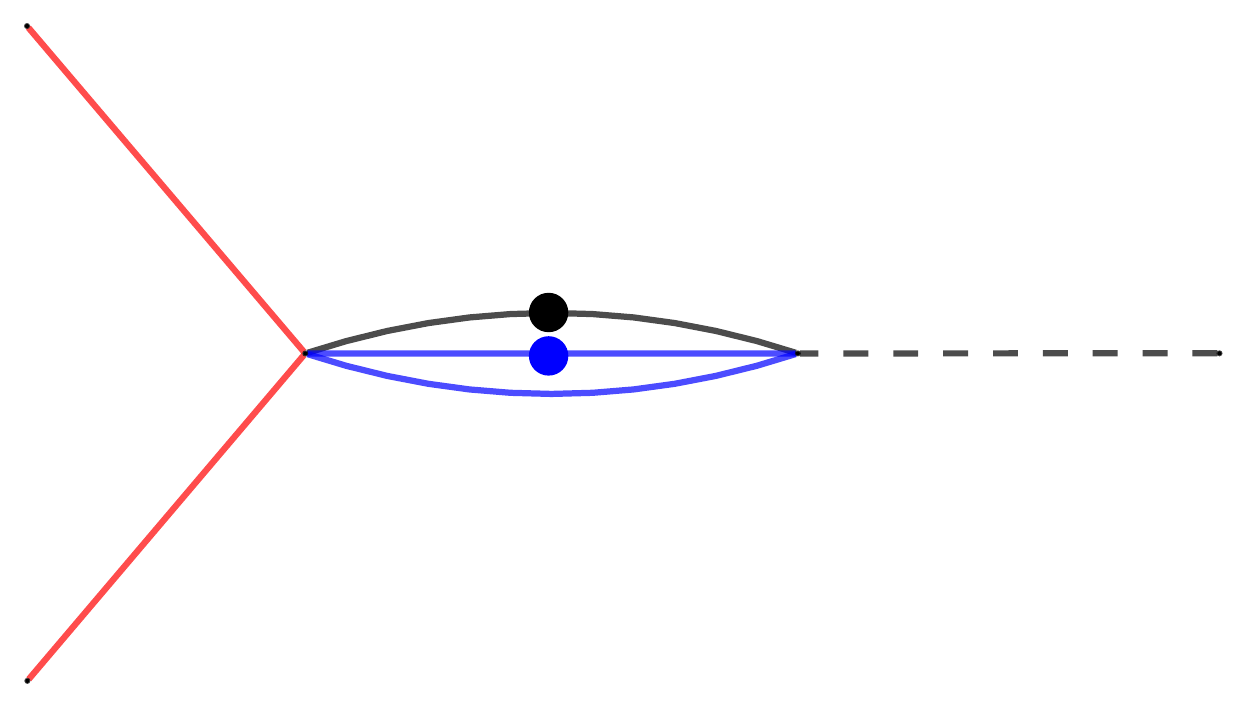}
	}
	\subfloat[\hspace{1.1cm}$\mathcal{T}_6$]{%
		\includegraphics[width=0.14\textwidth]{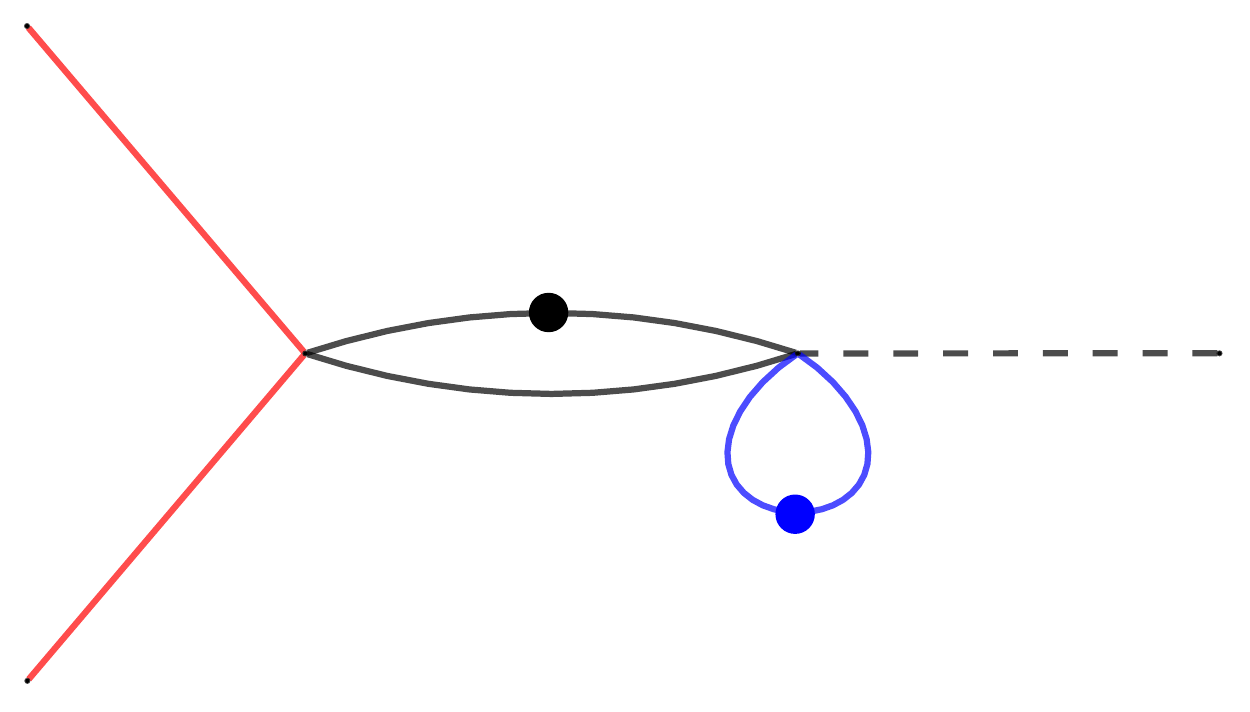}
	}
	\\
	\subfloat[\hspace{1.1cm}$\mathcal{T}_7$]{%
		\includegraphics[width=0.14\textwidth]{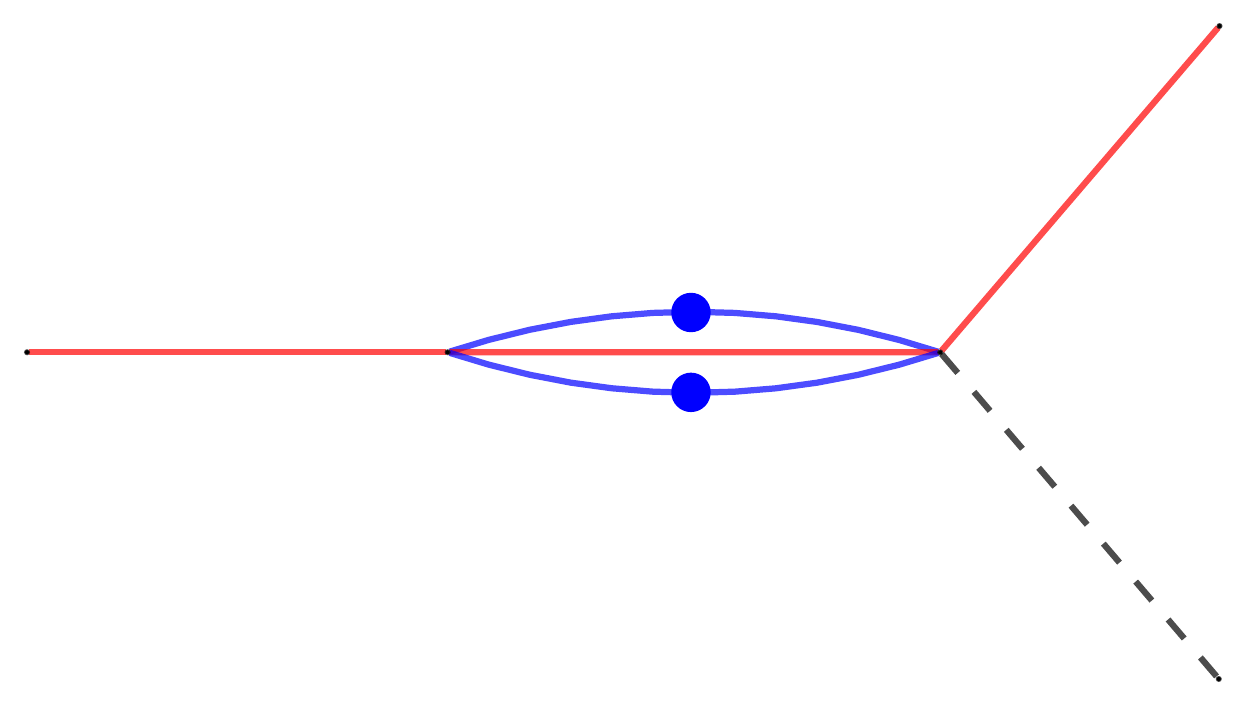}
	}
	\subfloat[\hspace{1.1cm}$\mathcal{T}_8$]{%
		\includegraphics[width=0.14\textwidth]{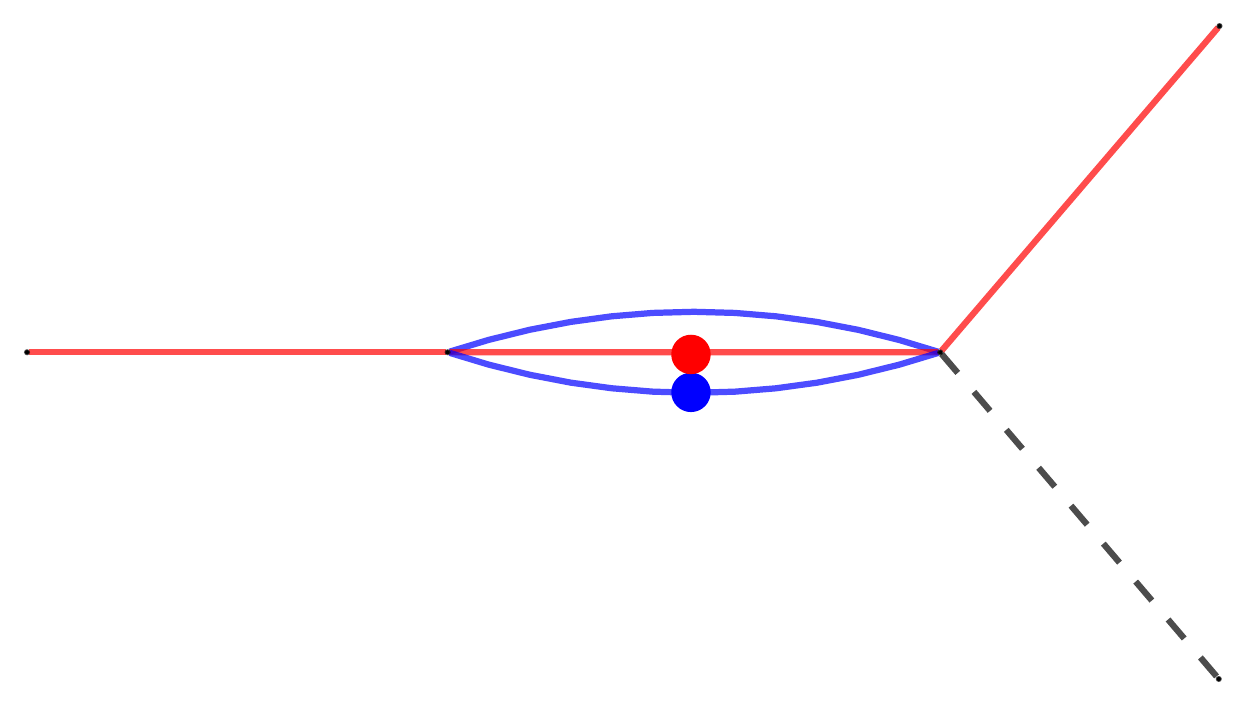}
	}
	\subfloat[\hspace{1.1cm}$\mathcal{T}_9$]{%
		\includegraphics[width=0.14\textwidth]{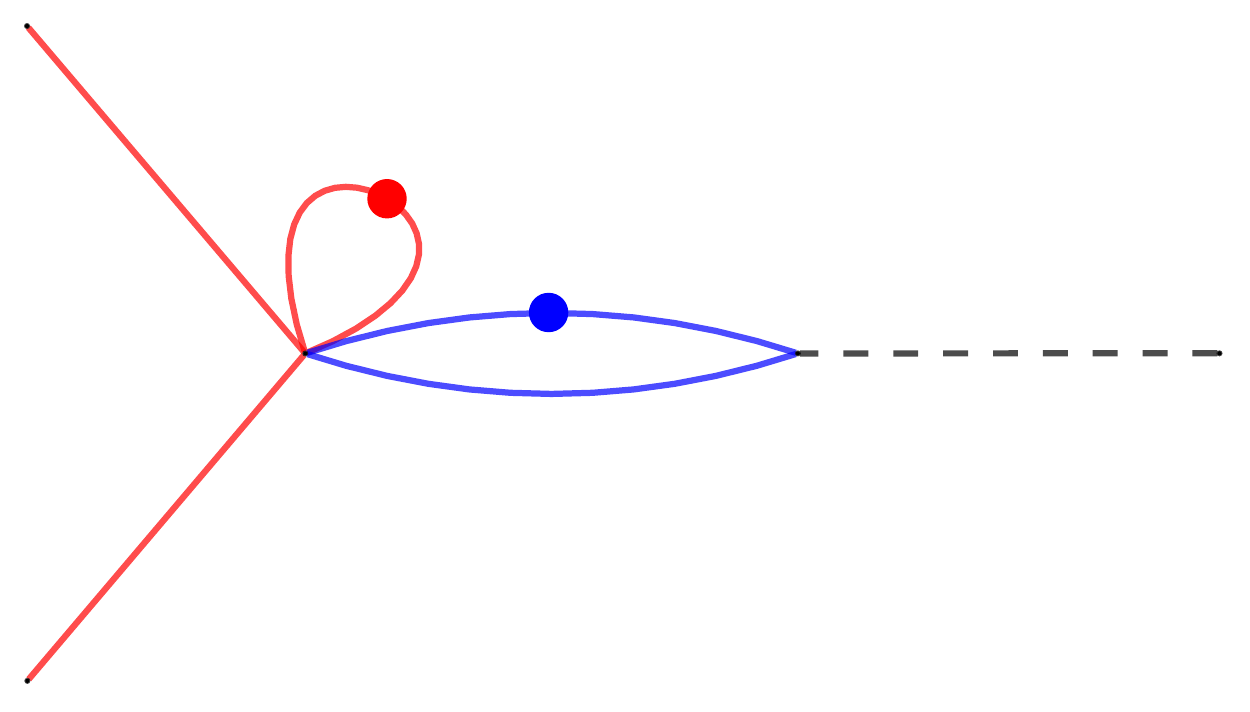}
	}
	\subfloat[\hspace{1.1cm}$\mathcal{T}_{10}$]{%
		\includegraphics[width=0.14\textwidth]{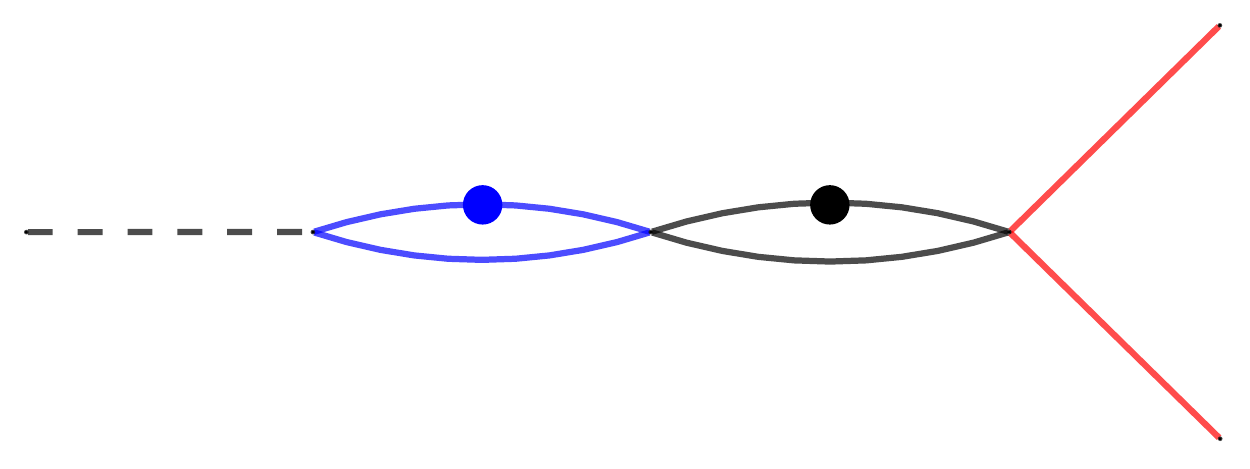}
	}
	\subfloat[\hspace{1.1cm}$\mathcal{T}_{11}$]{%
		\includegraphics[width=0.14\textwidth]{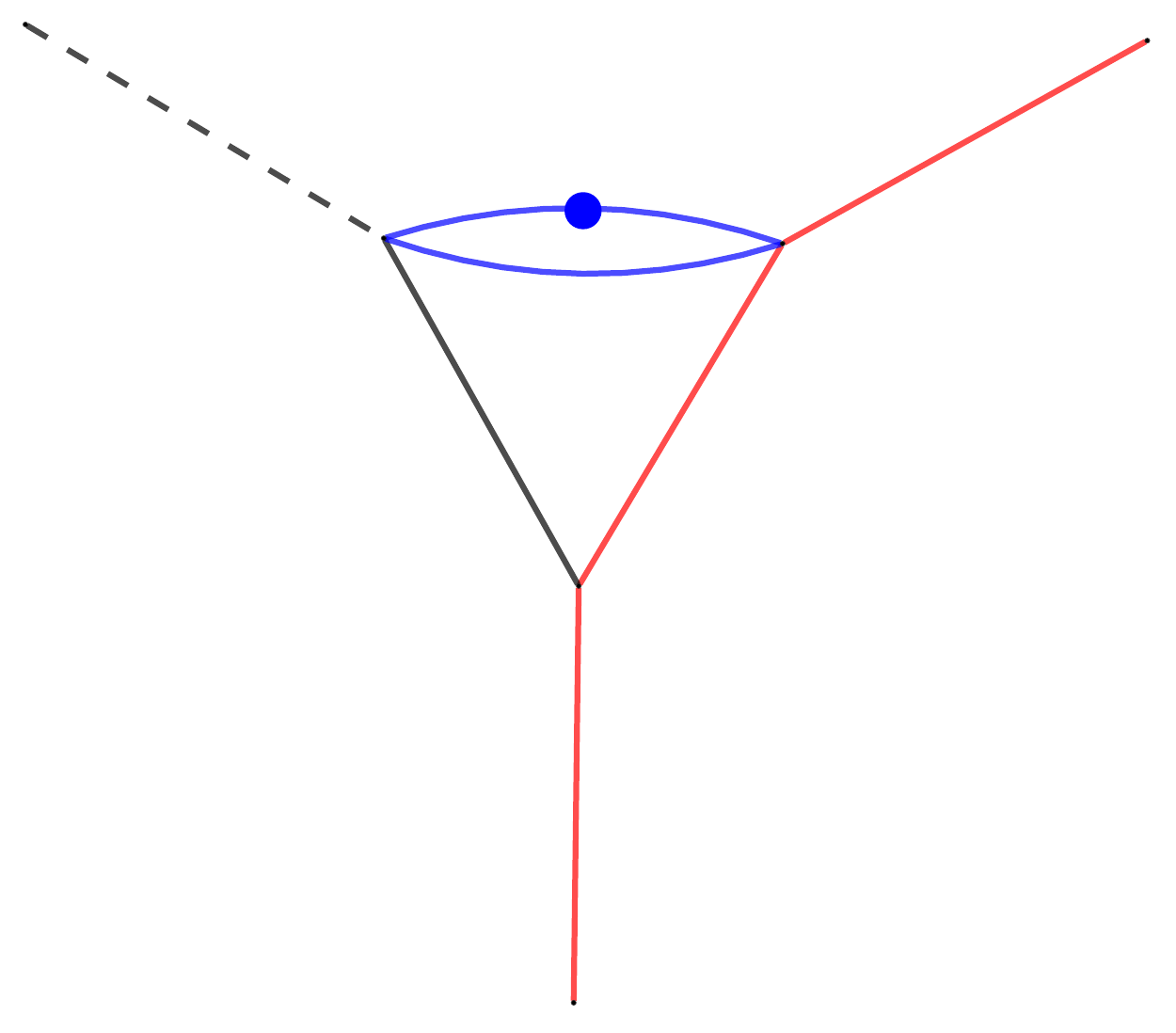}
	}
	\subfloat[\hspace{1.1cm}$\mathcal{T}_{12}$]{%
		\includegraphics[width=0.14\textwidth]{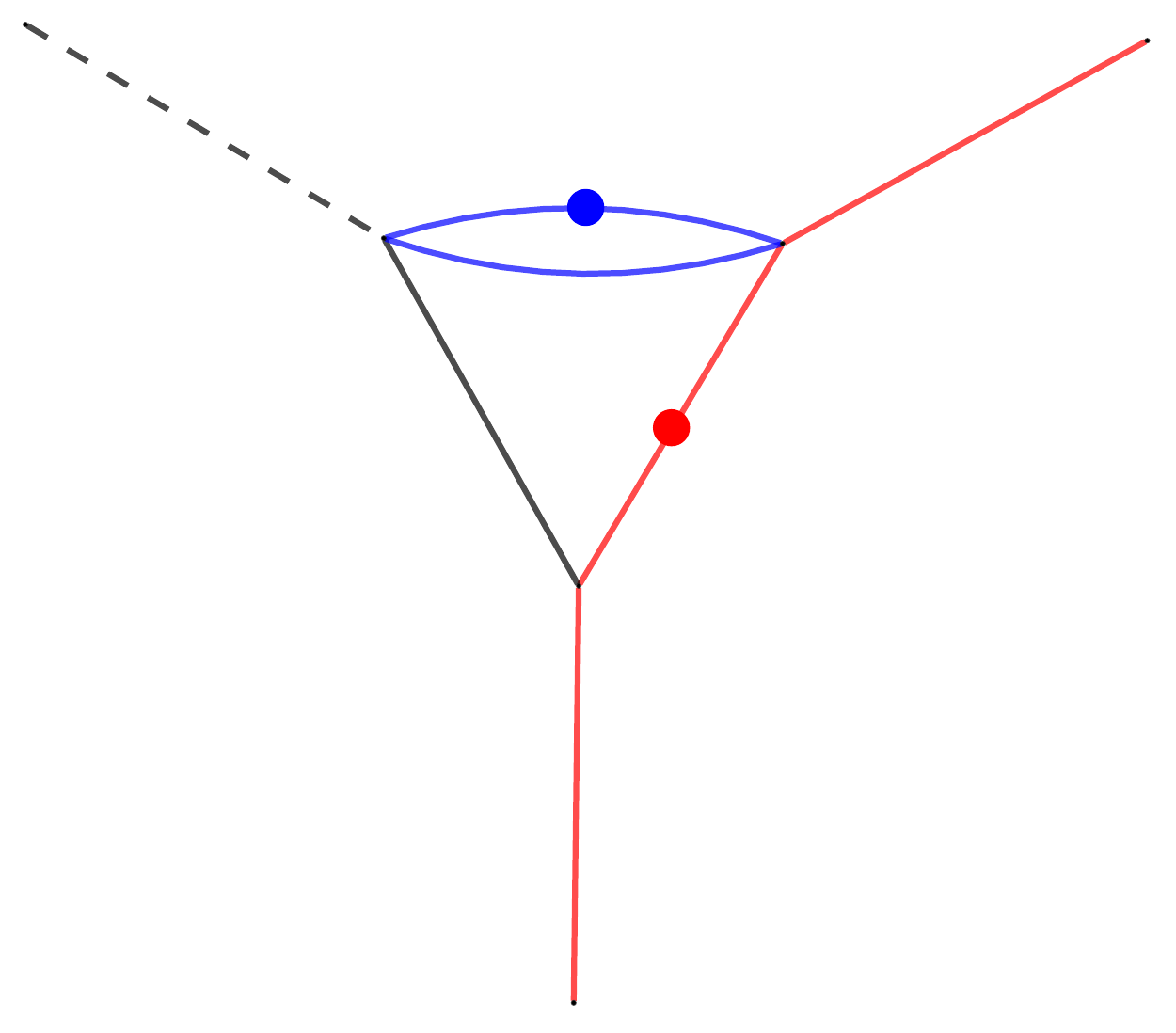}
	}
	\\
	\subfloat[\hspace{1.1cm}$\mathcal{T}_{13}$]{%
		\includegraphics[width=0.14\textwidth]{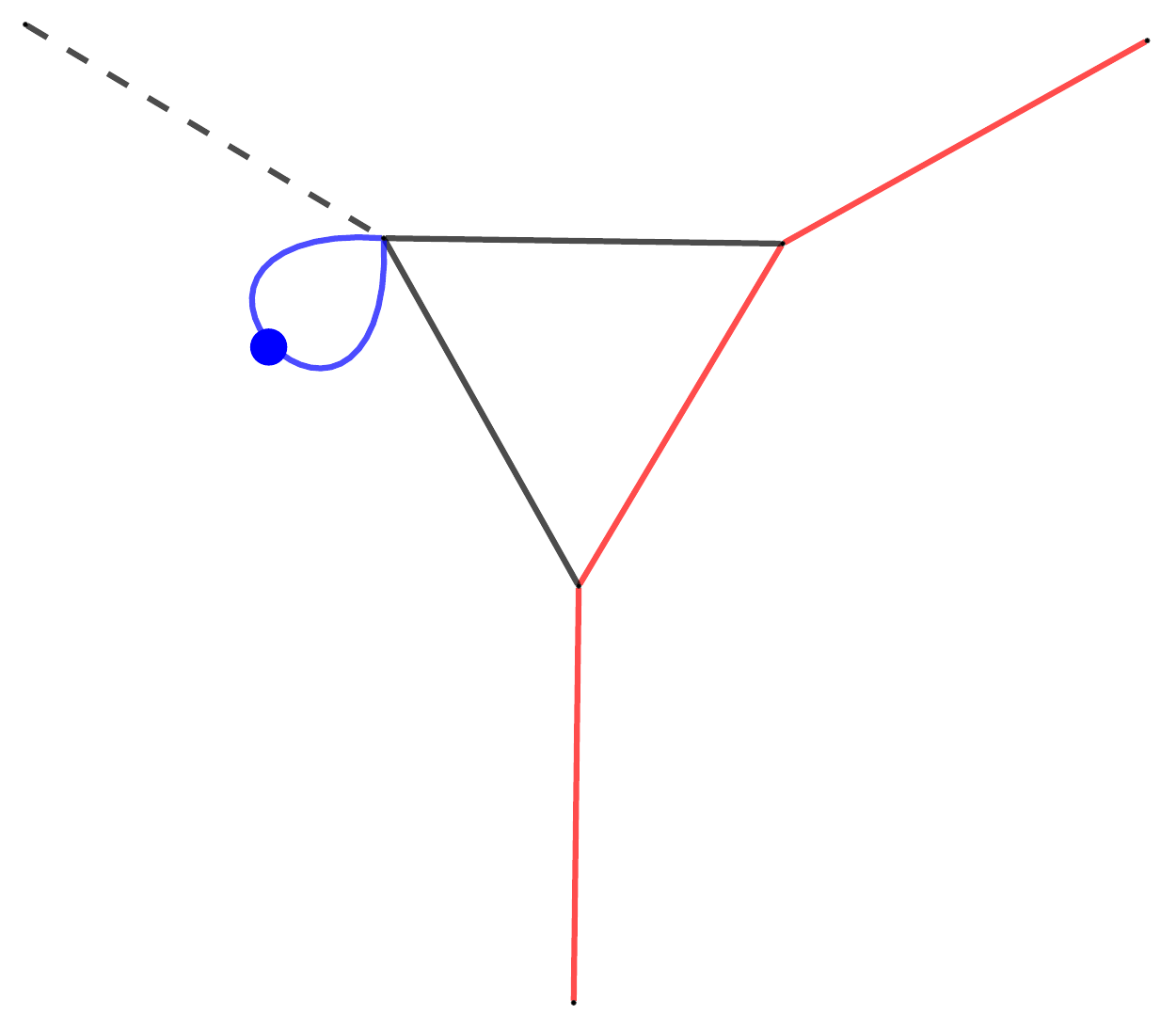}
	}
	\subfloat[\hspace{1.1cm}$\mathcal{T}_{14}$]{%
		\includegraphics[width=0.14\textwidth]{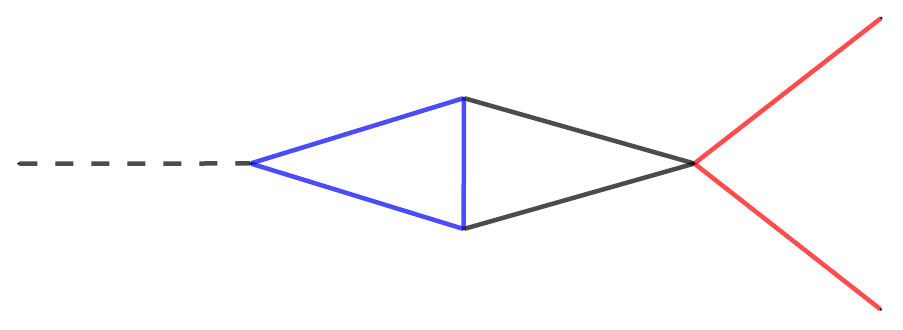}
	}
	\subfloat[\hspace{1.1cm}$\mathcal{T}_{15}$]{%
		\includegraphics[width=0.14\textwidth]{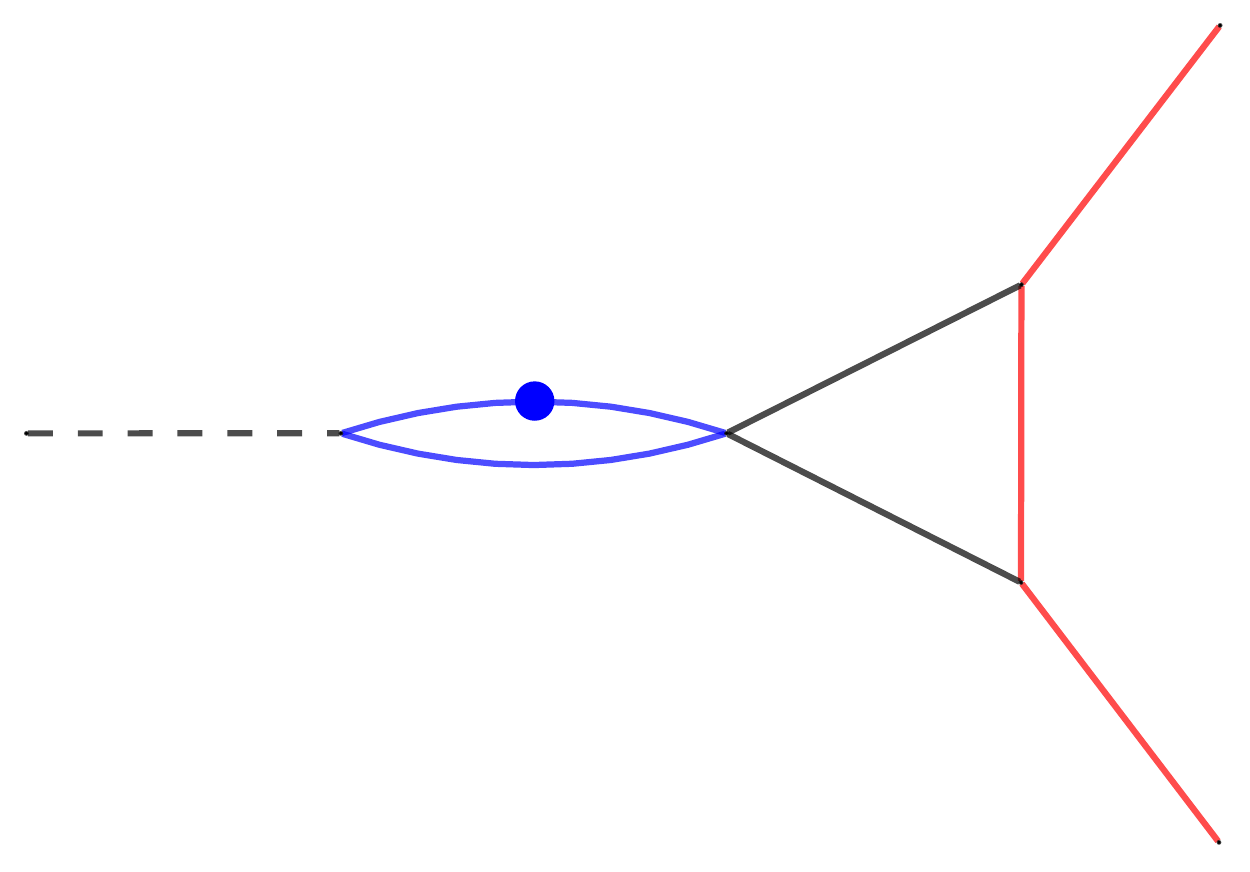}
	}
	\subfloat[\hspace{1.1cm}$\mathcal{T}_{16}$]{%
		\includegraphics[width=0.14\textwidth]{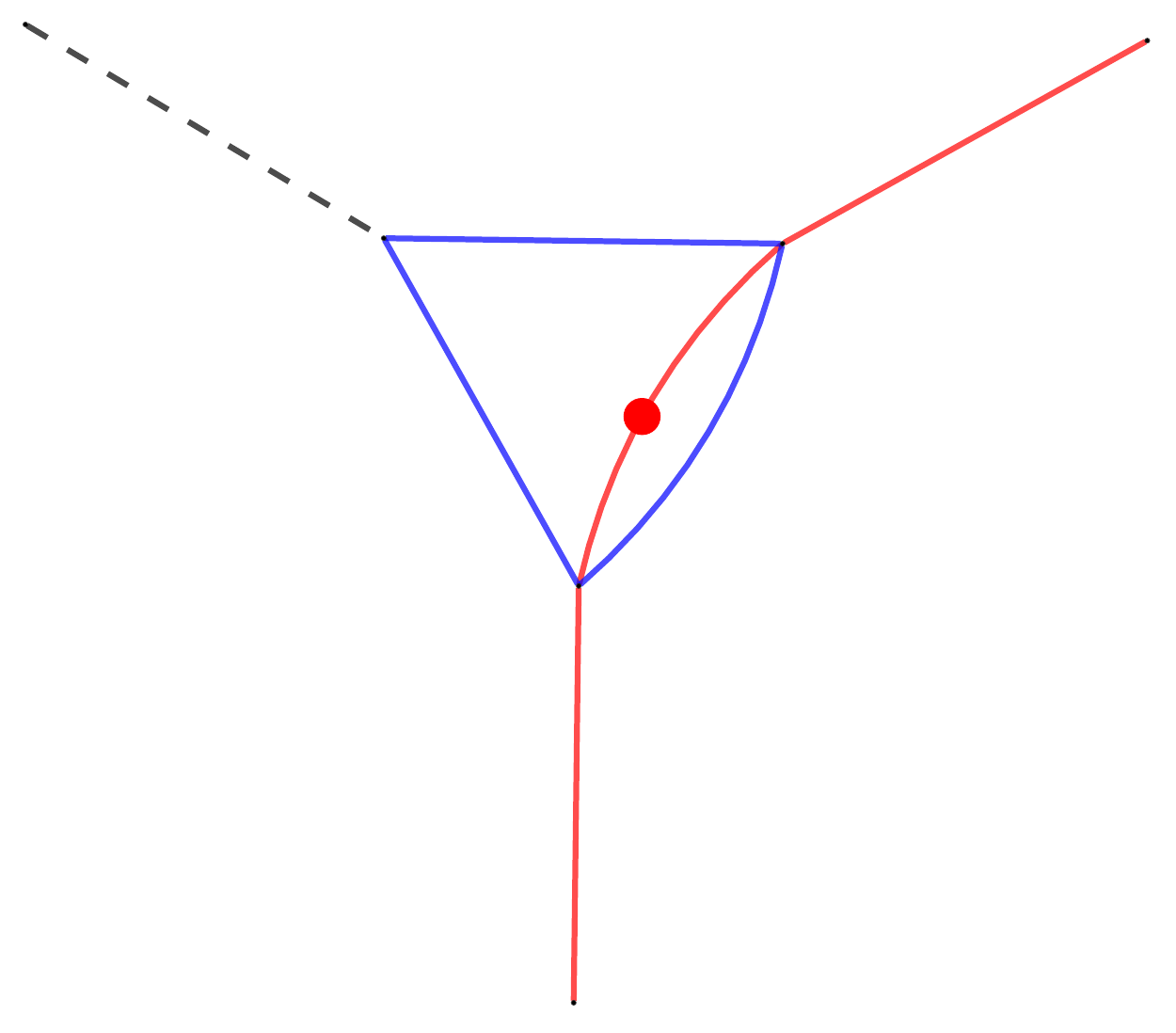}
	}
	\subfloat[\hspace{1.1cm}$\mathcal{T}_{17}$]{%
		\includegraphics[width=0.14\textwidth]{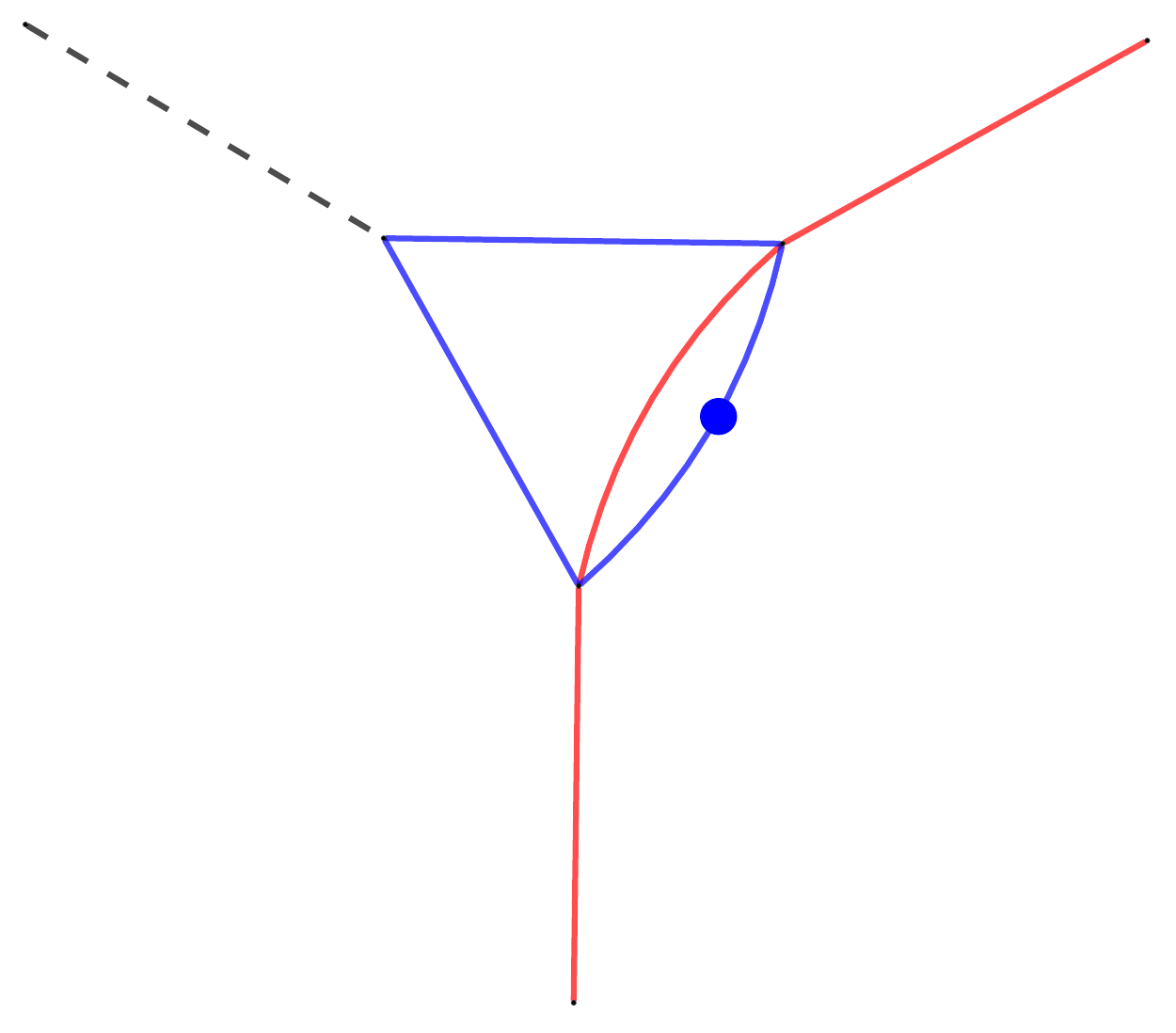}
	}
	\subfloat[\hspace{1.1cm}$\mathcal{T}_{18}$]{%
		\includegraphics[width=0.14\textwidth]{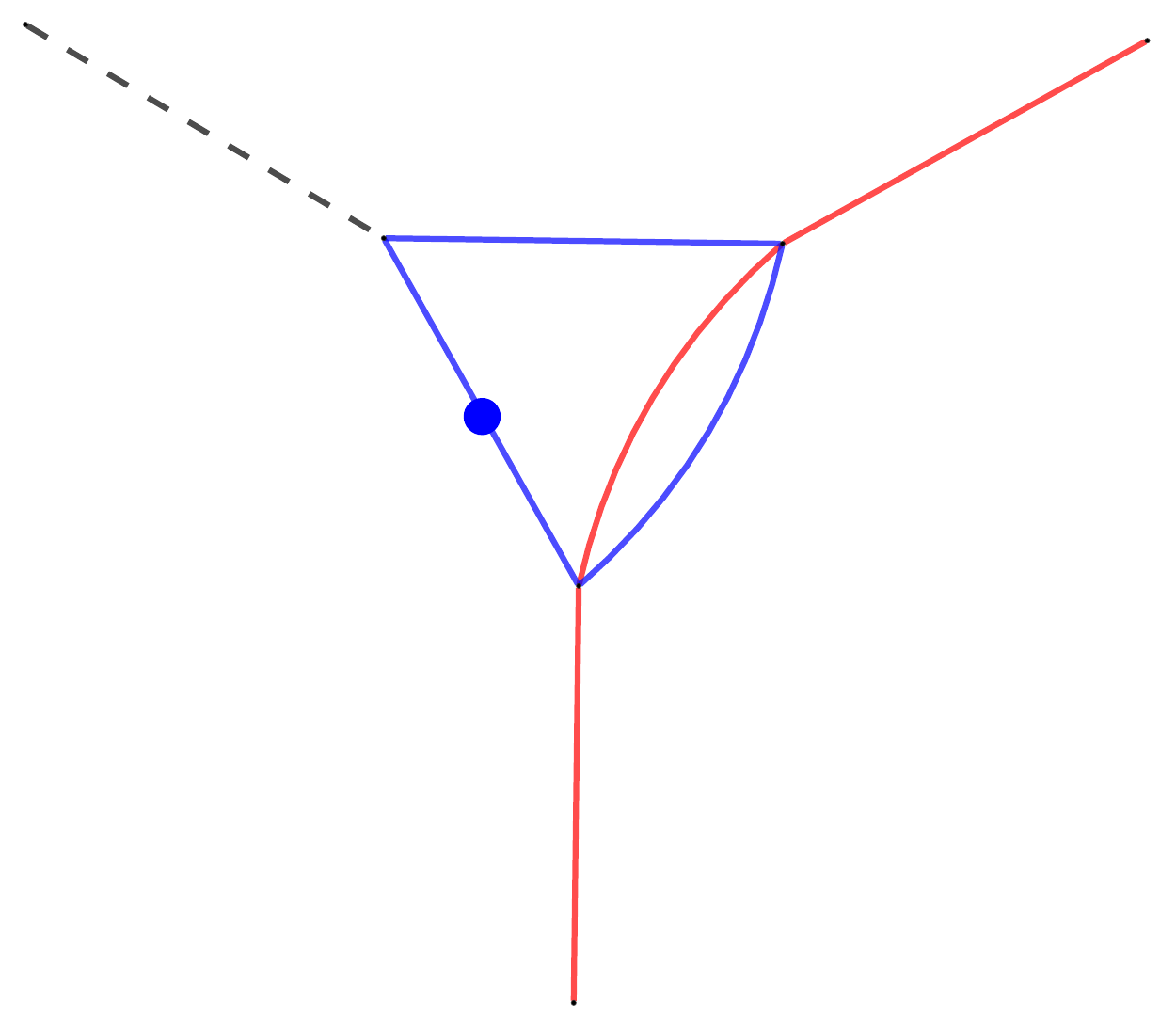}
	}
	\\
	\subfloat[\hspace{1.1cm}$\mathcal{T}_{19}$]{%
		\includegraphics[width=0.14\textwidth]{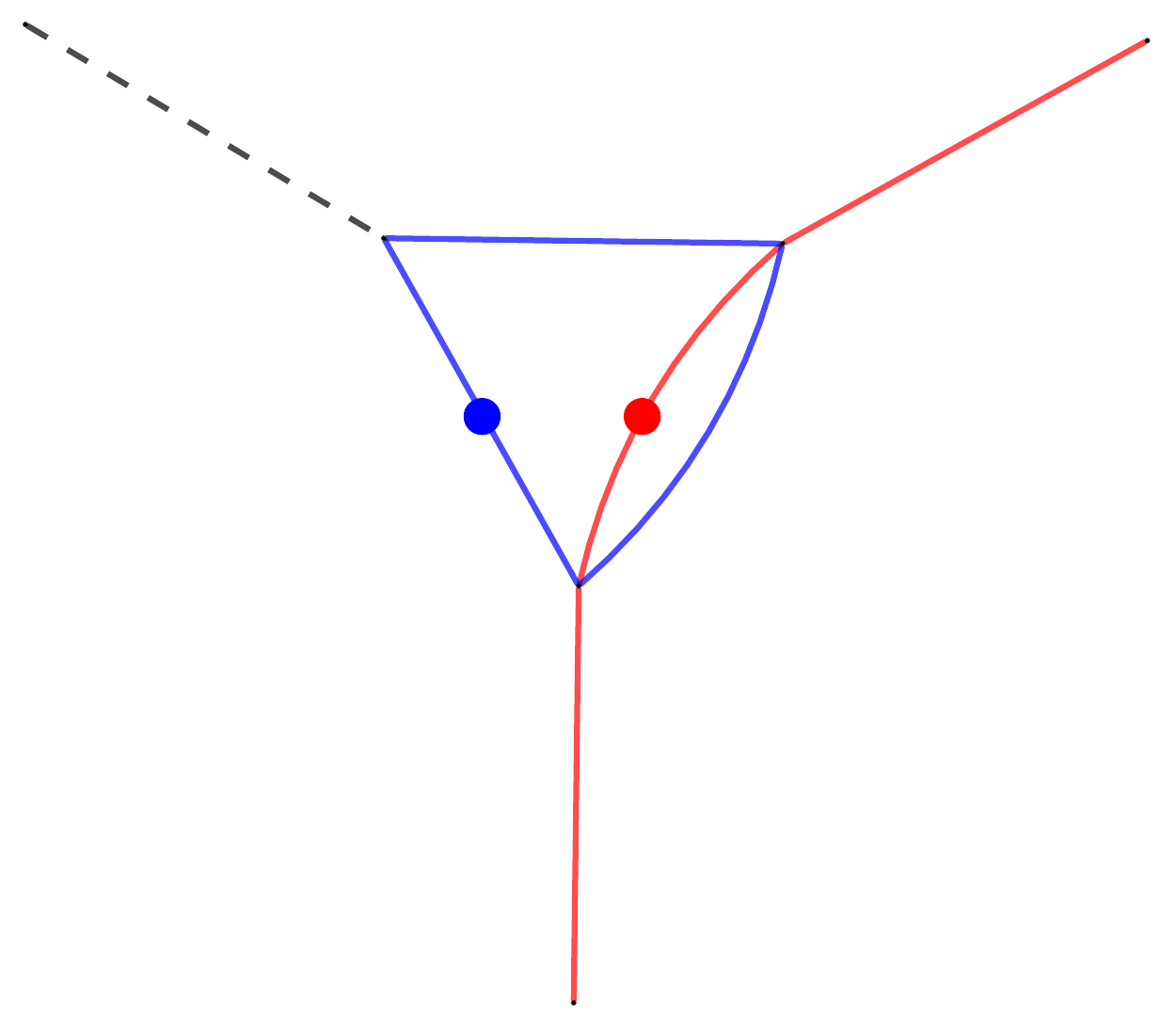}
	}
	\subfloat[\hspace{1.1cm}$\mathcal{T}_{20}$]{%
		\includegraphics[width=0.14\textwidth]{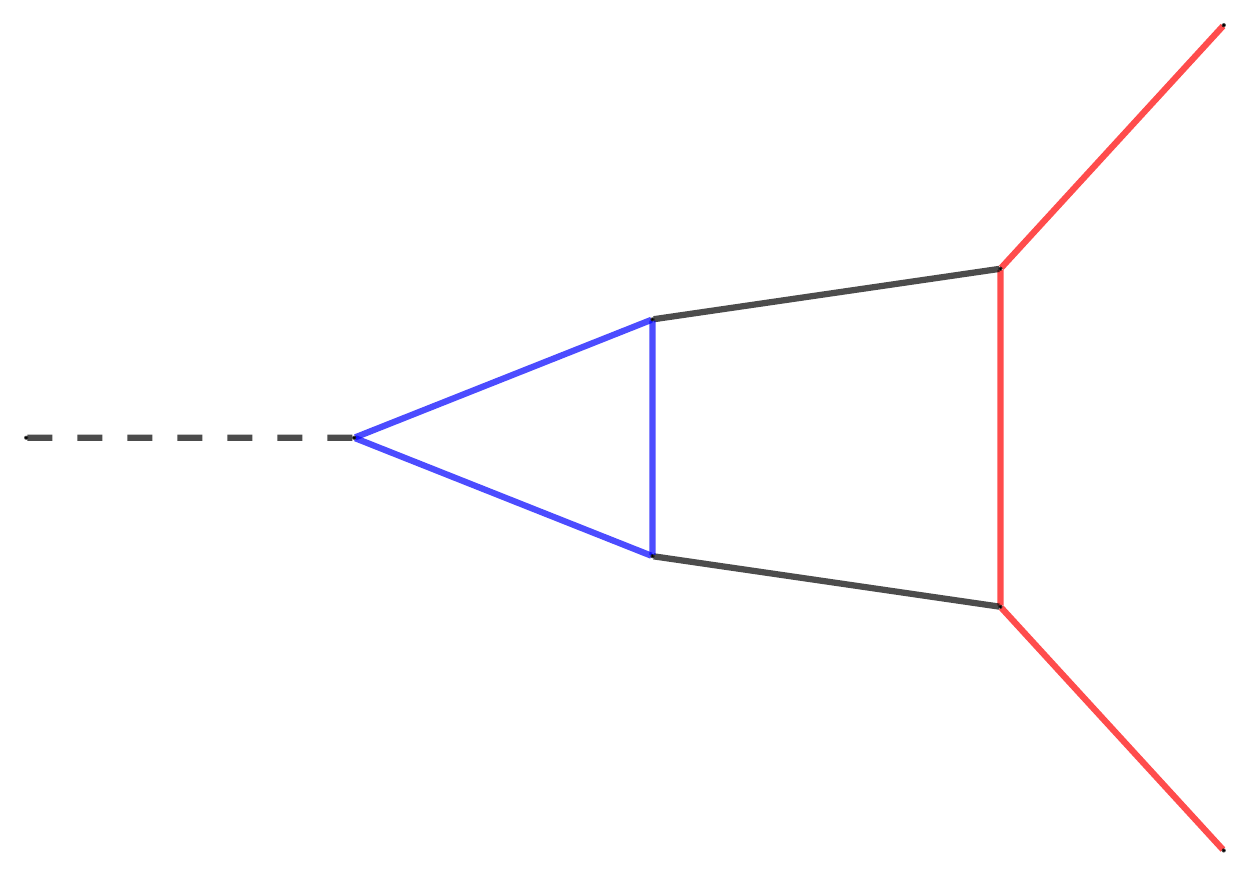}
	}
	\caption{\textbf{\textit{Master integrals required for the evaluation of the scalar and pseudo-scalar form factors}}, corresponding to the two-loop diagrams in Fig.~\ref{fig:2photon}. In each diagram, we indicate the off-shell external mediator (dashed lines), fermions with mass $m_\ell$ (blue lines) and $m_N$ (red lines), massless photons (black lines) and squared denominators (dots).}
	\label{fig:MIsTI}
\end{figure}

In this subsection, we summarize the calculation of the MIs required for the determination of the one-body form
factors $\mathcal{F}_{S}^{1b}(t; m_N,m_\ell)$ and $\mathcal{F}_{P}^{1b}(t; m_N, m_\ell)$ through the differential equations method along the lines of Ref.~\cite{Primo:2018zby}. As discussed in the main text, all the integrals of the type of Eq.~(\ref{eq:family}) can be reduced via IBPs to linear combinations of 20 independent MIs, which obey coupled first order differential equations in the kinematic variables $x$ and $y$ introduced in Eq.~(\ref{eq:vars}). We do this using the IBPs code {\sc Reduze2}~\cite{vonManteuffel:2012np} and {\sc LiteRed}~\cite{Lee:2012cn}. The two systems of differential equations of Eq.~(\ref{eq:sysxy}) can be combined into the total differential of the vector $\mathbf{I}$ of basis MIs,
\begin{equation}
	d \, \mathbf{I}(\epsilon,x,y) = \epsilon \,  d \mathbb{A}(x,y) \,  \mathbf{I}(\epsilon,x,y) \qquad \text{with} 
	\qquad df = dx \frac{\partial f}{\partial x} + dy \frac{\partial f}{\partial y} \ .
	\label{eq:diff_system_appendix}
\end{equation}
The factorization of the dimensional regulator $\epsilon$ from the kinematics dependence of Eq.~(\ref{eq:diff_system_appendix}), which is encoded in the matrix $\mathbb{A}_j$, greatly simplifies the task of determining the expression of the MIs as a Laurent series around $\epsilon=0$ (i.e., $d=4$), since it decouples the solution of the differential equations order by order in $\epsilon$. This factorization is, however, not a general property of the differential equations for Feynman integrals (which usually exhibit a rational dependence on $\epsilon$) and it is rather a consequence of a specific choice of the MIs. In fact, for a given family of Feynman integrals, the basis of MIs is not unique, as any complete set of integrals, which are linearly independent under IBPs, can be taken as a valid basis. Although different basis of MIs are completely equivalent, only certain ones fulfill canonical $\epsilon$-factorized differential equations of the type of Eq.~(\ref{eq:diff_system_appendix}). For the problem under consideration, the following basis has been identified through the Magnus-Dyson algorithm~\cite{Argeri:2014qva},
\begin{alignat}{2}
	\ensuremath \text{I}_1 & = \epsilon^2 \, \mathcal{T}_1 ~, \qquad && 
	\ensuremath \text{I}_2 = \epsilon^2 \, \mathcal{T}_2 ~, \nonumber\\  
	\ensuremath \text{I}_3 & = \epsilon^2 \, \lambda_{\ell} \, \mathcal{T}_3 ~, \qquad && 
	\ensuremath \text{I}_4 = - \epsilon^2 \, t \, \mathcal{T}_4 ~, \nonumber\\
	\ensuremath \text{I}_5 & = \epsilon^2 \, \left(\frac{1}{2} \, \left(-t + \lambda_{\ell}\right) \, \mathcal{T}_4 + \lambda_{\ell} \, \mathcal{T}_5\right) ~, \qquad  &&
	\ensuremath \text{I}_6 = - \epsilon^2 \, t \, \mathcal{T}_6 ~, \nonumber\\
	\ensuremath \text{I}_7 & = \frac{\epsilon^2 \, m_N^2 \, \left(t + \lambda_{\ell}\right) \, \rho_N}{\left(\lambda_N + t\right) \, \rho_{\ell}} \, \left(\mathcal{T}_7 + 2 \, \mathcal{T}_8\right) ~, \qquad && 
	\ensuremath \text{I}_8 = \epsilon^2 \, m_N^2 \, \mathcal{T}_8 ~, \nonumber\\
	\ensuremath \text{I}_9 & = \epsilon^2 \, \lambda_{\ell} \, \mathcal{T}_9 ~, \qquad && \ensuremath \text{I}_{10} = - \epsilon^2 \, t \, \lambda_{\ell} \, \mathcal{T}_{10} ~, \nonumber\\
	\ensuremath \text{I}_{11} & = \epsilon^3 \, \lambda_N \,   \mathcal{T}_{11} ~, \qquad && 
	\ensuremath \text{I}_{12} = \frac{\epsilon^2 \, \lambda_{\ell}}{4 \, t} \, \Big(\left(t - \lambda_N\right) \, \left(\mathcal{T}_4 + 2 \, \mathcal{T}_5\right) - 4 \, m_N^2 \, \lambda_N \, \mathcal{T}_{12}\Big) ~, \nonumber\\
	\ensuremath \text{I}_{13} & = \epsilon^3 \, \lambda_N \, \mathcal{T}_{13} ~, \qquad && 
	\ensuremath \text{I}_{14} = \epsilon^3 \, (-1 + 2 \, \epsilon) \, t \, \mathcal{T}_{14} ~, \nonumber\\
	\ensuremath \text{I}_{15} & = \epsilon^3 \, \lambda_{\ell} \,  \lambda_N \, \mathcal{T}_{15} ~, \qquad && 
	\ensuremath \text{I}_{16} = \epsilon^3 \, \lambda_N \, \mathcal{T}_{16} ~, \nonumber\\ 
	\ensuremath \text{I}_{17} & = \epsilon^3 \, \lambda_N \, \mathcal{T}_{17} ~, \qquad && 
	\ensuremath \text{I}_{18} = \frac{\epsilon^2}{t} \, \Big(\lambda_{\ell} \, \left(-t  + \lambda_N\right) \, \mathcal{T}_9 + \epsilon \, \left(t - \lambda_{\ell}\right) \, \lambda_N \, \mathcal{T}_{17} + (-1 + 2 \, \epsilon) \, \lambda_{\ell} \,\lambda_N \, \mathcal{T}_{18} \Big) ~, \nonumber\\
	\ensuremath \text{I}_{19} & = \frac{\epsilon^2}{2 \, t} \, \Big( t \, \left(\lambda_{\ell} - t\right) \, \mathcal{T}_3 - 2 \, t \, m_{\ell}^2 \, \left(\mathcal{T}_7 + 2 \, \mathcal{T}_8\right) && + \left(4 \, t \, m_{\ell}^2 + \lambda_{\ell} \, \left(\lambda_N - t\right)\right) \, \mathcal{T}_9 + \epsilon \, \frac{4 \, t^2 \, m_N^2}{\lambda_N + t} \, \mathcal{T}_{16} \nonumber\\
	& \qquad + \epsilon \, \left(\lambda_N \, \left(t - \lambda_{\ell}\right) - 4 \, t \, m_{\ell}^2\right) \, \, \mathcal{T}_{17} \, \, \, && +  (2 \, \epsilon - 1) \, \left(4 \, t \, m_{\ell}^2 + \lambda_{\ell} \, \lambda_N - t^2\right) \, \mathcal{T}_{18} + 2 \, t^2 \, \left(m_{\ell}^2 - m_N^2\right) \, \mathcal{T}_{19} \Big) ~, \nonumber\\
	\ensuremath \text{I}_{20} & = -\epsilon^4 \, t \, \lambda_N \, \mathcal{T}_{20} ~, 
	\label{eq:canonical_MIs_appendix}
\end{alignat}
with $\mathcal{T}_i$'s depicted in Fig.~\ref{fig:MIsTI}. In Eqs. (\ref{eq:canonical_MIs_appendix}) we have introduced the notation
\begin{equation}
	\lambda_i = \sqrt{-t} \, \sqrt{4 \, m^2_i - t} ~, \quad \rho_i = \sqrt{\frac{2 \, m^2_i - t - \lambda_i}{m^2_i}}, \quad \text{with: }i = N, \ell ~.
\end{equation}

A few comments are in order. First, the $\epsilon$ dependence of the coefficients in Eq.~(\ref{eq:canonical_MIs_appendix}) has been adjusted in such a way that all integrals are finite in the four dimensional limit. Therefore, they admit, rather than a Laurent series around  $\epsilon= 0$, a Taylor expansion of the type
\begin{equation}
	\mathbf{I}(\epsilon,x,y) = \sum_{k=0}^{\infty} \, \mathbf{I}^{(k)}(x,y) \, \epsilon^k ~.
	\label{eq:canonical_MIs_Taylor_expansion_appendix}
\end{equation}
Second, we redefine at this stage of the computation the loop integration measure as
\begin{equation}
	\int \widetilde{d^d k_i} = \left( \frac{m^2_l}{\mu^2}\right)^{\epsilon} \, \int \frac{d^d k_i}{i \, \pi^{d/2} \, \Gamma_{\epsilon}} ~,
	\label{eq:integral_measure_definition_appendix}
\end{equation}
with $\mu$ being the ’t Hooft scale of dimensional regularization and $\Gamma_{\epsilon}=\Gamma(1+ \epsilon)$ is the $\Gamma$ function. In this way, all $\ensuremath \text{I}_i$
are dimensionless functions of the variables $x$ and $y$ and the two-loop tadpole integral is conveniently normalized to one, namely $\ensuremath \text{I}_1 (\epsilon, x, y) = 1$. The standard normalization $1/(2 \pi)^d$ can be then be restored in the final expression of $\mathcal{F}_S^{1b}$ and $\mathcal{F}_P^{ 1b}$. Finally, it is worth noticing that, while all the $20$ listed MIs appear in the decomposition of the scalar form factor $\mathcal{F}_S^{1b}$, the pseudo-scalar one $\mathcal{F}_P^{1b}$ turns out to be decomposed as a linear combination of only $10$ MIs, namely $\{\text{I}_4,\text{I}_5,\text{I}_7,\text{I}_8,\text{I}_{11},\text{I}_{12},\text{I}_{14},\text{I}_{16},\text{I}_{17},\text{I}_{20}\}$.

Owing to the $\epsilon$-factorization of the differential equation, it is straightforward to write a formal expression of the Taylor coefficients $\ensuremath \mathbf{I}^{(k)}(x, y)$ introduced in Eq.~(\ref{eq:canonical_MIs_Taylor_expansion_appendix}) in terms of iterated integrals over the kernel matrix $\mathbb{A}(x, y)$,
\begin{equation}
	\mathbf{I}^{(k)}(x,y) = \sum_{i=0}^{k} \, \int_{\gamma} \underbrace{d \mathbb{A} \dots d \mathbb{A}}_{\text{$i$ times}} \, \mathbf{I}^{(k-i)}(x_0,y_0)
	\label{eq:I_k_iterated_integrals_appendix}
\end{equation}
where $\gamma$ is a regular path in the $xy$-plane and $\mathbf{I}^{(i)}(x_0 , y_0)$ is an integration constant. The dependence of the canonical differential equations on $x$ and $y$ turns out to be also in the most suitable form to reduce the repeated integrals of Eq.~(\ref{eq:I_k_iterated_integrals_appendix}) to well-known transcendental functions. In fact, the total differential of $\mathbb{A}_j (x, y)$ is a $d$log-form of the type
\begin{equation}
	d \mathbb{A}= \sum_{i=1}^{12} \mathbb{M}_i \,  d \log  (\eta_i)
	\label{eq:differential_A_matrix_appendix} ~, 
\end{equation}
with
\begin{equation}
	\begin{split}
		& \eta_1 = x ~,\qquad\qquad \;\,
		\eta_2  = 1 + x ~, \qquad \;\;\;\,
		\eta_3  = 1 - x ~, \nonumber \\
		& \eta_4 = 1 + x^2 ~, \qquad \;
		\eta_5 = y ~, \qquad\qquad \;\;\,
		\eta_6 = 1 + y ~,\nonumber \\
		& \eta_7 = 1 - y ~, \qquad \;\;\;
		\eta_8 = 1 + y^2 ~, \qquad \;\;
		\eta_9 = x + y ~, \nonumber \\
		& \eta_{10} = x - y ~, \qquad
		\eta_{11} = 1 + xy ~,   \qquad
		\eta_{12} = 1 - xy ~.
	\end{split}
	\label{eq:alphabet_appendix}
\end{equation}
Upon factorization over the complex numbers of $\eta_i$, it is then possible to express $\mathbf{I}^{(k)}(x, y)$ in terms of GPLs,
\begin{equation}
	G \left( \Vec{a}_n;t \right)=G \left(a_1,\Vec{a}_{n{-}1};t \right)= \int_{0}^{t} \frac{dt'}{t'-a_1} \, G \left(\Vec{a}_{n{-}1};t' \right), \qquad G\left( \Vec{0}_{n};t \right) = \frac{1}{n!} \, \log^{n} \left( t \right) \, .
	\label{eq:GPLs_def_appendix}
\end{equation}

The iterative integration of Eq.~(\ref{eq:diff_system_appendix}) by means of Eq.~(\ref{eq:I_k_iterated_integrals_appendix}) leaves, at every order in the $\epsilon$ expansion, an undetermined boundary constant for each MI. In principle, these constants can be determined by exploiting the knowledge of the analytic expression of the MIs in specific kinematic limits. Unfortunately, for the most complicated integrals, this information is generally not easy to obtain through direct integration techniques. Nevertheless, the differential equations are shown to also encode enough information to determine most of the integration constants and, therefore, minimize the needed amount of external independent information. In most cases, the differential equations exhibit singularities, not only at the physical thresholds of the MIs, but also in kinematics points which are regular points of Feynman integrals (and, for this reason, often referred to as pseudo-thresholds). The requirement of the solution of the differential equations being regular at such nonphysical pseudo-thresholds can be translated into linear equations for the integration constants. In particular, the boundary constants have been determined in the following way:
\begin{itemize}
	\item The tadpoles $\ensuremath \text{I}_{1,2}$ and the factorized integral $\ensuremath \text{I}_{6}$ are obtained by direct integration and provided as an external input to the system of differential equations:
	\begin{equation}
		\begin{split}
			& \ensuremath \text{I}_{1} =1 \, , \nonumber \\
			& \ensuremath \text{I}_{2}(x,y,\epsilon) = \left( \frac{(1 - x^2)^2 \, y^2}{(1 - y^2)^2 \, x^2} \right)^{-\epsilon} \, , \nonumber \\
			& \ensuremath \text{I}_{6}(x,\epsilon)= \left( \frac{(1 - x^2)^2}{x^2} \right)^{-\epsilon} \, \left(1 - \frac{\pi^2}{6} \, \epsilon^2 - 2 \, \zeta_3 \, \epsilon^3 - \frac{\pi^4}{40} \, \epsilon^4 + \mathcal{O}(\epsilon)^5 \right) ~.
		\end{split}
	\end{equation}
	
	\item The boundary constants of the integrals $\ensuremath \text{I}_{3,4,5,9,10,11,12,14,16,17,18,19}$ are determined by imposing their regularity at the pseudo-threshold $t \to 0$. In particular, due to the prefactors that appear in the definition of the canonical MIs of Eq.~(\ref{eq:canonical_MIs_appendix}), $\ensuremath \text{I}_{3,4,5,9,10,14,16,17}$ vanish in this limit.\footnote{The fact that $\ensuremath \text{I}_{14}$ vanishes in the $t \to 0$ limit can be inferred from the results of Ref.~\cite{Anastasiou:2006hc}.} The same conclusion is inferred for $\ensuremath \text{I}_{11,12,18,19}$, analyzing the differential equation in this limit. $\ensuremath \text{I}_{10}$ results to be vanishing in the $t \to 0$ limit, due to the faster convergence of the factorized (canonical) massive bubble, with respect to the massless one.
	
	\item The boundary constants of $\ensuremath \text{I}_{7,8}$ are determined thanks to the regularity at $m^2_N \to 0$; specifically, due to the prefactors in Eq.~(\ref{eq:canonical_MIs_appendix}), they vanish in this limit.
	
	\item The boundary constants of $\ensuremath \text{I}_{13,15,20}$ are determined by demanding regularity in the $t \to 4 m^2_N$ limit. Thanks to the prefactors in Eq.~(\ref{eq:canonical_MIs_appendix}), they vanish in this limit.
\end{itemize}

The solution of Eq.~(\ref{eq:diff_system_appendix}) is originally derived in the region $0 < x < 1$ $\wedge$  $0 < y < x$. In such region, all the polynomials $\eta_{i}$ are real and positive and yield real-valued expression of the MIs. The definition of the kinematic variables given in Eq.~(\ref{eq:vars}) can be inverted by choosing the solutions
\begin{equation}
	x= \frac{1}{2} \left( \sqrt{4 - \sigma_{\ell}} - \sqrt{- \sigma_{\ell}} \right) ~, \qquad
	y = \frac{1}{2} \left( \sqrt{4 - \sigma_{N}} - \sqrt{- \sigma_{N}} \right) ~, \qquad  
	\sigma_i = \frac{t}{m^2_i} ~, \qquad i = N,\ell  ~,
	\label{eq:x_y_inversion_appendix}
\end{equation}
so that the region $0 < x < 1$ $\wedge$ $0 < y < x$ corresponds to the kinematic configuration $t < 0$ $\wedge$ $0 < m^2_N <m^2_{\ell}$. At negative $t$, the same definition of $x$ and $y$ is also valid in the reversed mass hierarchy $0 < m^2_{\ell}< m^2_N$ and therefore, can be used over the whole kinematic region of interest for our the study of the form factors. However, once the analytical expressions of the MIs are known in one region, analytic continuation can be used to obtain the values of the MIs in any other kinematic configuration of physical interest. In practice, this can be achieved by propagating the Feynman prescription for positive values of the square momentum transfer, $t \to t + i 0^{+}$ to the $x$ and $y$ variables, so that they acquire the proper imaginary part. All the GPLs expressions in this work are manipulated via {\sc PolyLogTools}~\cite{Duhr:2019tlz} and an \emph{in-house} code. They are numerically evaluated using {\sc GiNaC}~\cite{Vollinga:2004sn}. In some cases, high precision numerical values for the boundary constants are fitted to analytic expressions thanks to the {\sc PSLQ} algorithm~\cite{PSLQ}.

\subsection{Master integrals for the equal-mass case}
\label{sec:AppendixA:equalmass}

In this case ($m_N = m_\ell \equiv m$), all the scalar integrals are reduced via IBPs to linear combinations of $15$ MIs. They obey a system of differential equations which reads
\begin{equation}
	d \, \mathbf{I}(\epsilon,w) = \epsilon \, d\mathbb{A}(w) \,  \mathbf{I}(\epsilon,w)
	\qquad \text{with} \qquad df = dw \, \frac{\partial f}{\partial w} ~.
	\label{eq:DEQ_Isomass_appendix}
\end{equation}
where the variable $w$ is related to the only scales present in the problem, namely $t$ and $m^2$, via the relation
\begin{equation}
	- \frac{t}{m^2} = \frac{(1 - w)^2}{w} ~.
	\label{eq:landau_vars_appendix}
\end{equation}
A particular choice of MIs, for which the factorization in Eq.~(\ref{eq:DEQ_Isomass_appendix}) holds, is given by
\begin{alignat}{2}
	\ensuremath \text{I}_1 & = \epsilon^2 \, \mathcal{T}_1 ~, \qquad && 
	\ensuremath \text{I}_2  = \epsilon^2 \, \lambda_m \, \mathcal{T}_2 ~, \nonumber\\ 
	\ensuremath \text{I}_3 & = - \epsilon^2 \, t \, \mathcal{T}_3 ~, \qquad && 
	\ensuremath \text{I}_4 = \epsilon^2 \, \left(\frac{1}{2} \, \mathcal{T}_3 \, \left(\lambda_m + t\right) + \mathcal{T}_4 \, \lambda _m\right) ~, \nonumber\\ 
	\ensuremath \text{I}_5 & = - \epsilon^2 \, t \, \mathcal{T}_5 ~, \qquad && 
	\ensuremath \text{I}_6 = \epsilon^2 \, m^2 \, \mathcal{T}_6 ~, \nonumber\\
	\ensuremath \text{I}_7 & = - \epsilon^2 \, t \, \lambda_m \,  \mathcal{T}_7 ~, \qquad && 
	\ensuremath \text{I}_8 = \epsilon^3 \, \lambda_m \, \mathcal{T}_8 ~, \nonumber\\
	\ensuremath \text{I}_9 & = \epsilon^2 \, \left(\frac{1}{4} \, \left(4 \, m^2 - \lambda_m - t\right) \, \left(\mathcal{T}_3 + 2 \, \mathcal{T}_4\right) + m^2 \, \left(4 \, m^2 - t\right) \, \mathcal{T}_9\right) ~, \qquad && 
	\ensuremath \text{I}_{10} = \epsilon^3 \, \lambda_m \, \mathcal{T}_{10} ~, \nonumber\\
	\ensuremath \text{I}_{11} & = \epsilon^3 \, (1 - 2 \, \epsilon) \, t \, \mathcal{T}_{11} ~, \qquad && 
	\ensuremath \text{I}_{12} = \epsilon^3 \, t \, \left(t - 4 \, m^2\right) \, \mathcal{T}_{12} ~, \nonumber\\   
	\ensuremath \text{I}_{13} & = \epsilon^3 \, \lambda_m \, \mathcal{T}_{13} ~, \nonumber\\
	\ensuremath \text{I}_{14} & = \epsilon^2 \, \Big(\left(4 \, m^2 - \lambda_m - t\right) \, \left(\mathcal{T}_2 - \epsilon \, \mathcal{T}_{13}\right) + (2 \, \epsilon - 1) \, \left(4 \, m^2 - t\right)\, \mathcal{T}_{14} \Big) ~, \nonumber\\ 
	\ensuremath \text{I}_{15} & = - \epsilon^4 \, \lambda_m \, t \, \mathcal{T}_{15} ~,
	\label{eq:canonical_MIs_equal_mass_appendix}
\end{alignat}
where we have introduced
\begin{equation}
	\lambda_m = \sqrt{-t} \, \sqrt{4 \, m^2 - t} ~.
\end{equation}

\begin{figure}
	\centering
	\captionsetup[subfigure]{labelformat=empty}
	\subfloat[\hspace{1.2cm}$\mathcal{T}_1$]{%
		\includegraphics[width=0.14\textwidth]{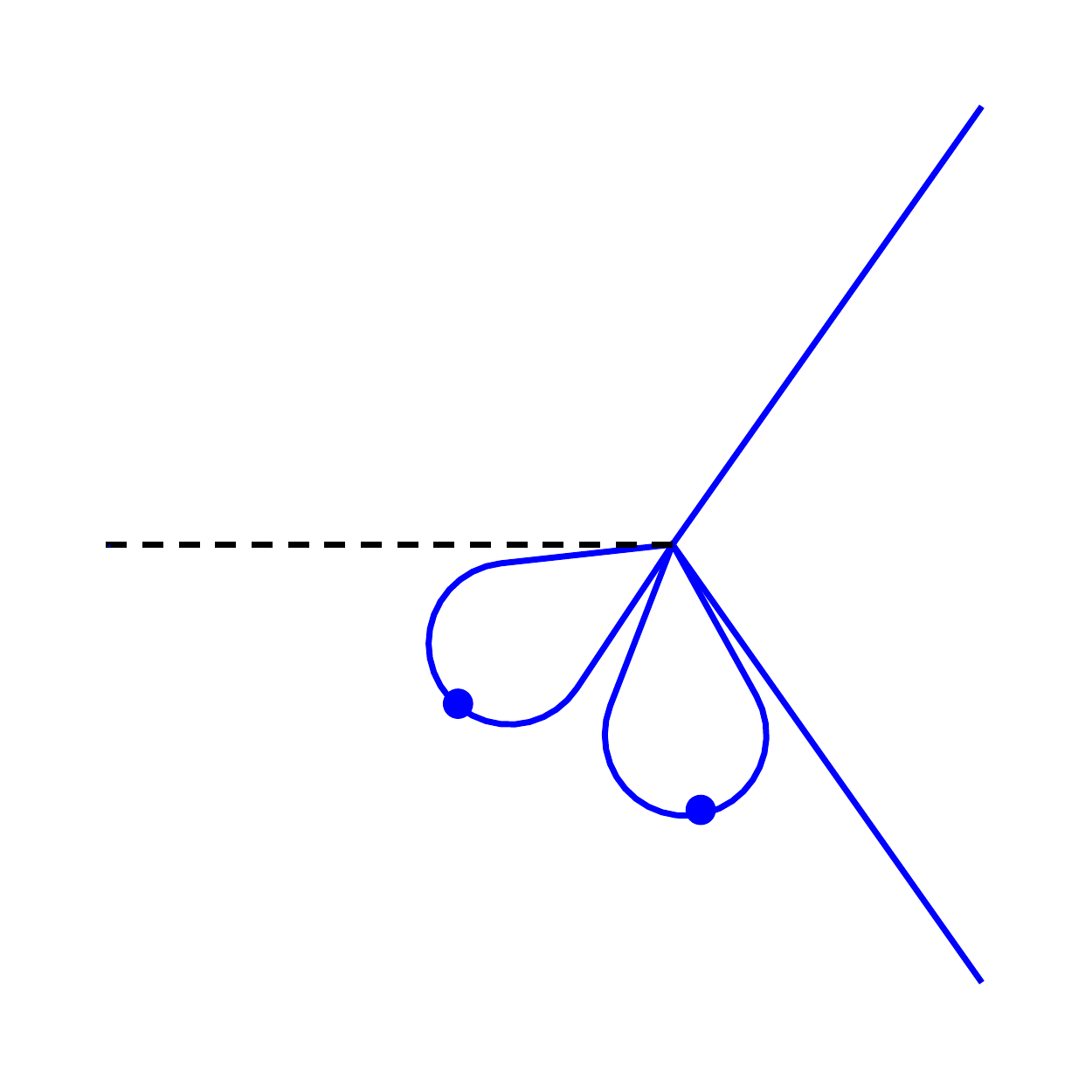}
	}
	\subfloat[\hspace{1.2cm}$\mathcal{T}_2$]{%
		\includegraphics[width=0.14\textwidth]{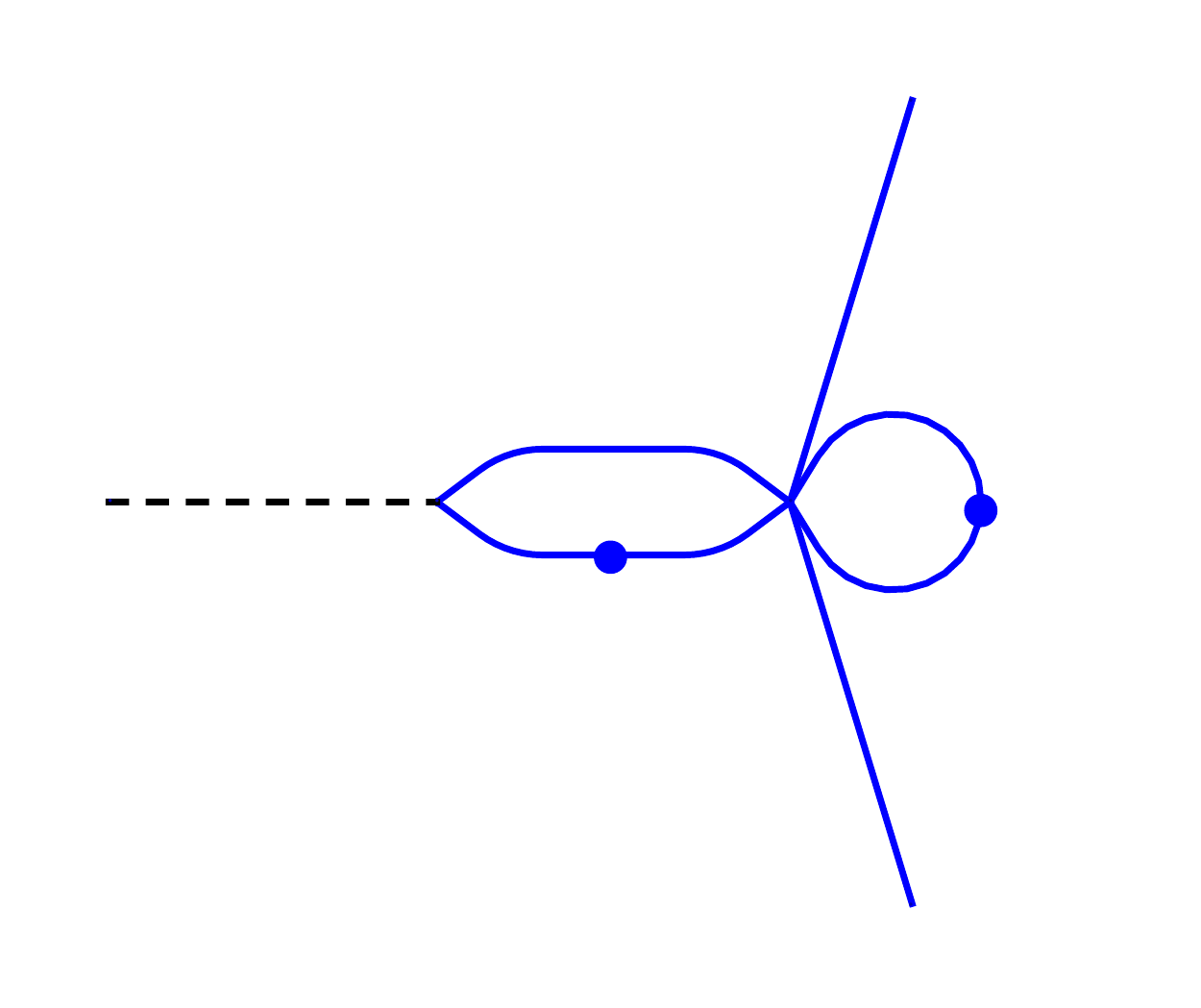}
	}
	\subfloat[\hspace{1.2cm}$\mathcal{T}_3$]{%
		\includegraphics[width=0.14\textwidth]{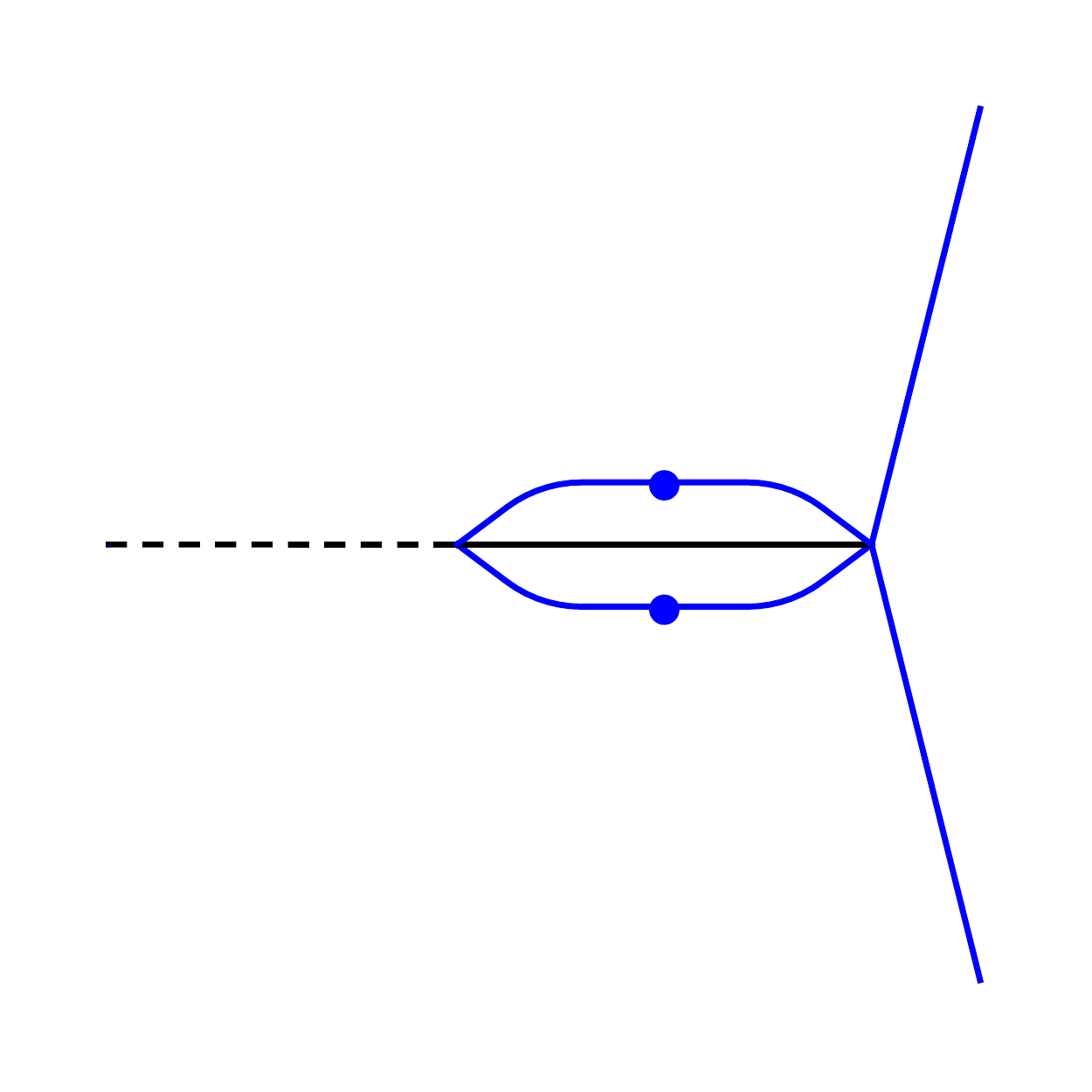}
	}
	\subfloat[\hspace{1.2cm}$\mathcal{T}_4$]{%
		\includegraphics[width=0.14\textwidth]{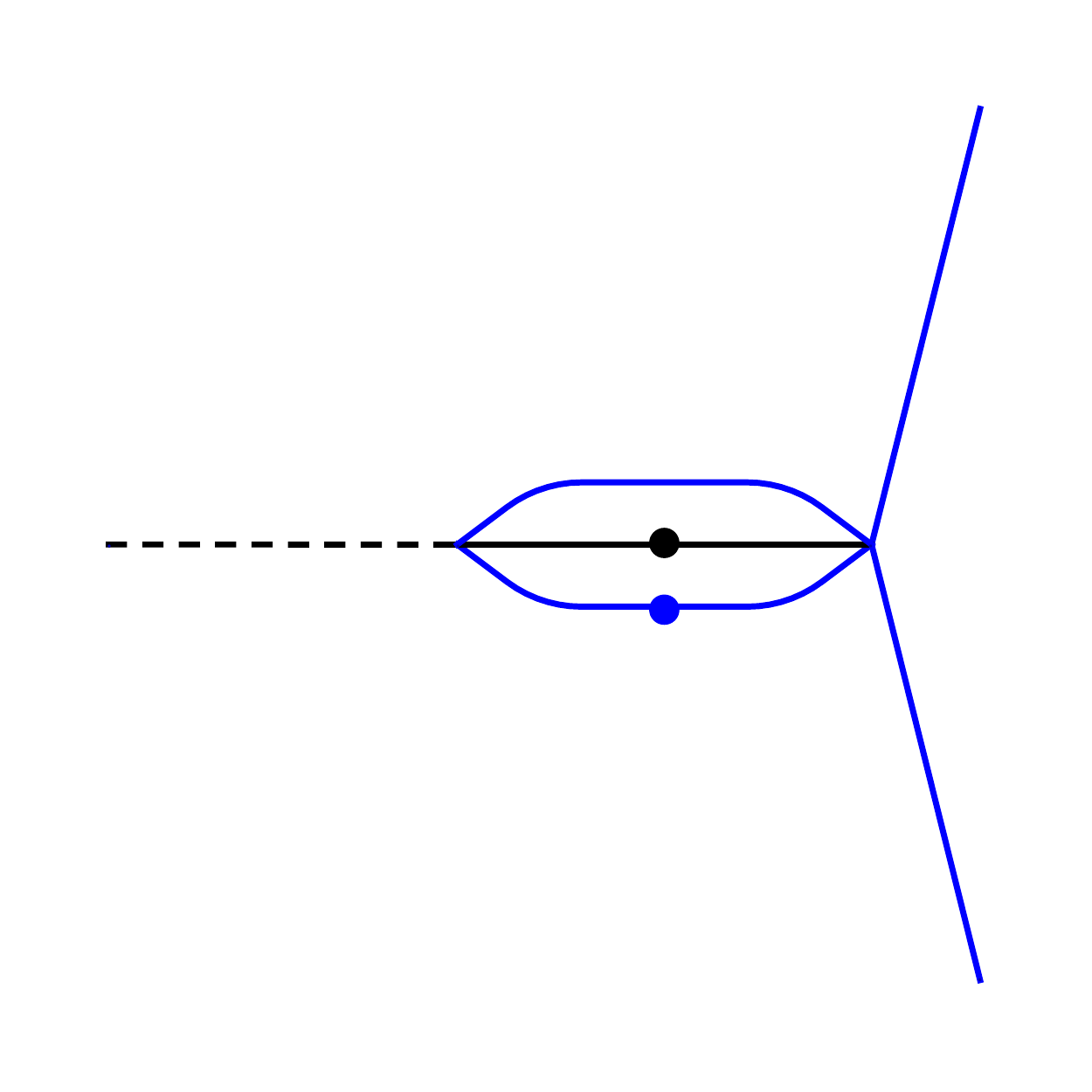}
	}
	\subfloat[\hspace{1.2cm}$\mathcal{T}_5$]{%
		\includegraphics[width=0.14\textwidth]{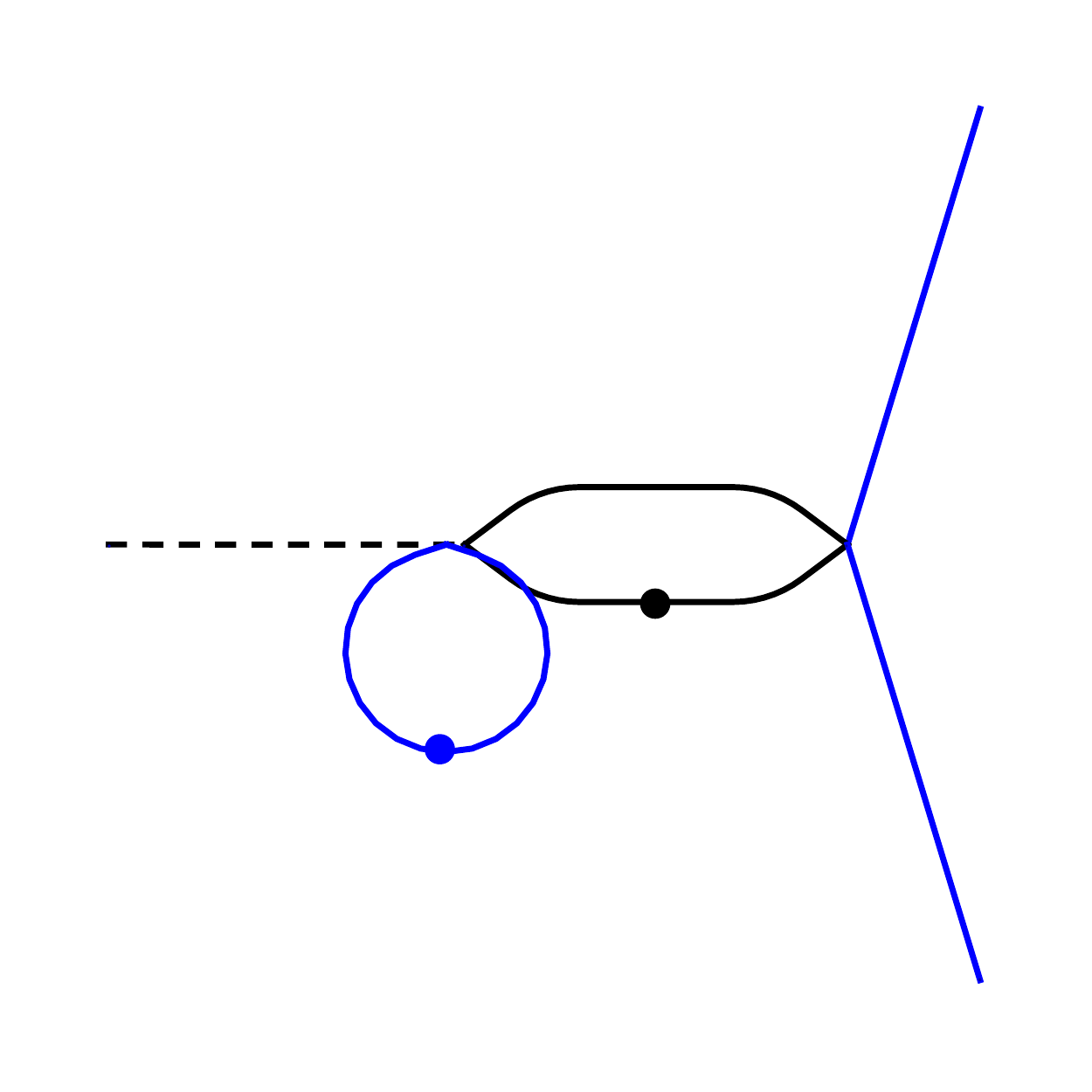}
	}
	\subfloat[\hspace{1.2cm}$\mathcal{T}_6$]{%
		\includegraphics[width=0.14\textwidth]{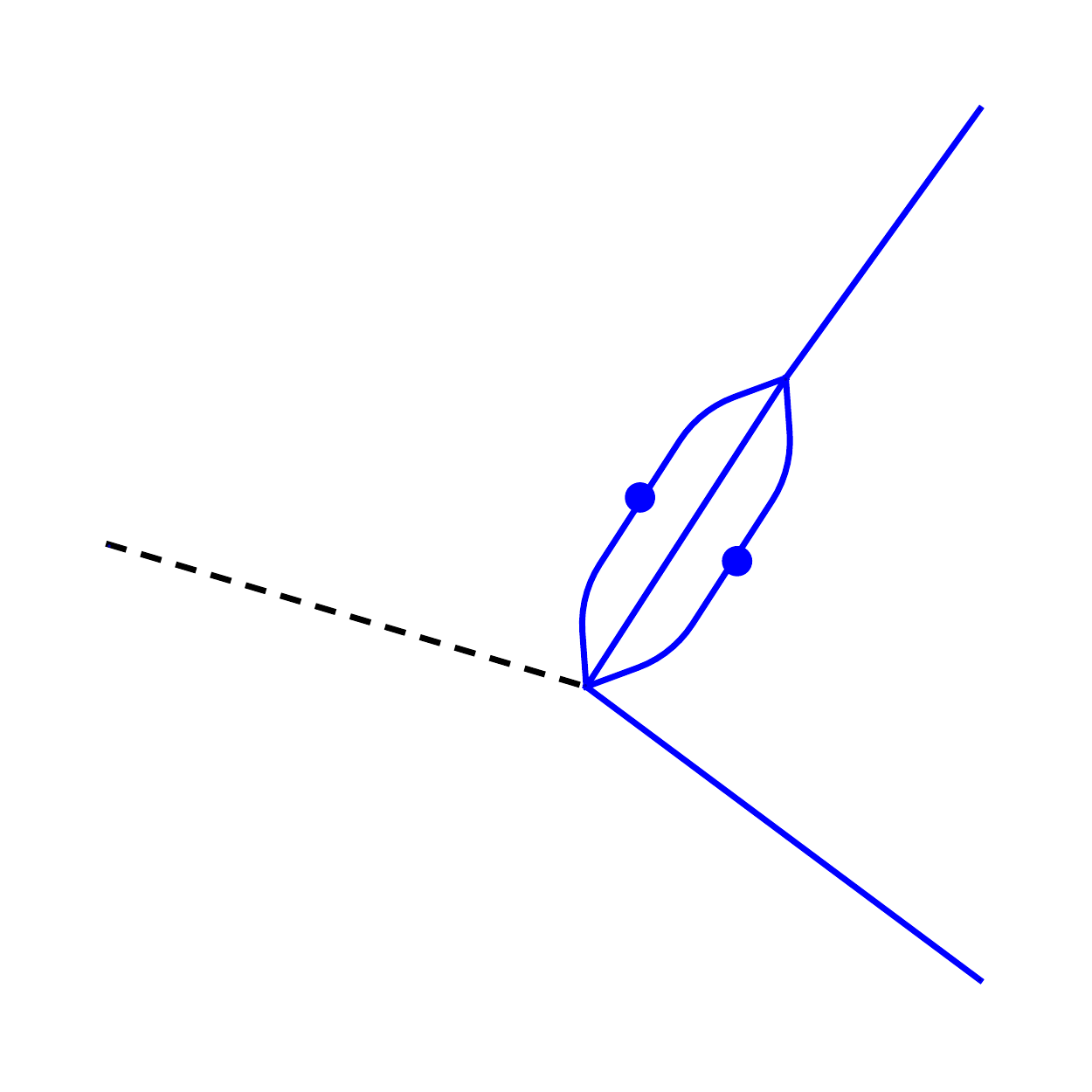}
	}
	\\
	\subfloat[\hspace{1.2cm}$\mathcal{T}_7$]{%
		\includegraphics[width=0.14\textwidth]{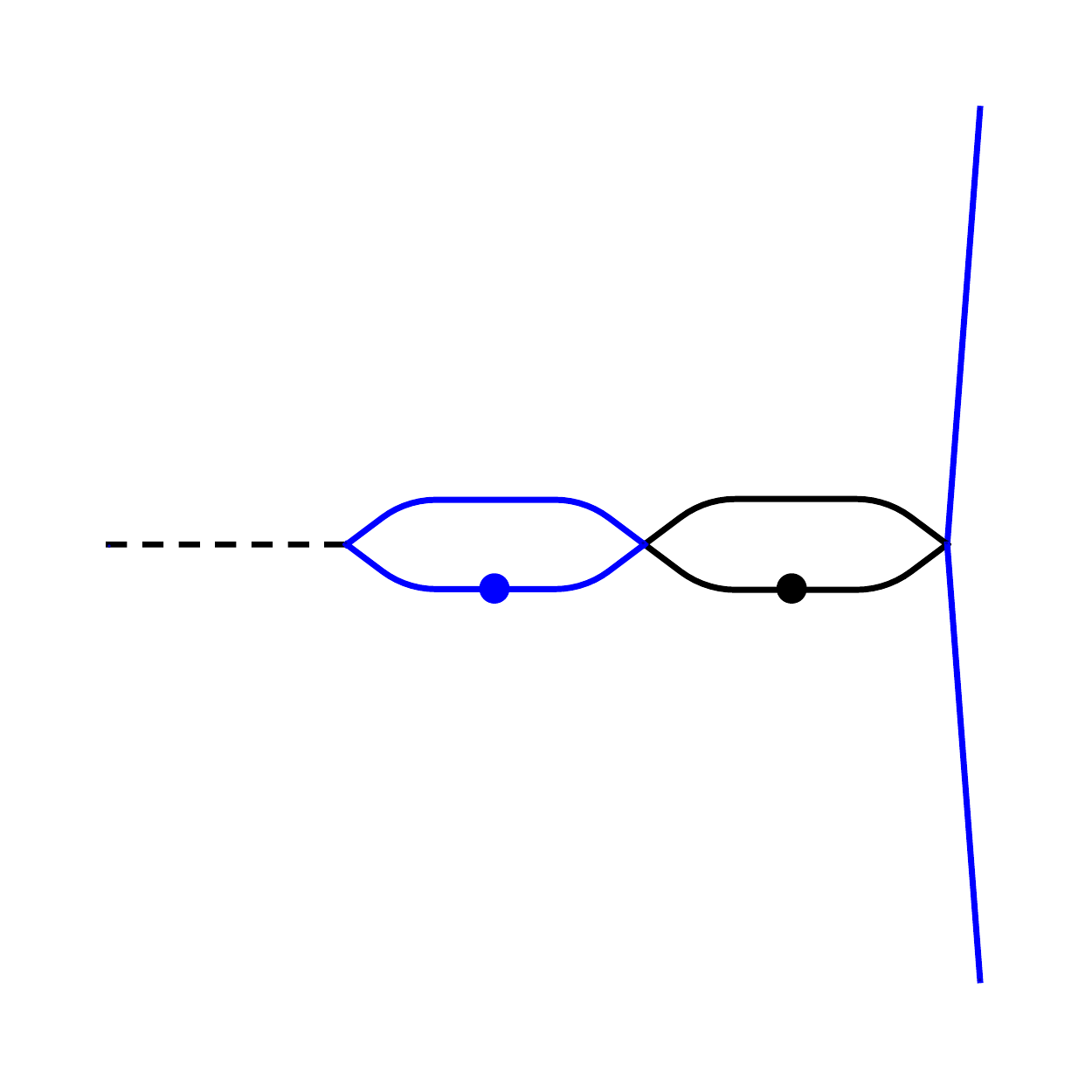}
	}
	\subfloat[\hspace{1.2cm}$\mathcal{T}_8$]{%
		\includegraphics[width=0.14\textwidth]{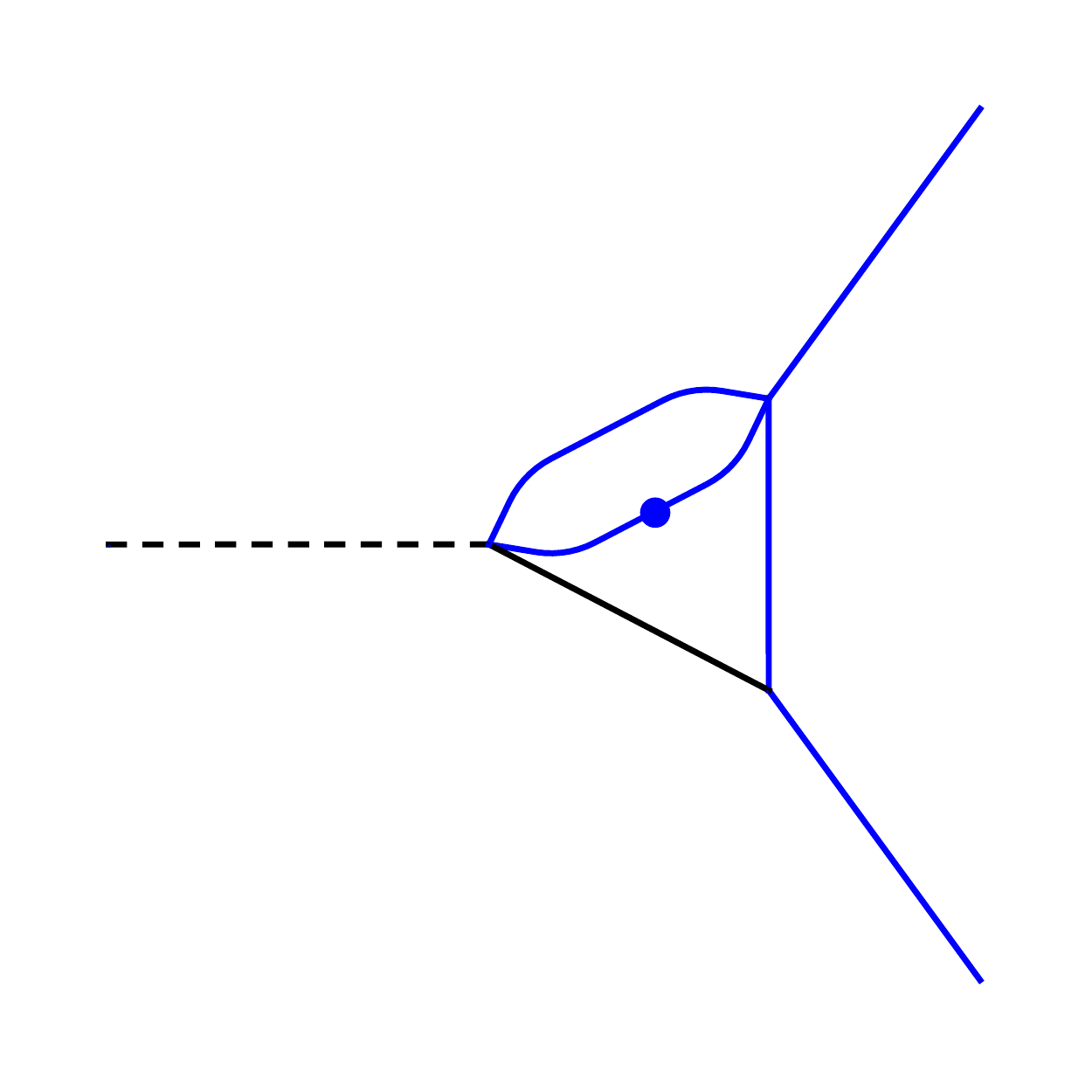}
	}
	\subfloat[\hspace{1.2cm}$\mathcal{T}_9$]{%
		\includegraphics[width=0.14\textwidth]{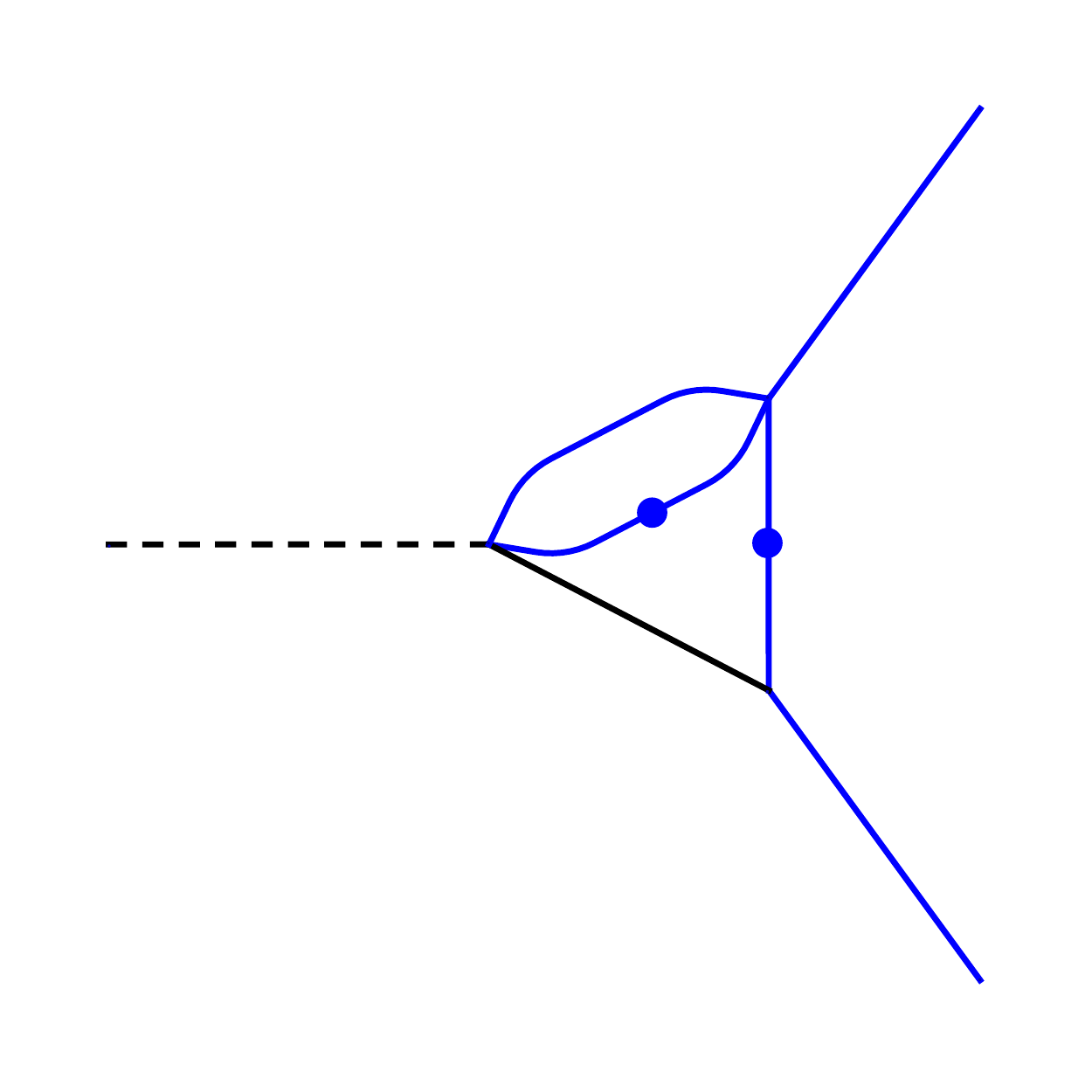}
	}
	\subfloat[\hspace{1.2cm}$\mathcal{T}_{10}$]{%
		\includegraphics[width=0.14\textwidth]{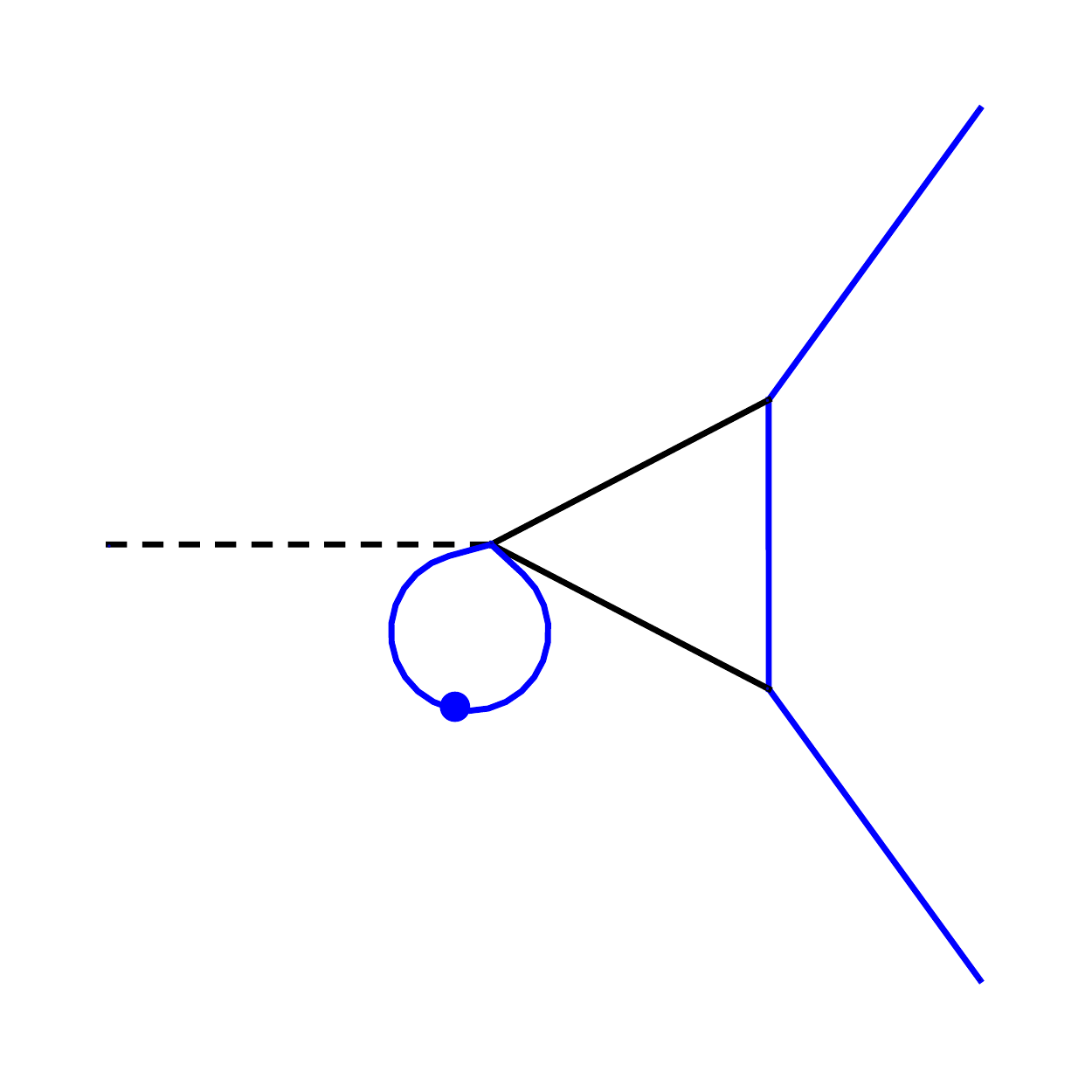}
	}
	\subfloat[\hspace{1.2cm}$\mathcal{T}_{11}$]{%
		\includegraphics[width=0.14\textwidth]{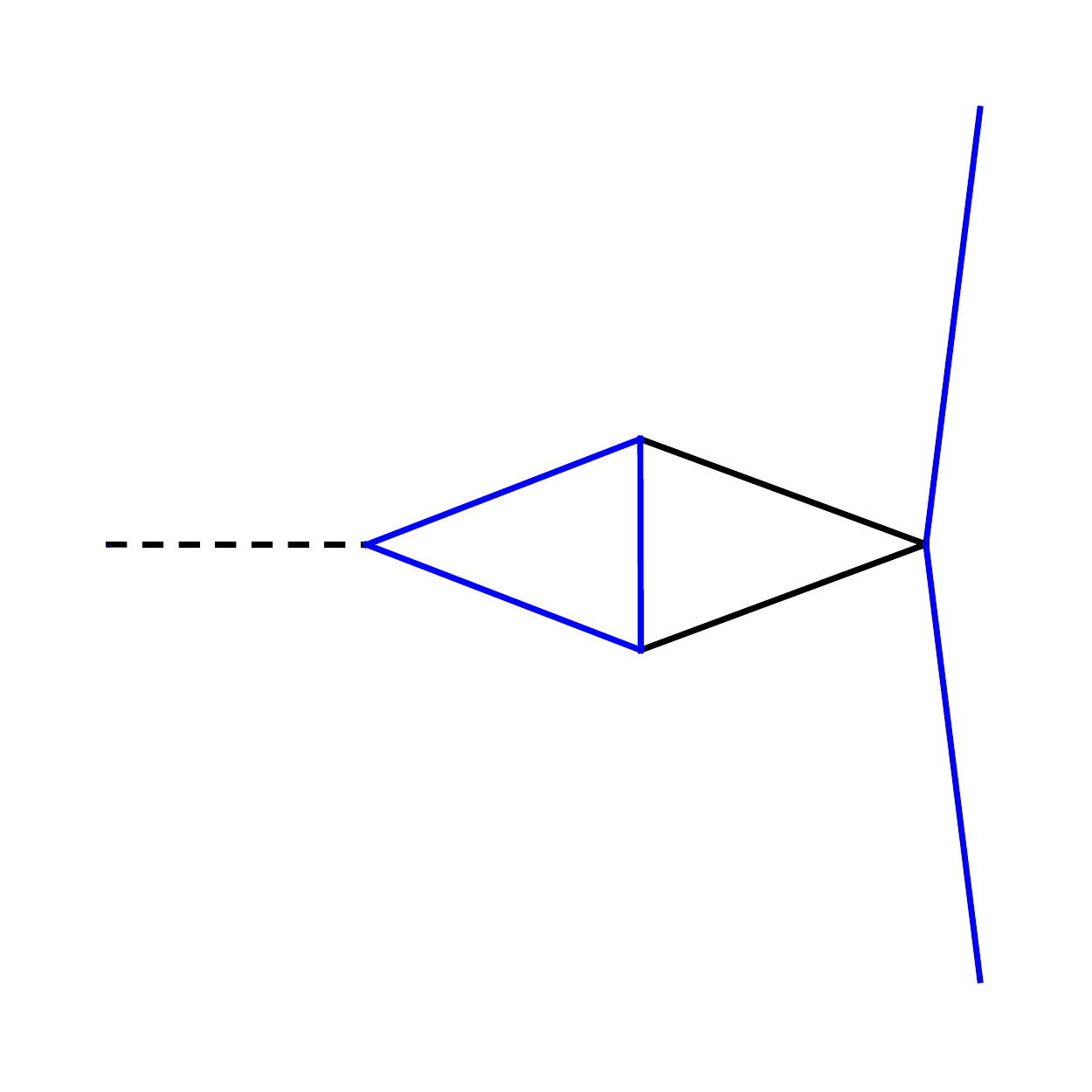}
	}
	\subfloat[\hspace{1.2cm}$\mathcal{T}_{12}$]{%
		\includegraphics[width=0.14\textwidth]{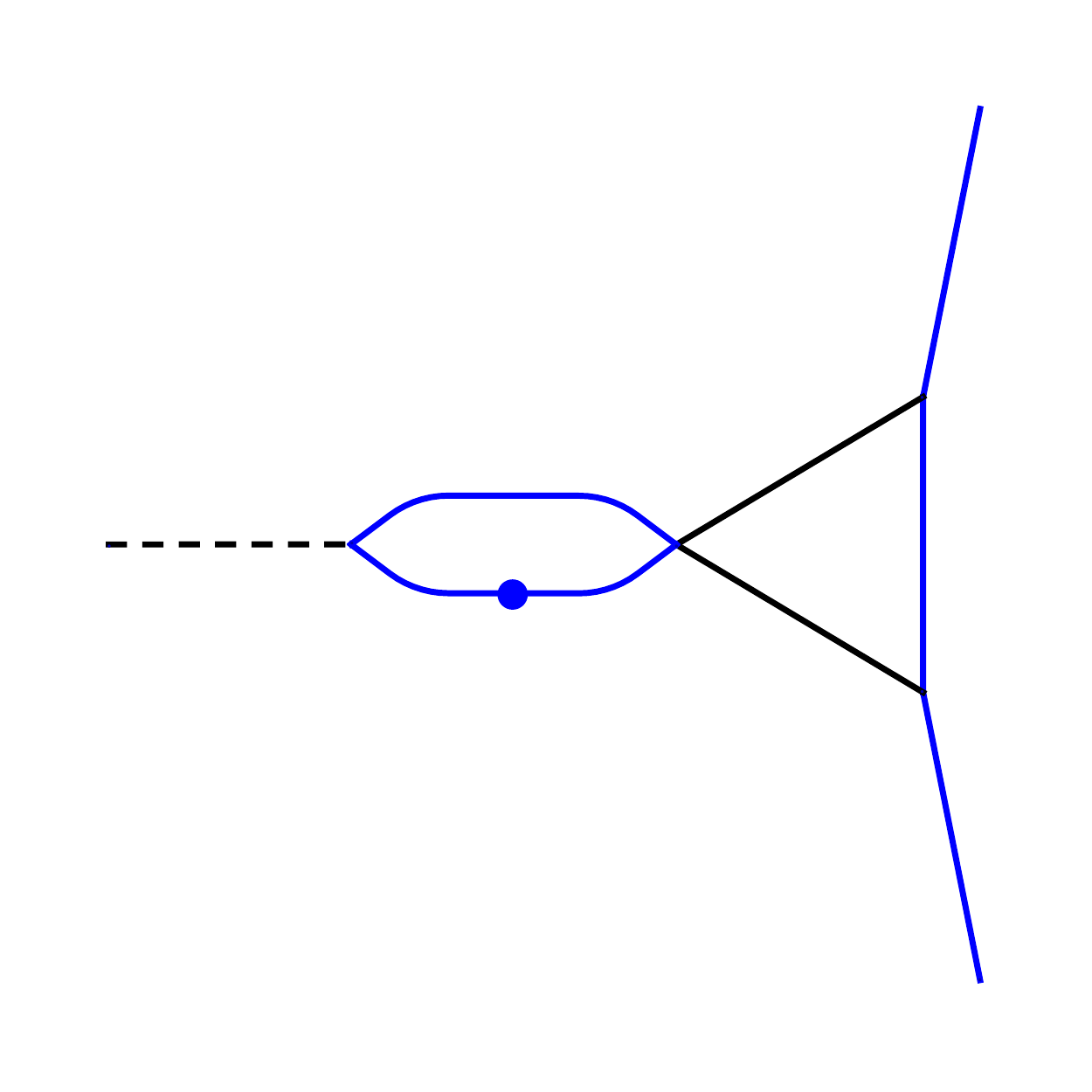}
	}
	\\
	\subfloat[\hspace{1.2cm}$\mathcal{T}_{13}$]{%
		\includegraphics[width=0.14\textwidth]{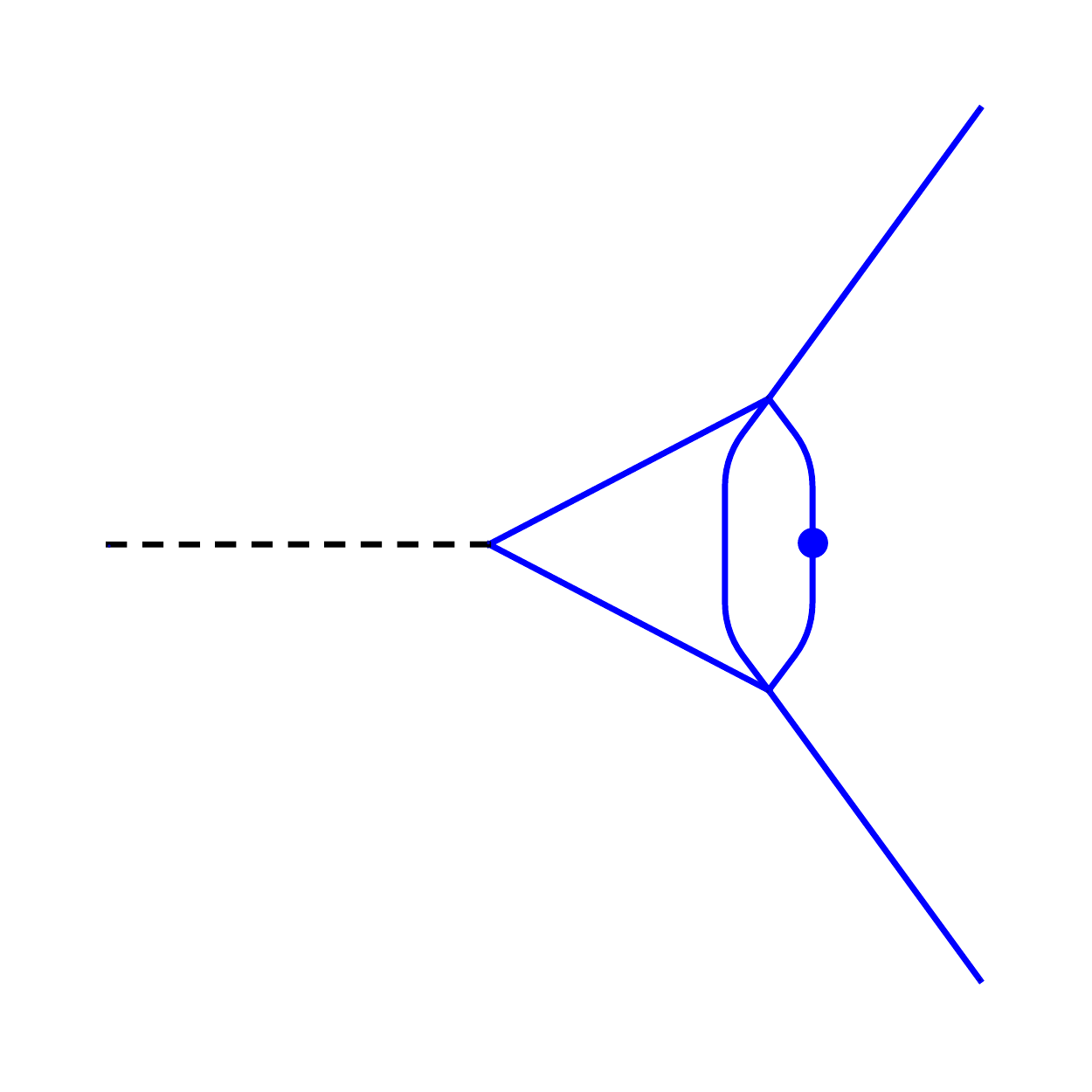}
	}
	\subfloat[\hspace{1.2cm}$\mathcal{T}_{14}$]{%
		\includegraphics[width=0.14\textwidth]{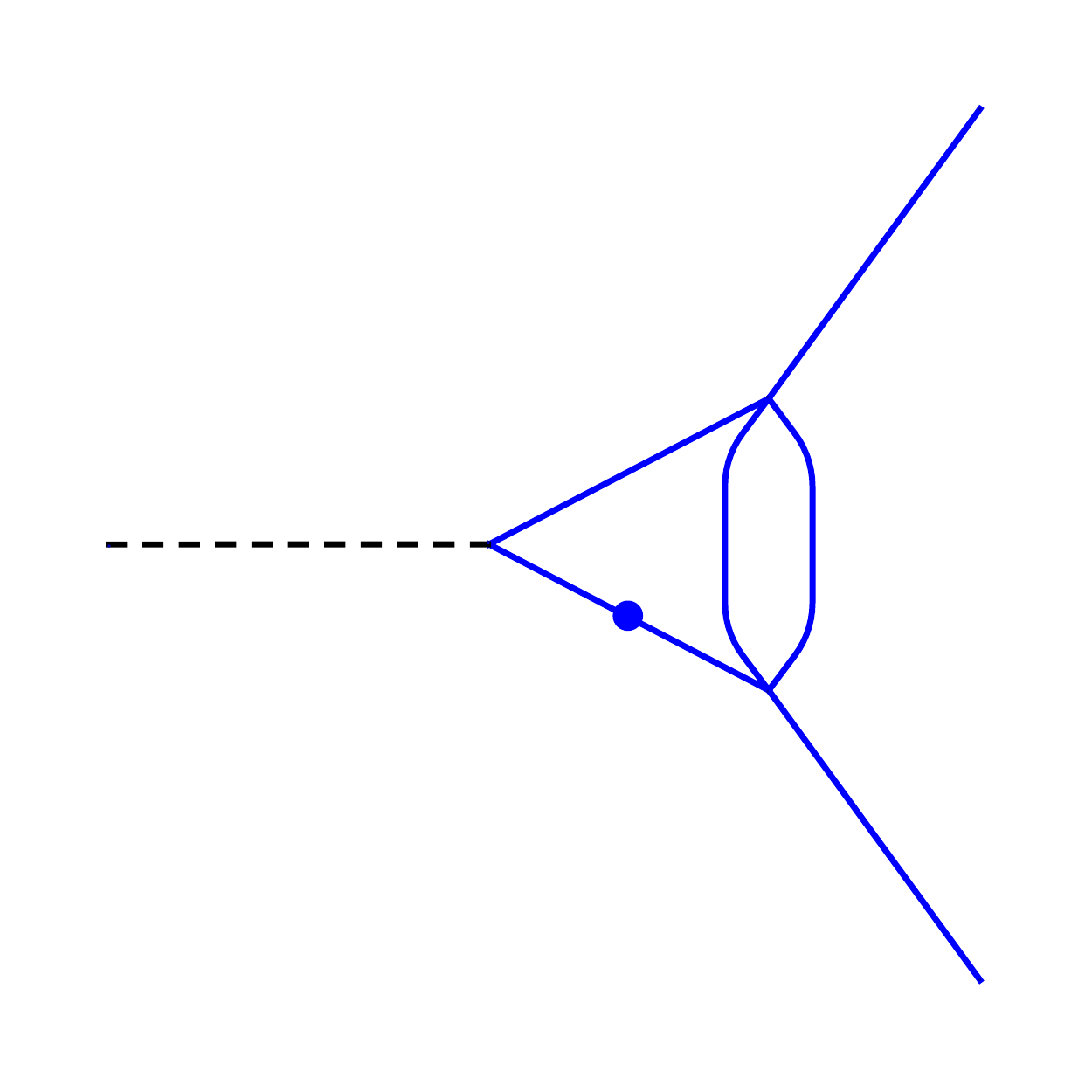}
	}
	\subfloat[\hspace{1.2cm}$\mathcal{T}_{15}$]{%
		\includegraphics[width=0.14\textwidth]{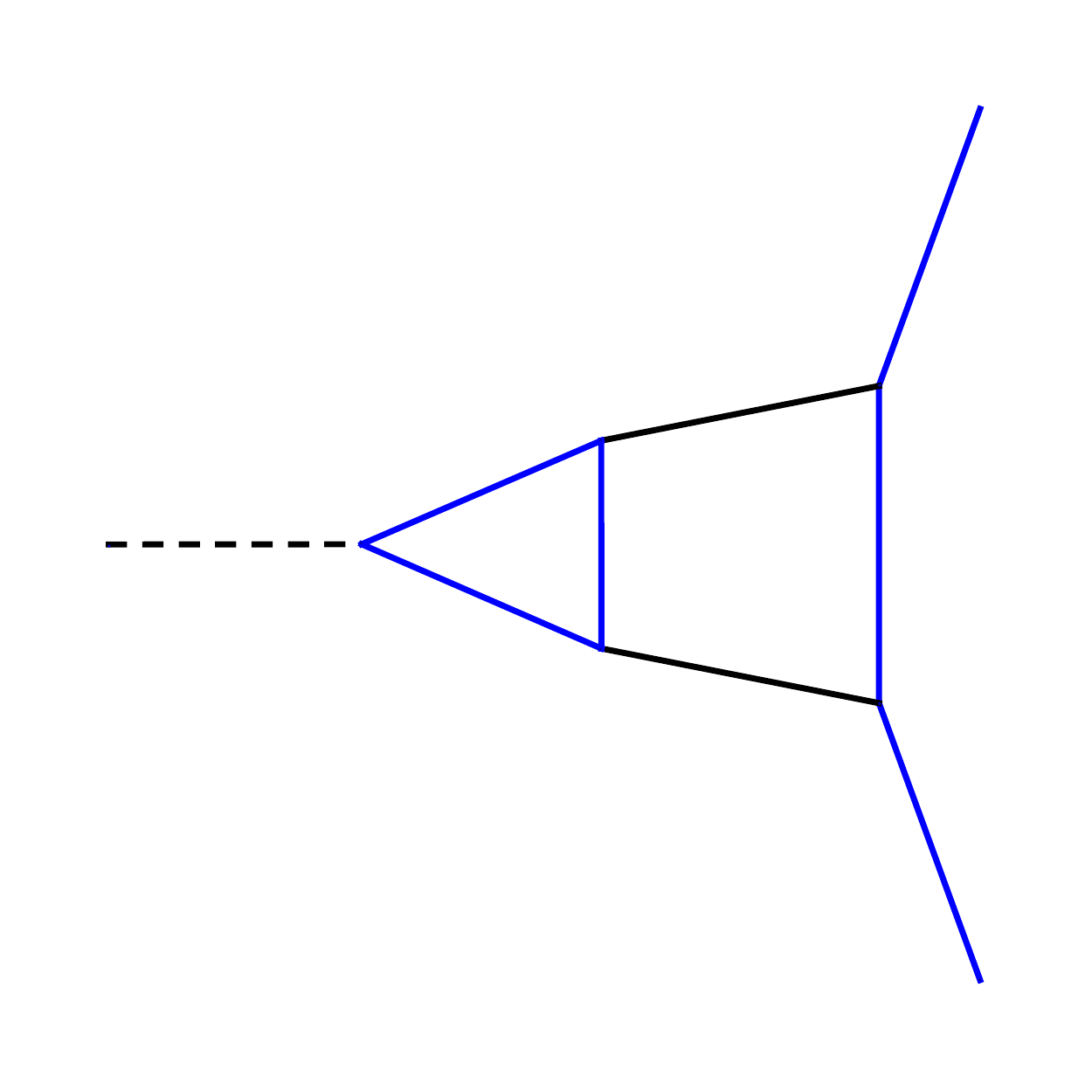}
	}
	\caption{\textbf{\textit{Master integrals required for the evaluation of the scalar and pseudo-scalar form factors}}, corresponding to the two-loop diagrams in Fig.~\ref{fig:2photon}, in the equal-mass case. In each diagram, we indicate the off-shell external mediator (dashed lines), massive fermions with mass $m$ (blue lines), massless photon (black lines) and squared denominators (dots).}
	\label{fig:MIsTI_isomass}
\end{figure}

The integral measure is defined as in Eq.~(\ref{eq:integral_measure_definition_appendix}), with the replacement $m^2_{\ell} \to m^2$. The Taylor coefficients of the MIs, $\mathbf{I}^{(k)}(w)$, can be obtained as described in Eqs.~(\ref{eq:canonical_MIs_Taylor_expansion_appendix}) and~(\ref{eq:I_k_iterated_integrals_appendix}). In this case, the total differential $d \mathbb{A}(w)$ reads
\begin{equation}
	d \,  \mathbb{A} = \sum_{i=1}^{3} \mathbb{M}_i \, d \, \log \left( \eta_i \right) ~,
\end{equation}
where, once again, $\mathbb{M}_i$ are constant matrices and $n_i$ reduce to 
\begin{equation}
	\eta_1 = w ~, \qquad \eta_2 = 1 + w ~, \qquad \eta_3 = 1 - w ~.
\end{equation}
Thus, the iterated integrals introduced in Eq.~(\ref{eq:GPLs_def_appendix}) are written in terms of $a_i = \pm 1, 0$, yielding a more restricted class of functions, the \emph{Harmonic PolyLogarithms} (HPLs) \cite{Remiddi:1999ew}. As in the general case, the fixing of the boundary terms in Eq.~(\ref{eq:I_k_iterated_integrals_appendix}) deserves a few comments.

\begin{itemize}
	\item The integrals $\ensuremath \text{I}_{1,5,6}$ are provided as an external input. In particular, $\ensuremath \text{I}_{1,5}$ are obtained by direct integration
	\begin{equation}
		\begin{split}
			& \ensuremath \text{I}_{1} = 1 ~, \nonumber\\
			& \ensuremath \text{I}_{5}(w, \epsilon) = \left( \frac{(1 - w)^2}{w} \right)^{-\epsilon} \, \left(1 - \frac{\pi^2}{6} \, \epsilon^2 - 2 \, \zeta_3 \, \epsilon^3 - \frac{\pi^4}{40} \, \epsilon^4 + \mathcal{O}(\epsilon)^5 \right) ~, 
		\end{split}
	\end{equation}
	while $\ensuremath \text{I}_{6}$ is obtained from the general case as~\cite{Argeri:2002wz}
	\begin{equation}
		\text{I}_{6}(\epsilon) = 
		-\frac{\pi^2}{12} \, \epsilon^2 + \frac{1}{4} \, \left(2 \, \pi^2 \, \log(2) - 7 \, \zeta_{3}\right) \, \epsilon^3 + \frac{1}{360} \, \left(31 \, \pi^4 - 180 \, \log^4(2) - 360 \, \pi^2 \, \log^2(2) - 4320 \, \text{Li}_4\left(\frac{1}{2}\right) \right) \, \epsilon^4 + \mathcal{O}(\epsilon^5) ~.
	\end{equation}
	
	\item The integral $\ensuremath \text{I}_{12}$ is obtained as the product of $\ensuremath \text{I}_{2}$ and $\ensuremath \text{I}_{10}$.
	
	\item The boundary constants for $\ensuremath \text{I}_{2,3,4,8,9,11,13,14}$ are determined by the regularity at the pseudo-threshold $t\to 0$. In particular $\ensuremath \text{I}_{2,3,4,8,11,13}$ vanish in this limit due to the prefactors in Eq.~(\ref{eq:canonical_MIs_equal_mass_appendix}). Analyzing the differential equation in the $t \to 0$ limit, we obtain relations among MIs in this limit, namely $\ensuremath \text{I}_{9} \left. \right|_{t \to 0} = 3 \, \ensuremath \text{I}_{6}$ and $\ensuremath \text{I}_{14} \left. \right|_{t \to 0} = 6 \, \ensuremath \text{I}_{6}$; the latter are sufficient to determine the boundary constants for $\ensuremath \text{I}_{9,14}$. $\ensuremath \text{I}_{7}$ vanishes due to the factorized canonical bubble.
	
	\item The boundary constants for $\ensuremath \text{I}_{10}$ are determined from the regularity at the pseudo-threshold $t \to 4 \, m^2$. In particular due to the prefactor in Eq.~(\ref{eq:canonical_MIs_equal_mass_appendix}), $\ensuremath \text{I}_{10}$ vanishes in this limit.
	
	\item The boundary constants for $\ensuremath \text{I}_{10}$ are obtained by comparing our results with those in Ref.~\cite{Bonciani:2003hc}.
\end{itemize}

Finally, we stress that Eq.~(\ref{eq:DEQ_Isomass_appendix}) is solved in the region $0 < w < 1$, which implies $t < 0 \, \wedge \, m^2>0$, where $\eta_i$'s are positive and the MIs are real-valued. In such a regime, Eq.~(\ref{eq:landau_vars_appendix}) can be inverted, yielding 
\begin{equation}
	w = \frac{\sqrt{4 \, m^2 - t} - \sqrt{-t}}{\sqrt{4 \, m^2 - t} + \sqrt{-t}} ~.
\end{equation}
\section{Limits of the scalar form factor,  $\boldsymbol{\mathcal{F}_{\mathcal{S}}^{\rm 1b}(t;m_N, m_{\ell})}$}
\label{sec:FS_limit}

The expressions of the form factors for arbitrary kinematics are too long to be shown here. Nevertheless, there are two limits which render manageable analytical expressions. We hereby present the results for these two limiting behaviors: the soft (i.e., $t \to 0$) and the equal-mass ($m_N \to m_\ell$) limits.

\subsection{Soft limit: $\boldsymbol{\mathcal{F}_{\mathcal{S}}^{\rm 1b}(t=0;m_N, m_{\ell})}$}
\label{subsection_FS_t_to_0}

\begin{figure}[t]
\centering
\includegraphics[scale=0.3]{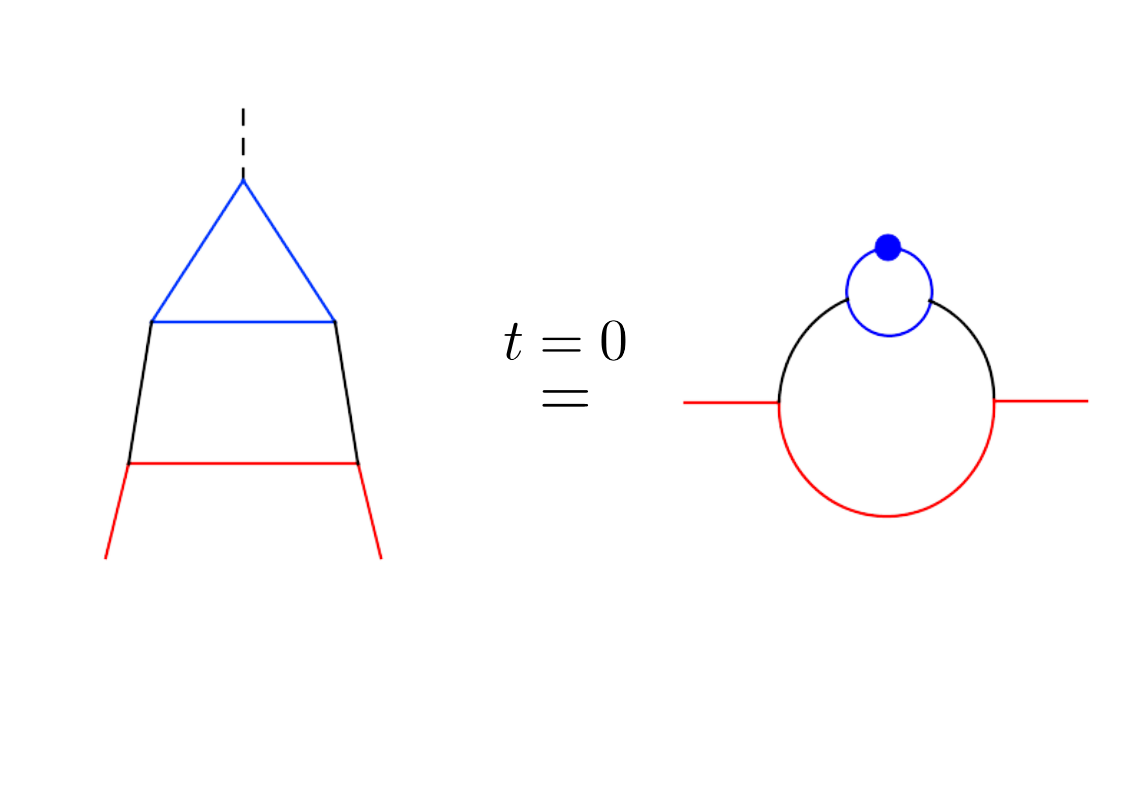}
\caption{\textbf{\textit{Diagrammatic representation of the $\boldsymbol{q^\mu \to 0}$ limit for the one-body scalar form factor, $\boldsymbol{{\cal F}_S^{\rm 1b}}$}}. At zero momentum transfer, no momentum flows into the diagram through the dashed line, which just corresponds to the insertion of $\Lambda_{S} = \mathbb{I}$ in the fermionic trace.}
\label{fig:t0diag}
\end{figure}

\begin{figure}
  \centering
  \captionsetup[subfigure]{labelformat=empty}
  \subfloat[\hspace{1.1cm}$\mathcal{T}_1$]
  {%
    \includegraphics[width=0.14\textwidth]{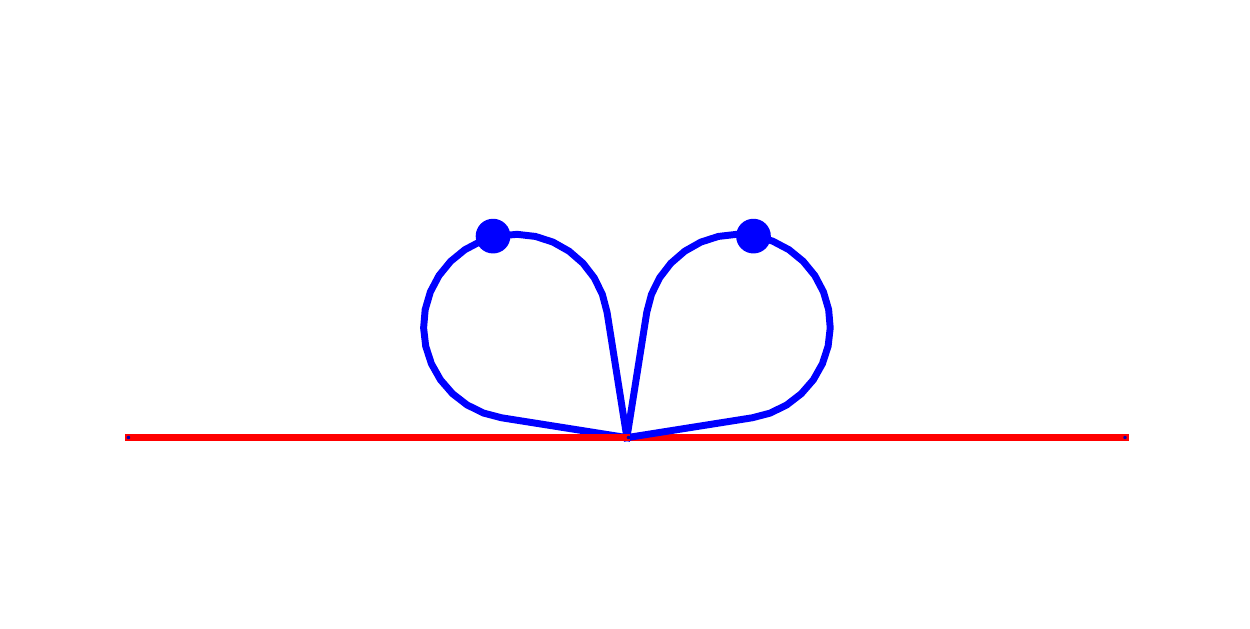}
  }
  \subfloat[\hspace{1.1cm}$\mathcal{T}_2$]{%
    \includegraphics[width=0.14\textwidth]{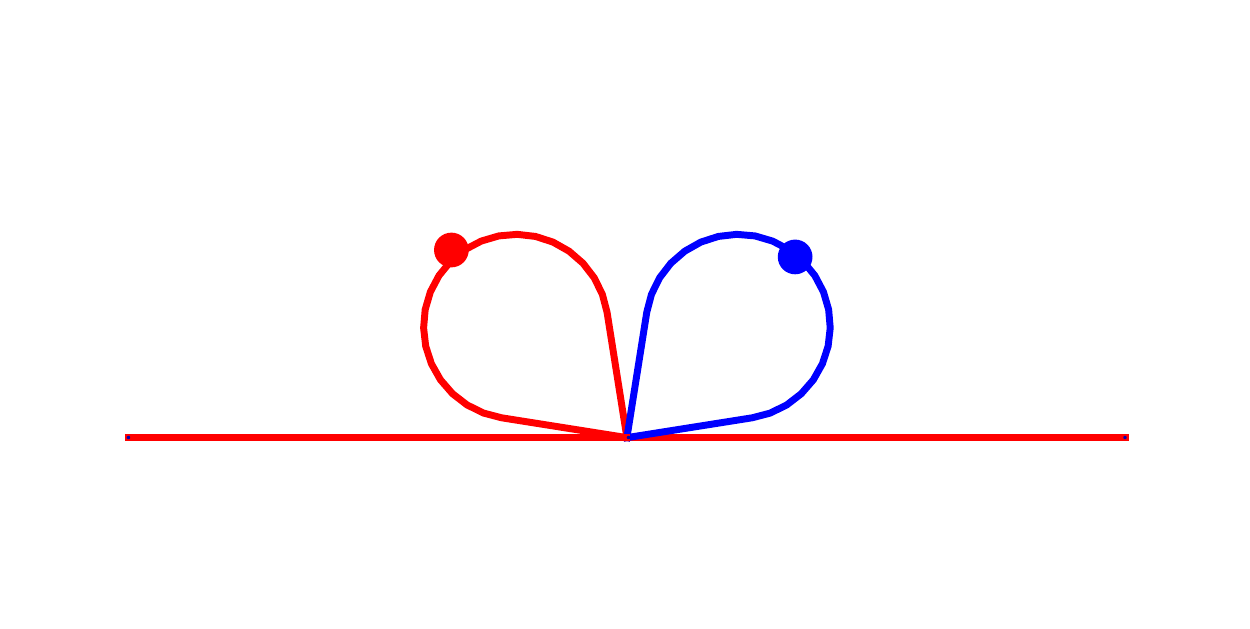}
  }
  \subfloat[\hspace{1.35cm}$\mathcal{T}_3$]{%
    \includegraphics[trim={0 2cm 0 0},clip,  width=0.17\textwidth]{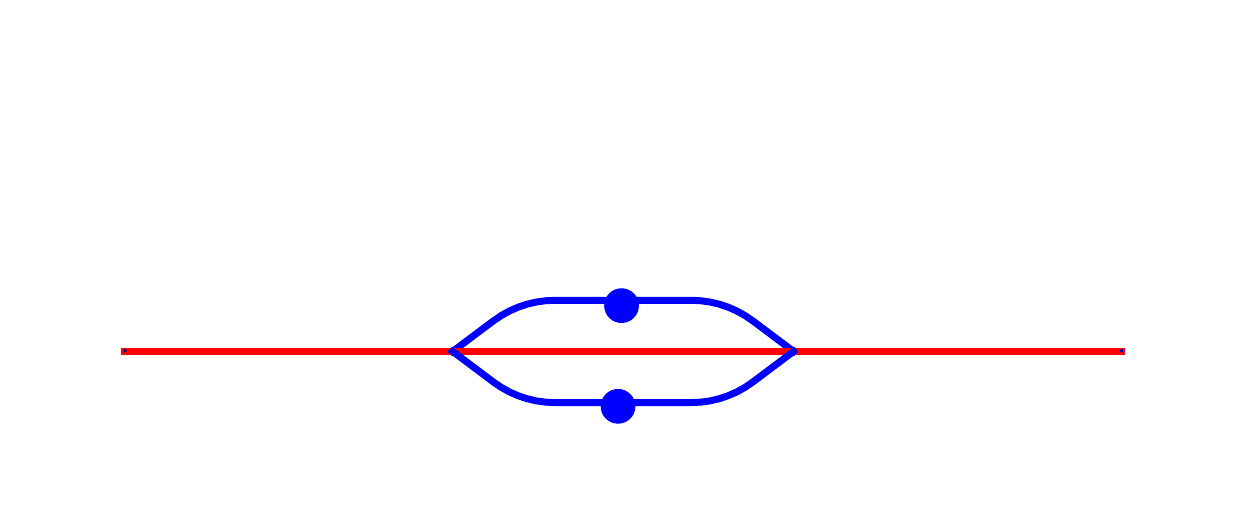}
  }
  \subfloat[\hspace{1.55cm}$\mathcal{T}_4$]{%
    \includegraphics[trim={0 1.8cm 0 0},clip,width=0.19\textwidth]{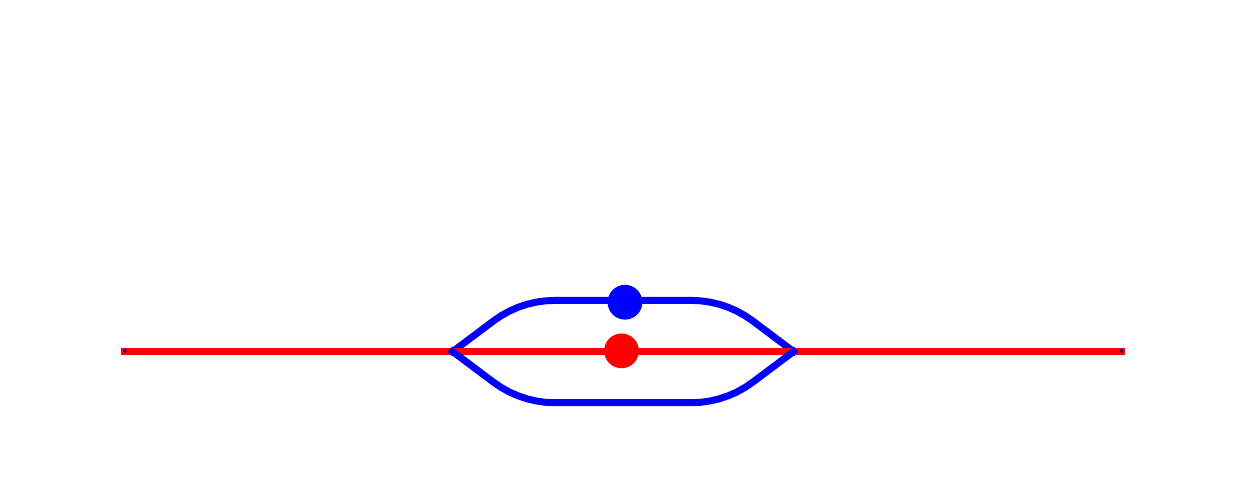}
  }
\caption{
  \textbf{\textit{Master integrals required for the evaluation of the scalar form factors in the soft limit, $\boldsymbol{t \to 0}$}}. The diagrams contain fermions with mass $m_\ell$ (blue lines) and $m_N$ (red lines). External fermions are on-shell.}
 \label{fig:MIsTI_FS_tat0}
\end{figure}

Setting $q^{\mu} = 0$ before the integration of the loop momenta has the advantage of drastically simplifying the whole IBPs reduction to MIs, as it removes the dependence on $t$ from very the beginning. In order to understand how the limit at the integrand level simplifies the reduction procedure, we observe that the limit $q^{\mu} \to 0$ of Eq.~\eqref{eq:Ds} leaves only four independent inverse propagators $D_i$ out of the six of the two-loop topology, since $D_2 \to D_1$ and $D_5 \to D_4$. This implies that, at $q^\mu = 0$, the decomposition of the form factor into Feynman integrals given in Eq.~\eqref{eq:family} reduces to
\begin{equation}
{\cal F}_S^{\rm 1b}\left(t = 0; m_N, m_\ell \right) = \sum_{\vec{n}} c_{\vec{n}\,j}(m_i^2,\eps)\int \frac{{\dd^d k_1}}{(2\pi)^d} \frac{{\dd^d k_2}}{(2\pi)^d} \,
\frac{D_7^{n_7}}{\Den_{1}^{n_1}\Den_{3}^{n_3}\Den_{4}^{n_4}\Den_{6}^{n_6}}\,, \quad n_i\in \mathbb{N} ~,
\label{eq:familyap}
\end{equation}
with
\begin{gather}
\Den_1 = k_1^2 - m_\ell^2 ~,\quad \Den_3 = k_2^2 - m_N^2 ~, \quad \Den_4 = (k_2 + p)^2 ~, \quad \Den_6 = (k_1 + k_2+p)^2 - m_\ell^2 ~,
\label{eq:Dsap}
\end{gather}  
and where, as for arbitrary $t$, the auxiliary denominator $\Den_7 = (k_1 - p)^2$ has been introduced so that all the five possible scalar products between the loop momenta $k_{1,2}$ and the external momentum $p$ can be expressed as linear combinations of $D_i$. Therefore, the $q^{\mu} \to 0$ limit can be understood, at a diagrammatic level, as the reduction of the two-loop vertex function to the self-energy-like diagram depicted in Fig.~\ref{fig:t0diag}. For this case we identify $4$ MIs, which can be chosen as
\begin{alignat}{2}
\ensuremath \text{I}_1 = & \epsilon^2 \, \mathcal{T}_1 ~, \hspace{2cm} && 
\ensuremath \text{I}_2 = \epsilon^2 \, \mathcal{T}_2 ~, \nonumber\\[2ex] 
\ensuremath \text{I}_3 = & \epsilon^2 \, m_{\ell} \, m_{N} \left(\mathcal{T}_3+2 \, \mathcal{T}_4 \right) ~, \hspace{2cm} && 
\ensuremath \text{I}_4 = \epsilon^2 \, m^2_N \mathcal{T}_4 ~,
\label{eq:FS0masters}
\end{alignat}
where $\mathcal{T}_{i}$'s are depicted in Fig.~\ref{fig:MIsTI_FS_tat0}. These MIs obey a canonical system of differential equations with respect to the variable $z_{N \ell} \equiv m_{N} / m_{\ell}$, which can be solved in terms of HPLs. Normalizing the MIs as in Eq.~(\ref{eq:integral_measure_definition_appendix}), the boundary conditions are obtained in the following way:
\begin{itemize}
    \item $\ensuremath \text{I}_1$ and $\ensuremath \text{I}_2$ are obtained by direct integration
    \begin{eqnarray}
    \ensuremath \text{I}_1 & =& 1 ~, \nonumber\\[2ex] \text{I}_2(z,\epsilon) & = & z^{-2 \, \epsilon} ~. \nonumber
    \end{eqnarray}
    
    \item The boundary conditions for $\ensuremath \text{I}_3$ and $\ensuremath \text{I}_4$ are determined from to their regularity at $m_N \to 0$ \cite{Argeri:2002wz}. In particular,  thanks to the prefactors in Eq.~(\ref{eq:FS0masters}), they vanish in this limit.
\end{itemize}

The form factor ${\cal F}_S^{\rm 1b}(t = 0; m_N, m_\ell)$ is then reduced via IBPs to the following linear combination of MIs,
\begin{align}
{\cal F}_S^{\rm 1b}\left(t = 0; m_N, m_\ell \right) = - \frac{128 \, \pi^2 \, \alpha_{\text{em}}^2 \, m_{\ell}}{\epsilon^2 \, \left(\epsilon + 1\right) \, \left(1 - 2 \, \epsilon\right) \, \left(1 - 3 \, \epsilon\right) \, m_N} \, \Big[& \epsilon \,  \left(\epsilon \, \left(2 \, \epsilon - 1\right) + 1\right) \, \GG_1 + \epsilon \, \left(\epsilon \, \left(2 \, \epsilon - 3\right) - 1\right) \, \GG_2 \nonumber\\
& + \left(\frac{m_N}{m_{\ell}} \, \left(2 \, \epsilon + 1\right) \, \left(3 \, \epsilon - 1\right) - \frac{m_{\ell}}{m_N} \, \left(4 \, \epsilon^3 - 4 \, \epsilon^2 - \epsilon + 3\right) \right) \, \GG_3 \nonumber\\
& + \left(\frac{m_{\ell}^2}{m_N^2} \, 4 \, \left(\epsilon - 1\right) \, \left(\epsilon + 1\right) \, \left(2 \, \epsilon - 1\right) - 2 \, \epsilon \, \left(\epsilon \, \left(2 \, \epsilon + 3\right) - 3\right) \right) \, \GG_4 \Big] ~.
\label{eq:FS0masters}
\end{align}

After the insertion of the analytic expression of the four MIs, rewritten as functions of the mass ratio $z_{N\ell} \equiv m_N/m_\ell$ only, restoring the original integral measure and  expanding Eq.~\eqref{eq:FS0masters} around four dimensions, we obtain 
\begin{eqnarray}
{\cal F}_S^{\rm 1b}\left(t = 0; z_{N\ell} \equiv \frac{m_N}{m_\ell}\right)  &= & -\frac{2 \, \alpha_{\rm em}^2}{\pi^2\, z_{N\ell}} \Big\{ 1  - \frac{\ln (z_{N\ell})}{2} + f_S(z_{N\ell}) + f_S(- z_{N\ell}) \Big\} ~, \\[2ex]
\textrm{with} \hspace{7mm} 
f_S(\tau) & \equiv & \frac{1}{4 \,\tau^2} \left(4 + 3 \, \tau + \tau^3\right) \Big[ \ln|\tau| \ln(1 + \tau) + {\rm Li}_2(-\tau) \Big] ~,
\label{eq:formSt0ap}
\end{eqnarray}
as reported in Eq.~\eqref{eq:formSt0}.

\subsection{Equal-mass limit: $\boldsymbol{\mathcal{F}_{\mathcal{S}}^{\rm 1b}(t;m,m)}$}

In this limit, the results are expressed in terms of GPLs, $G( \vec{a},w) \equiv G_{\vec{a}}(w)$, where $w$ was introduced in Eq.~(\ref{eq:landau_vars_appendix}),
\begin{align}
\frac{\pi^2}{\aem} \, \mathcal{F}^{\rm 1b}_{\mathcal{S}}(w) = & - \frac{ (20 \, w^2 + 4 \, w)}{ \, (w-1) \, (w+1)^2} \,  G_0(w) 
\; + \; \frac{ w}{ \, (w-1) \, (w+1)} \, \big[24 \, G_{-1,0}(w) + 4 \, G_{1,0}(w) \big] \nonumber\\ 
& + \frac{ w}{ \, (w-1)^2 \, (w+1)^3} [-12 \, w^3 - 38 \, w^2 + 8 \, w + 2] \, G_{0,0}(w) 
\; - \; \frac{w}{\pi^2 \, (w-1)^2} \, \big[2 \, G_{0,0,0}(w) + 4 \, \zeta_3\big] \nonumber\\ 
& + \frac{ w \, \zeta_3}{ \, (w-1)^3 \, (w+1)} \, [5 \, w^2 - 6 \, w + 5] \, G_0(w)
\; + \; \frac{w}{\pi^2 \, (w+1)^2} \, \left[12 \, G_{0,0,1}(w) - \frac{4 \, \pi^2}{3} \, G_0(w) + 16 \, G_1(w) \right] \nonumber\\ 
& + \frac{ w}{ \, (w-1)^2 \, (w+1)^2} \, \Big[(-16 \, w^2 + 64 \, w - 16) \, G_{0,-1,0}(w) + (8 \, w^2 - 48 \, w + 8) \, G_{1,0,0}(w) + 16 \, w \, G_{0,1,0}(w) \Big] \nonumber\\ &
+ \frac{ w}{ \, (w-1) \, (w+1)^3} \Bigg[(-14 \, w^2 + 4 \, w - 14) \, G_{0,1}(w) + \left(\frac{w^2}{3} + 2 \, w + \frac{1}{3}\right) \, G_{1,0}(w) \, \pi^2 \nonumber\\ 
& + (2 \, w^2 + 12 \, w + 2) \, \big(G_{1,0,0,0}(w) \, - \, G_{0,0,1,0}(w)\big) + \left(2 \, w^2 - \frac{20 \, w}{3} + 2 \right) \pi^2 \Bigg] \nonumber\\
& + \frac{ w}{ \, (w-1)^3 \, (w+1)^3} \Bigg[\left(\frac{5 \, w^4}{6} - 2 \, w^3 + \frac{23 \, w^2}{3} - 2 \, w + \frac{5}{6}\right) \, G_{0,0}(w) \, \pi^2 \nonumber\\
& + \left(8 \, w^4 - 32 \, w^3 + 112 \, w^2 - 32 \,  w + 8\right) \, G_{0,0,-1,0}(w) + \left(\frac{3 \, w^4}{2} - 2 \, w^3 + 9 \, w^2 - 2 \, w + \frac{3}{2}\right) \, G_{0,0,0,0}(w) \nonumber\\
& + \left(-9 \, w^4 + 12 \, w^3 - 54 \, w^2 + 12 \, w - 9\right) \, G_{0,0,0,1}(w) + \left(-2 \, w^4 + 24 \, w^3 - 76 \, w^2 + 24 \, w - 2\right) \, G_{0,1,0,0}(w) \nonumber\\
& + \left(\frac{w^4}{18} - \frac{2 \, w^3}{45} + \frac{11 \, w^2}{45} - \frac{2 \, w}{45} + \frac{1}{18} \right) \, \pi^4 \Bigg] ~.
\end{align}

This form factor is related, by crossing symmetry, to the one appearing in the amplitude for a (scalar) Higgs boson decaying into a heavy-quark pair~\cite{Bernreuther:2005gw}, which contains the expression of the total form factor, due to the contribution of several vertex diagrams, in $s$-channel. 
The expression given above is the contribution of a single diagram, namely the vertex diagram with a closed triangle-loop of a massive fermion, in $t$-channel, and it is shown here, for the first time.

\section{Limits of the pseudo-scalar form factor,  $\boldsymbol{\mathcal{F}_{\mathcal{P}}^{\rm 1b}(t;m_N, m_{\ell})}$}
\label{sec:FP_limit}

As for the scalar case, we provide here the expressions for the soft (i.e., $t \sim 0$) and the equal-mass (i.e., $m_N \to m_{\ell}$) limits for the pseudo-scalar form factor.

\subsection{Soft asymptotic limit: $\boldsymbol{\mathcal{F}_{\mathcal{P}}^{\rm 1b}(t \sim 0;m_N, m_{\ell})}$}

We consider an asymptotic expansion around $t \sim 0$ for the full integrated result, by exploiting the algebraic properties and functional relations among GPLs. The results are expressed in terms of the dimensionless variables $(x,y)$ introduced in Eq.~(\ref{eq:vars}). Notice that $t \sim 0$ and fixed $m_{N, \ell}$ corresponds to $(x,y) \sim (1,1)$. The ordering $x<y$ or $y<x$ is kept fixed in this limit, and depends on the fermion mass hierarchy. 

For the mass hierarchy $m_l^2 < m_N^2$ (i.e., $0 < x < y$):
\begin{align}
\frac{\pi^2}{\aem}\, \mathcal{F}_{\mathcal{P}}^{\rm 1b}(t \sim 0; m_N > m_{\ell}) = & - \frac{\pi^2}{3} + \frac{(1-x)}{ \, (1-y)} \,\big[- 2 \, \log (1-x) - 2 \, \log(2) + 3 \big] + \frac{(1-x)^2}{ \, (1-y)} \, \left[- \log(1-x) - \log(2) + \frac{1}{2} \right] \nonumber\\ 
& + (1-x)^2 \, \left[\frac{\log(1-x)}{2} + \frac{\log(2)}{2} + \frac{37 \, \pi^2}{360} - \frac{1}{4} \right] + \frac{(1-x)}{\pi^2} \, \left[ \log(1-x) + \log(2) + \frac{\pi^2}{4} - \frac{3}{2}\right] \nonumber\\
&  + \mathcal{O}\big((1-y) , \; (1-x)^3 \big) ~. 
\end{align}
Here, $\mathcal{O}\big((1-y)^n, \; (1-x)^m \big)$ indicates that the expression was first expanded in $y$ to order $n-1$, and then expanded in $x$ to order $m-1$.

For the mass hierarchy $m_N^2 < m_l^2$ (i.e., $0 < y < x$):
\begin{align}
\frac{\pi^2}{\aem} \, \mathcal{F}_{\mathcal{P}}^{\rm 1b}(t \sim 0; m_N < m_{\ell}) = & \frac{(1-x)}{ \, (1-y)} \, \left[- \frac{3\log (1-x)}{2} -\frac{\log(1-y)}{2} - 2 \, \log(2) + \frac{17}{8} \right] \nonumber\\ 
& + (1-x) \, \left[ \frac{3 \, \log (1-x)}{4} + \frac{\log(1-y)}{4} + \log(2) + \frac{\pi^2}{4} - \frac{21}{16} \right] + \mathcal{O}\big((1-x)^2 , \; (1-y) \big) ~. 
\end{align}
Analogously, $\mathcal{O}\big((1-x)^n, \; (1-y)^m \big)$ indicates that the expression was first expanded in $x$ to order $n-1$, and then expanded in $y$ to order $m-1$.

\subsection{Equal-mass limit: $\boldsymbol{\mathcal{F}_{\mathcal{P}}^{\rm 1b}(t;m,m)}$}

The results for the equal-mass limit are expressed in terms GPLs: $G( \vec{a},w) \equiv G_{\vec{a}}(w)$, where $w$ was introduced in Eq.~(\ref{eq:landau_vars_appendix}),

\begin{align}
\frac{\pi^2}{\aem} \, \mathcal{F}^{\rm 1b}_{\mathcal{P}}(w) = & - \frac{ w^2}{ \, (w-1)^3 \, (w+1)} \, \left[ \frac{2\pi^2}{3} \, G_0(w) + 4 \, G_{0,0,0}(w) \right] \nonumber\\ 
& + \frac{ w}{ \, (w-1)^2} \, \left[ G_{0,0}(w) - 4 \, G_{0,0,1}(w) - 4 \, G_{0,1,0}(w) + 8 \, G_{1,0,0}(w) + 12 \, \zeta_3 + \frac{\pi^2}{3} \right] \nonumber\\ 
& + \frac{ w}{ \, (w-1) \, (w+1)} \, \left[ - 3 \, \zeta_3 \, G_0(w) - \frac{\pi^2}{6} \, G_{0,0}(w) - \frac{\pi^2}{3} \, G_{1,0}(w) - \frac{1}{2} \, G_{0,0,0,0}(w) \right. \nonumber \\ 
& \hspace{3.15cm} \left. + 3 \, G_{0,0,0,1}(w) + 2 \, G_{0,0,1,0}(w) - 2 \, G_{0,1,0,0}(w) - 2 \, G_{1,0,0,0}(w) - \frac{\pi^4}{45} \right] ~. 
\end{align}

The calculation of the two-loop pseudo-scalar form factor in the $s$-channel was considered previously, for the two-loop QCD corrections to the heavy quark form factors (see the expression of $A_R$ in Ref.~\cite{Bernreuther:2005rw}). We have reproduced it using our MIs, which are, instead, defined for a $t$-channel process. The result shown above refers to a $t$-channel process, and it is presented here for the first time.
\newpage

\bibliography{biblio}

\end{document}